\documentclass[twocolumn]{aastex631}
\usepackage{hyperref}
\usepackage{graphicx}
\usepackage{natbib}
\usepackage{float}
\usepackage{amsmath}

\newcommand{\nt}{\textcolor{black}}

\shorttitle{Implementation of Mie Scattering Aerosols in \texttt{POSEIDON}}
\shortauthors{Mullens, Lewis, \& MacDonald}

\begin{document}

\title{Implementation of Aerosol Mie Scattering in \texttt{POSEIDON} with Application to the hot Jupiter HD~189733~b's Transmission, Emission, and Reflected Light Spectrum}

\author[0000-0003-0814-7923]{Elijah Mullens}
\affiliation{Department of Astronomy and Carl Sagan Institute, Cornell University, 122 Sciences Drive, Ithaca, NY 14853, USA}

\author[0000-0002-8507-1304]{Nikole K. Lewis}
\affiliation{Department of Astronomy and Carl Sagan Institute, Cornell University, 122 Sciences Drive, Ithaca, NY 14853, USA}

\author[0000-0003-4816-3469]{Ryan J. MacDonald}
\altaffiliation{NHFP Sagan Fellow}
\affiliation{Department of Astronomy, University of Michigan, 1085 S. University Ave., Ann Arbor, MI 48109, USA}

\begin{abstract}
\nolinenumbers
\noindent 
Aerosols are a ubiquitous feature of planetary atmospheres and leave clear spectral imprints in exoplanet spectra. Pre-JWST, exoplanet retrieval frameworks mostly adopted simple parametric approximations. With JWST, we now have access to mid-infrared wavelengths where aerosols have detectable composition-specific resonance features. Here, we implement new features into the open-source atmospheric retrieval code \texttt{POSEIDON} to account for the complex scattering, reflection, and absorption properties of Mie scattering aerosols. 
We provide an open-source database of these Mie scattering cross sections and optical properties. We also extend the radiative transfer and retrieval functionality in \texttt{POSEIDON} to include multiple scattering reflection and emission spectroscopy. We demonstrate these new retrieval capabilities on archival \emph{Hubble} and \emph{Spitzer} transmission and secondary eclipse spectra of the hot Jupiter HD~189733~b. We find that a high-altitude, low-density, thin slab composed of sub-micron particles is necessary to fit HD~189733~b's transmission spectrum, with multiple aerosol species providing a good fit. We additionally retrieve a sub-solar H$_2$O abundance, a sub-solar K abundance, and do not detect CO$_2$. Our joint thermal and reflection retrievals of HD~189733~b's secondary eclipse spectrum, however, finds no evidence of dayside aerosols, a sub-solar dayside H$_2$O abundance, enhanced CO$_2$, and \nt{slighty sub-}solar alkali abundances. We additionally explore how retrieval model choices, such as cloud parameterization, aerosol species and properties, and thermal structure parameterization affect retrieved atmospheric properties. Upcoming JWST data for hot Jupiters like HD~189733~b will be well suited to enable deeper exploration of aerosol properties, allowing the formulation of a self-consistent, multi-dimensional picture of cloud formation processes. 
\end{abstract}

\section{Introduction}\label{sec:Introduction}


Aerosols are solid particles formed from the gas phase via photochemistry, nucleation, or condensation and have been found in all four gas giants and multiple rocky bodies in our Solar System \citep[e.g.,][]{Sromovsky2009, Sayanagi2016, Sromovsky2012, Romani1988, Sagan1985}, brown dwarfs \citep[e.g.,][]{Burningham2021}, and exoplanets \citep[e.g.,][]{Pont2013,Grant2023}. Aerosol formation and dissipation plays an integral role in  determining the local and global thermochemical properties of atmospheres by depleting and replenishing the gas phase of refractory elements, cooling down an atmosphere by reflecting incident starlight, or heating it via latent-heat release or preventing internal flux from escaping \citep[e.g.,][]{Fortney2005,Visscher2010, Helling2006, HellingWoitke2006, Lee2016, Parmentier2016, Morley2014, Deitrick2022}. 

Once formed, aerosols in planetary atmospheres attenuate radiation via absorption and scattering. Solid particles absorb incoming radiation differently than gas since solid aggregates have no well-defined energetic transitions \citep[e.g.,][]{WakefordSing2015}, and scatter radiation depending on their size. `Small' particles in the Rayleigh/dipole regime symmetrically scatter short-wavelength radiation \nt{in the forward and backward directions} while `large' particles in the geometric-optics regime have a strong forward directional dependence. Particles in between the two limiting cases, where particle sizes are on-par with incident wavelengths, are in the Mie regime \citep{Mie1908, Marley2013, Gao2021}. For a thorough introduction to Mie scattering and its applications to planetary atmospheres, see \citet{GoodyYung1989}.


Aerosols in the Mie scattering regime leave strong spectral imprints in observed transmission and emission spectra of exoplanets. In transmission geometries light passes through the upper-atmospheric \nt{(0.1 to 1e-6 bars \citep{Kataria2016,Madhusudhan2019}, where the pressures probed depend on wavelength of light)} terminator of a planet where aerosols absorb and scatter light away from the beam. In this geometry, starlight traverses a large portion of a planet's atmosphere before being transmitted to the observer. This `grazing geometry' amplifies any scattering and absorption features and has been leveraged to significantly detect atmospheric chemical composition and aerosols \citep[e.g.,][]{SeagerSasselov2000,Fortney2005,Sing2018}, even if their optical depth is negligible at normal viewing geometries. Scattering processes imprint a slope in short wavelengths where higher energy photons are scattered away from the beam and can be attributed to both Rayleigh and Mie scattering processes \citep[e.g.,][]{Sing2016}.


The first transmission spectrum of the hot Jupiter HD~209458~b indicated the presence of aerosols due to atomic absorption lines being muted by an amplified scattering slope \citep{Charbonneau2002}. At visible to near-IR (0.3-1.0 \textmu m) wavelengths, Mie scattering is primarily sensitive to the particle size rather than specific aerosol composition except for a handful of species (e.g., MnS, ZnS, TiO$_2$, Fe$_2$O$_3$, organic hazes) \citep[][]{Pinhas2017,He2023}.
Absorption features due to vibrational modes of compositionally-specific, sub-micron sized aerosols can be observed in longer IR (5-25 \textmu m) wavelengths \citep{WakefordSing2015}. Aerosol composition and size can be inferred by leveraging observations of both the short wavelength scattering slope and longer wavelength absorption features \citep[e.g][]{WakefordSing2015,Wakeford2017,Mai2019,Fairman2024}, which is now possible in the era of JWST.




Aerosols also play a significant role in emission geometries where light from substellar objects is directly observed via direct imaging or secondary eclipse. \nt{The observable emitted thermal emission from the top of the atmosphere originates from deeper, hotter atmospheric pressures than probed by transmission spectroscopy.} \nt{Aerosols interact with thermal flux by either scattering \textit{or} absorbing light. When light is absorbed, it is lost to the upwards propagating beam. When light is scattered, aerosols can back-scatter and forward-scatter light asymmetrically \citep{GoodyYung1989}.} This is in stark contrast to Rayleigh scattering where light is symmetrically scattered \nt{in the forward and backward direction}, and to transmission geometries where any absorbed or scattered light is lost completely to the observed beam. Aerosols can also have strong back-scattering cross sections which reflect light directly from the star itself \citep[e.g.,][]{Marley1999, Sudarsky2000, Sudarsky2003, Cahoy2010, Esteves2015}. The wavelengths a substellar object appears brighter or dimmer in is due to a complex interplay between aerosol and gas-phase scattering, multiple scattering, absorption, and blackbody emission \citep[e.g,.][]{deKok2011}. Modeling aerosols and their radiative properties is crucial to linking the observed flux from substellar objects to aerosol properties and inferred thermal structure, but including aerosols in the state-of-the-art observational interpretation tools requires increased model complexity and computational time. 



Exoplanet spectra have been historically characterized via theory-driven atmospheric models and data-driven atmospheric retrievals \citep[see review,][]{Madhusudhan2019}. Atmospheric retrievals have become an integral part of interpreting spectra from substellar objects \citep[][]{Madhusudhan2018,MacDonald2023Catalog}. This data-driven inverse method approach utilizes Bayesian MCMC retrieval frameworks in order to sample across an agnostic uniform parameter space. Retrieval frameworks simplify \nt{the thermochemical atmospheric structure by assuming iso-thermal/parameterized pressure-temperature profiles and uniform chemical abundance profiles} in order to generate thousands of synthetic spectra to retrieve a `best-fit' spectra and posterior probability distributions for each model parameter. Retrievals such as these have the benefit of being agnostic to physically motivated models and explore a large parameter space, letting observations drive the modelling.

Because of the limitations of both observations and exoplanet retrieval codes, aerosols in exoplanet atmospheres have often been limited to simple gray parameterizations, such as a gray-opaque deck and scattering slope power law parameterization \citep[e.g.,][]{ MacDonald2017,Goyal2018}. These models could infer the presence of aerosols from data, but not their properties. In the era of JWST, we have access to data at mid-infrared wavelengths (5-25 \textmu m) and of needed precision to identify the signatures of specific aerosols and their properties in exoplanet atmospheres \citep[e.g.,][]{WakefordSing2015, Wakeford2017}. Recently, specific aerosol species have been utilized to explain the JWST~MIRI~LRS transit observation of WASP-17~b \citep{Grant2023} and WASP-107~b \citep{Dyrek2024}, direct imaging observation of VHS~1256~b \citep{Miles2023}, phase curve of GJ~1214~b \citep{Kempton2023}, and emission spectrum of WASP-69~b \citep{Schlawin2024}. Some retrieval codes have included compositionally specific, Mie scattering aerosols (e.g., \texttt{NEMESIS} \citep{Irwin2008} ,\texttt{BREWSTER} \citep{Burningham2017}, \texttt{PLATON} \citep{Zhang2019}, \texttt{petitRADTRANS} \citep{Molliere2019}, \texttt{SCARLET} \citep{Benneke2019}, \texttt{CHIMERA} \citep{Mai2019}\nt{, \texttt{PICASO} \citep{Schlawin2024}}), but typically treat aerosol influence on transmission, emission, and reflection spectra separately. Additionally, Mie theory calculations can take from 1-10 minutes per particle size, making it unfeasible for retrievals to compute properties `on-the-fly' while exploring a large parameter space.



For the entirety of this work we will be introducing and benchmarking new functionalities we developed to include compositionally specific aerosols and their radiative properties into the open-source atmospheric retrieval code \texttt{POSEIDON}\footnote{\url{https://github.com/MartianColonist/POSEIDON}} \citep{MacDonaldMadhusudhan2017, MacDonald2023}. In our implementation we aim to maximize computationally efficiency and explore a broad range of aerosol species and cloud configurations. The new aerosol implementation is compatible with transmission, emission (via direct emission or secondary eclipse), and reflection. We benchmark these new capabilities on available pre-JWST (\textit{HST} and \textit{Spitzer}) transmission, emission, and reflection data of the well studied hot Jupiter HD~189733~b.

HD~189733~b has a wealth of publicly available pre-JWST transmission, emission, and reflection data that has been studied extensively via forward models, retrievals, and general circulation models. The presence of aerosols inferred on the limb of HD~189733~b via transmission data \citep[e.g.,][]{Pont2013} and on the dayside via the albedo spectrum \citep[e.g.,][]{Evans2013, Barstow2014} provides the perfect test-bed for our enhancements to \texttt{POSEIDON}. In particular, the multitude of studies analyzing aerosols on HD~189733~b allow us to explore specific, physically-motivated aerosols species in our suite of retrievals. \citet{Pont2013} and \citet{WakefordSing2015} posit that enstatite (MgSiO$_3)$ or fosterite (Mg$_2$SiO$_4$) grains fit the transmission observations and are expected to form in HD~189733~b's atmosphere. \citet{Lavvas2017} find that soots are thermally stable and provide a good fit to the transmission data. \citet{Kataria2016} and \citet{Parmentier2016} use three-dimensional GCMs with equilibrium cloud formation to predict sulfide clouds on the dayside and terminator, similar to the T-Y dwarfs studied in \citet{Morley2012}. \citet{Lee2016}'s three-dimensional GCM with disequilibrium cloud formation predicts TiO$_2$ to dominate the dayside hot-spot, SiO and SiO$_2$ to dominate the equatorial regions, and enstatite+fosterite to dominate the mid-latitudes. \citet{HengDemory2013} and \citet{Barstow2014} found that there exists a strong degeneracy between sodium abundance and cloud particle size when fitting the albedo data, where solutions with and without clouds could sufficiently fit the data. \citet{Zhang2020} co-fit the dayside emission and transmission spectrum assuming equilibrium chemistry and retrieved a Rayleigh scattering haze on the limb and a clear dayside. We explore all of these possibilities in our work.

This paper is structured as follows: \S~\ref{sec:POSEIDON-improvements} presents enhancements to \texttt{POSEIDON}. In \S~\ref{sec:hd189-transmission-retrievals} we run a suite of retrievals on the full transmission spectrum of HD~189733~b with the inclusion of compositionally specific aerosols (MgSiO$_3$, SiO$_2$, MnS, Na$_2$S, flame soot, and Tholins) spanning the entire proposed condensate phase space in extant work. We benchmark our results and then explore different aerosol models. We then run retrievals spanning the entire aerosol database on just the scattering slope \textit{HST} data of HD~189733~b and extend out to longer wavelengths to help guide future transmission observations. In \S~\ref{sec:hd189-emission-retrievals} we run retrievals on the combined reflection and emission data of HD~189733~b. We first explore a clear atmosphere, as well as retrievals with MgSiO$_3$, SiO$_2$ and MnS. We then explore how choosing different pressure-temperature (P-T) profiles affects results, and how different cloud properties affect forward model spectra. In \S~\ref{sec:Dicussion} we discuss our results and compare with extant work. In \S~\ref{sec:Conclusion} we conclude.

\section{Enhancements to \texttt{POSEIDON}}
\label{sec:POSEIDON-improvements}

To date, the open-source atmospheric retrieval code \texttt{POSEIDON} has been utilized to interpret \textit{HST}, \textit{Spitzer}, JWST and ground-based telescope transmission and emission data \citep[e.g.,][]{MacDonald2017, Sedaghati2017,Lewis2020,Kirk2021, Coulombe2023}. However, the forward model's aerosol prescription could only include `gray' opacity clouds and power law hazes, while the emission scheme implemented a single-stream approach with no scattering. With JWST, we now have access to mid-infrared wavelengths where compositionally specific clouds must be utilized to fit aerosol absorption features and infer their properties \citep[e.g.,][]{WakefordSing2015, Grant2023, Dyrek2024}

We first introduce how Mie scattering is implemented into \texttt{POSEIDON} in \S~\ref{sec:Mie-scattering}, with particular emphasis on the open-source precomputed aerosol database  in \S~\ref{sec:aerosol_database}. In \S~\ref{sec:cloud-models} we introduce new aerosol models into the forward model. We then discuss how the radiative transfer model has been updated to include multiple scattering and reflection in \S~\ref{sec:rad_model_improvements}. \S~\ref{sec:other-improvements} presents other improvements to \texttt{POSEIDON}, such as new pressure-temperature profiles, contribution visuals, and opacity database updates.

\subsection{Mie Scattering Implementation}\label{sec:Mie-scattering}

Our implementation of scattering in exoplanet spectra calculations requires three wavelength-dependent quantities: the effective extinction cross section, $\sigma_{\mathrm{ext}, \, \mathrm{eff}}$, the asymmetry parameter, $g$, and the single-scattering albedo, $\omega$. We use Mie scattering theory to pre-compute these quantities from aerosol refractive index data. The effective extinction cross section, $\sigma_{\mathrm{ext}, \, \mathrm{eff}}$, is the combined scattering and absorption cross section of an aerosol species averaged over its particle size distribution. In transmission spectra calculations, we assume any absorbed or scattered photon is lost to the beam, and hence $\sigma_{\mathrm{ext}, \, \mathrm{eff}}$ represents an additional aerosol extinction term alongside gas opacity. Future work will consider forward scattering for transmission spectra \citep[e.g.,][]{Robinson2017,Jaiswal2023}. In other observing geometries, such as during secondary eclipse, scattering can significantly shape the observed radiation in a non-trivial manner. The relative importance of scattering to absorption is expressed by the single-scattering albedo, $\omega$, which ranges from 0--1(where $\omega$ is 0 for black, complete absorbers and 1 for white, complete scatterers). The asymmetry parameter, $g$, encodes the degree of forward-scattering from a particle and ranges nominally from 0--1 (where 0 represents the Rayleigh-limit of symmetric scattering, 1 represents the large particle limit of complete forward-scattering e.g., \citealt{GoodyYung1989}). We note that for some aerosol species, $g$ can be negative (e.g., Cr) due to large imaginary refractive indices \citep[][]{Rfaqat2022}.

The extinction cross section for a single particle of radius $r$ is given by
\begin{equation}
\sigma_{\mathrm{ext}} = \pi \; r^2 \; Q_{\mathrm{ext}}\left(m, \; \frac{2\pi r}{\lambda}\right)
\label{eq:ext_cross_section}
\end{equation}
where $Q_{\mathrm{ext}}$ is the extinction efficiency and $m$ is the complex refractive index of a particle at wavelength $\lambda$. Refractive indices are wavelength-dependent complex numbers, where the imaginary component encodes the attenuation properties of an aerosol species. The extinction efficiencies are calculated from these refractive index data using Mie theory. Here, we use the Mie algorithm from \texttt{LX-MIE} \citep{KitzmannHeng2018}, as implemented within the open-source atmospheric retrieval code \texttt{Platon} \citep{Zhang2020}. We outline here the salient points of the Mie scattering calculation, which assumes spherical particles for simplicity. The extinction efficiencies are given by
\begin{equation}
Q_{\mathrm{ext}}\left(m,\; \frac{2\pi r}{\lambda}\right) = \frac{2}{x^2} \; \sum^{n_{max}}_{n=1}(2n+1) \; Re(a_n + b_n)
\label{eq:efficiency_ext}
\end{equation}
where $x = 2 \pi r / \lambda$ is the size parameter and the Mie coefficients for spherical particles, $a_n$ and $b_n$, are calculated via Riccati-Bessel functions and their derivatives \citep[see][ Section 3 for a more complete description of the algorithm]{KitzmannHeng2018}. 

\begin{figure}[t!]
     \centering
     \includegraphics[width=1\linewidth]{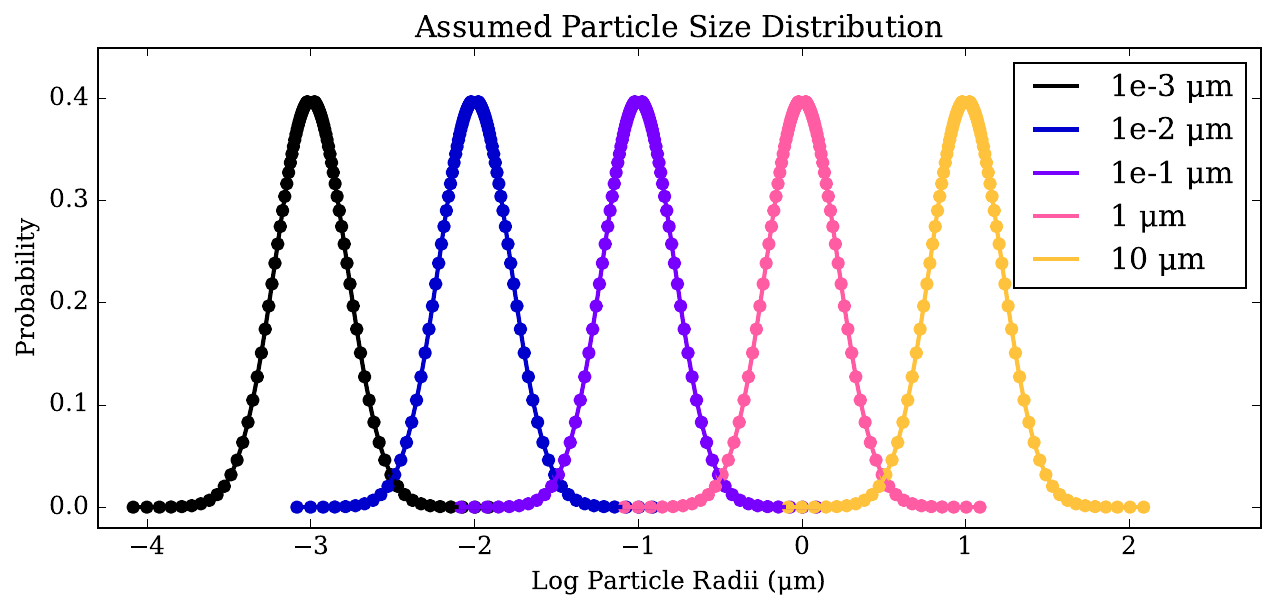}
     \caption{Assumed particle size distribution (in log radii) for a given mean particle size $r_m$. Dots represent specific particle sizes that are computed, with an over density near the mean particle size.}
     \label{fig:particle-size-distribution}
\end{figure}

\begin{figure*}[ht!]
     \centering
         \includegraphics[width=1.0\linewidth]{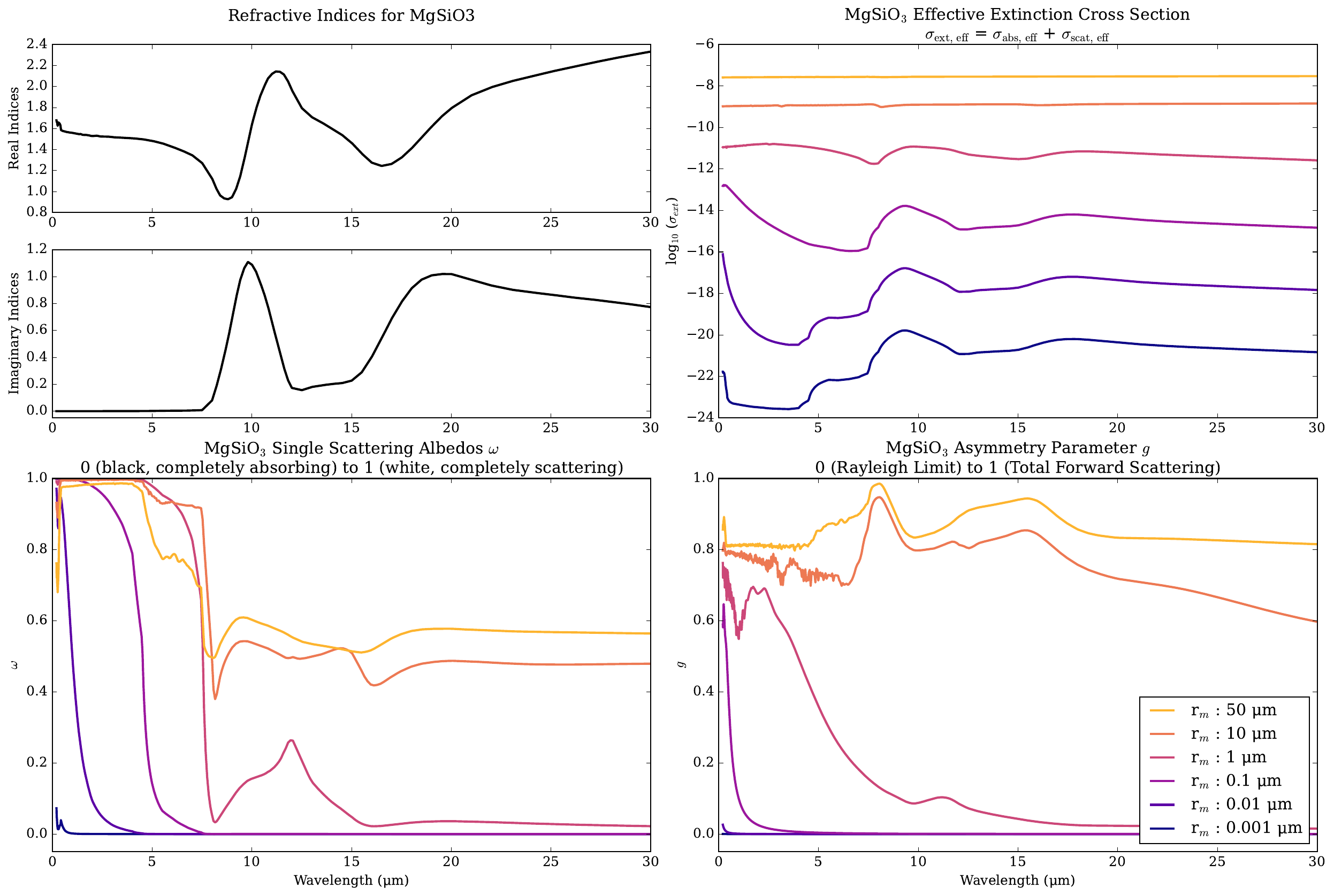}
     \caption{Graphic displaying an example aerosol's Mie properties. Top Left: refractive index lab data as it varies with wavelength. The effective extinction cross section $\sigma_{\mathrm{ext}, \, \mathrm{eff}}$ (Top Right), single scattering albedo $\omega$ (Bottom Left) and asymmetry parameter $g$ (Bottom Right) as they vary with mean particle size and wavelength. As a particle gets larger, $\sigma_{\mathrm{ext}, \, \mathrm{eff}}$ grows by orders of magnitude and particles preferentially forward scatter incident radiation ($\omega \sim 1$, $g \sim 1$,).  We note that the precomputed aerosol database for retrievals (Table \ref{table:aerosol_species}) only spans mean particle size from 0.001 \textmu m to 10 \textmu m, however the forward model algorithm can compute any mean particle size. Users can run forward models with their own refractive indices that will compute radiative properties `on-the-fly' and run an algorithm to include their own species into the precomputed aerosol database for retrievals. \nt{Similar plots are available for every aerosol in Table \ref{table:aerosol_species}, see the `aerosol\_database.pdf' on Zenodo (\S \ref{sec:Links}) or \href{https://poseidon-retrievals.readthedocs.io/en/latest/content/opacity_database.html}{\texttt{POSEIDON}'s Opacity Database}.}}
     \label{fig:mie_algorithm}
\end{figure*}

In planetary atmospheres condensate particles will not all be the same size and instead will follow a particle size distribution.  Mie theory predicts that the cross section for a single particle size exhibits a fine structure, but this is `washed out' for realistic aerosols with a range of particles sizes, causing absorption features to be broad in wavelength space \citep[e.g.,][]{Marley1999, WakefordSing2015}. Therefore, the {\it effective} Mie extinction cross section is expressed as an integral average over a log-normal particle radius distribution
\begin{equation}
\sigma_{\mathrm{ext}, \, \mathrm{eff}} =  \int \; P(r) \; \sigma_{\mathrm{ext}} \; dr = \int \; P(r) \; \sigma_{\mathrm{geo}} \; Q_{\mathrm{ext}} \; dr
\label{eq:eff_ext_cross_section}\end{equation}
\noindent where $P(r)$ is the particle size distribution (here assumed log-normal), and $\sigma_{\mathrm{geo}} = \pi r^2$ is the geometric cross section of particles defined by the distribution. We evaluate this integral via a change of variables 
\begin{equation}
z = \frac{\ln (r) - \ln (r_m)}{\ln \sigma_{r}}
\end{equation}
where  $r_m$ is the mean particle radius and $\sigma_r$ is the standard deviation. We assume throughout that $\ln \sigma_r = 0.5$. With this change of variables, the particle size distribution is given by
\begin{equation}
P(z) = \frac{1}{\sqrt{2\pi}} \; \exp \left(\frac{-z^2}{2}\right)
\end{equation}
and the geometric cross section is
\begin{equation}
\sigma_{\mathrm{geo}} = \pi \; r^2 = \pi \; \Bigl(r_m \; \exp (\ln \sigma_r \; z )\Bigr)^2 \; .
\end{equation}
We evaluate the integral in Equation~\ref{eq:eff_ext_cross_section} over $z$ with trapezoidal integration, where $z$ is spaced linearly in log space symmetrically from $\pm$ 0.1 to $\pm$ 5 with 50 points for each half of the distribution (100 total). This spacing produces an overdensity of points near 0 to better sample the most common particle radii near the mean particle radius, $r_m$, as shown in Figure \ref{fig:particle-size-distribution}.

Since we also require asymmetry parameters and single-scattering albedos for radiative transfer with scattering, we use formulae from \citet{GoodyYung1989} (see their Section 7.6 for derivations). The asymmetry parameter is given by
\begin{align}
g_{\mathrm{Mie}} = \frac{4}{Q_{\mathrm{scat}} \; x^2} \; \Bigr[ &\sum^{n_{max}}_{n=1} \frac{n(n+2)}{n+1} \;  Re(a_na_{n+1}^* + b_n b_{n+1}^*) \nonumber\\
    + &\sum^{n_{max}}_{n=1} \frac{2n+1}{n(n+1)} \; Re(a_nb_n^*) \Bigr] \; ,
\end{align}
and the scattering efficiency is given by
\begin{equation}
Q_{\mathrm{scat}}\left(m, \frac{2\pi r}{\lambda}\right) = \frac{2}{x^2}\sum^{n_{max}}_{n=1}(2n+1) \; (|a_n|^2 + |b_n|^2) \; .
\label{eq:scattering_efficiency }
\end{equation}
Given the scattering efficiency, the single scattering albedo is determined by 
\begin{equation}
\omega_{\mathrm{Mie}} = \frac{Q_{\mathrm{scat}}}{Q_{\mathrm{ext}}}
\end{equation}
We calculate $\omega$ and $g$ for the same range of $z$ and particle size distribution as in Equation~\ref{eq:eff_ext_cross_section}. The Mie algorithm we use iterates over refractive indices at each wavelength \textit{and} iterates Mie coefficients until the sums converge. Finally, we take the median of both quantities to find the effective asymmetry parameter and single scattering albedo.\footnote{\nt{For a tutorial on how to change the log-normal distribution width and other assumptions in the precomputed database, see `Making an Aerosol Database' tutorial in \href{https://poseidon-retrievals.readthedocs.io/en/latest/content/forward_model_tutorials.html}{Forward Model Tutorials}}}

We demonstrate the inputs and outputs of our Mie scattering implementation for MgSiO$_3$ in Figure \ref{fig:mie_algorithm}. The refractive indices show a feature near 9.6 \textmu m in the imaginary component, which can propagate into the effective extinction cross section for sufficiently small particle sizes. For larger particle sizes, $\sigma_{\mathrm{ext}, \, \mathrm{eff}}$ shows a muted resonance feature \citep[e.g.,][]{Marley1999, Min2004}, the particles become more efficient at scattering compared to absorption ($\omega \rightarrow 1$), and the particles become more forward scattering ($g \rightarrow 1$). We discuss how $g$ and $\omega$ are used for multiple scattering radiative transfer \S~\ref{sec:MS} and \S~\ref{sec:reflection}.

\subsubsection{Aerosol Database}\label{sec:aerosol_database}

Since Mie scattering calculations can prove computationally intensive, we pre-compute a database of scattering properties for a wide range of aerosol species. With nested loops and convergence criteria, the algorithm can be very time-intensive, especially for large particles, and is not feasible for retrievals where a large range of species and mean particle sizes are considered. Therefore, we pre-compute $\sigma_{\mathrm{ext}, \, \mathrm{eff}}$, $g$, and $\omega$ for specific aerosols that are expected to condense and form clouds in substellar atmospheres \citep[e.g.,][]{WakefordSing2015, KitzmannHeng2018, Burningham2021, Dominik2021, Lee2022}. Precomputing cross sections and formulating a database for \texttt{POSEIDON} to query and interpolate over expedites the computationally expensive process of deriving extinction efficiencies `on-the-fly'. 

We pre-compute these Mie scattering properties from laboratory-measured refractive index data. Our database spans mean particle sizes from 0.001\,$\mu$m--10 $\mu$m (spaced linearly with 1,000 points) and wavelengths from 0.2--30\,$\mu$m (at a spectral resolution of $R$ = 1,000). Table \ref{table:aerosol_species} summarizes all the aerosol species currently available in our open-source Mie scattering database. Our database includes condensates ranging from ices expected to form in the Y-dwarf and solar system gas giant regime \citep[e.g.,][]{Morley2014, Fletcher2023} to aluminum- and calcium-bearing condensates expected in super-hot exoplanets \citep[e.g.,][]{Wakeford2017}, allowing for retrieval explorations of clouds across many use cases. \nt{The `Notes' column in Table \ref{table:aerosol_species} includes the common name, specific polymorph, crystalline or amorphous, and specific temperature and polarization of refractive indices if one was used. Otherwise, aerosols were measured at room temperature and polarization-specific refractive indices for anisotropic aerosols were averaged.} Additionally, \texttt{POSEIDON} has the capability for users to compute Mie properties from their own refractive indices and add them directly to the database.\footnote{\nt{For a tutorial on how to add aerosols to the precomputed database, see `Making an Aerosol Database' tutorial in \href{https://poseidon-retrievals.readthedocs.io/en/latest/content/forward_model_tutorials.html}{Forward Model Tutorials}}}

We note that several aerosol species have special treatments in our database due to the varying sources of laboratory data in the literature. Several species drawn from \citet{WakefordSing2015} and \citet{KitzmannHeng2018} have different refractive indices and wavelength coverage (e.g., Al$_2$O$_3$ and CaTiO$_3$), so we include an entry in our database for both sources. \nt{Magnesium sulfide (MnS) has three separate entries: (i) \citet{WakefordSing2015} uses only measured refractive indices taken from Figure 5 of \citet{Huffman1967}, resulting in the characteristic non-uniform scattering slope; (ii) \citet{KitzmannHeng2018} uses the measured refractive indices from Table 1 and Figure 5 of \citet{Huffman1967}, and extended to longer wavelengths using the real indices of similar sulfide species (Na$_2$S) (where the imaginary indices were calculated with Kramers-Kronig analysis), where different methodologies resulted in a differently shaped, more uniform scattering slope; (iii) \citet{Morley2012} introduced a hybrid treatment, where the measured refractive indices are used in conjunction with a sulfide extrapolation. For this entry, we combine the refractive indices from \citet{WakefordSing2015} up to 13\,$\mu$m and the refractive indices from \citet{KitzmannHeng2018} from 13-30 \textmu m (this entry has both the non-uniform scattering slope and sulfide absorption at longer wavelengths, so we adopt this entry in our retrievals in \S~\ref{sec:hd189-transmission-retrievals}).} We additionally note two other entries: Ammonium dihydrogen phosphate (ADP) and the Saturn Phosphorus Haze. ADP utilizes the short-wavelength (0.2-1.9 \textmu m) measured absolute, real refractive indices (where the ordinary and extraordinary rays \nt{were weighted by 2/3 and 1/3 \citep{Mogli2007,Reed2017}}) found in \citet{Zernike1965} with an assumed imaginary index of 0 (this assumption is valid for ADP, which is a popular crystal used to make lasers, and therefore has negligible attenuation in short wavelengths) in combination with the measured real and imaginary indices (2-20 \textmu m) of liquid ADP found in \citet{Querry1974}. We preform the same weighting for the ordinary and extraordinary indices of $\alpha$ SiO$_2$ from Philipp in \citet{Palik1985}. The Saturn Phosphorus Haze uses the proposed imaginary refractive indices of the `mystery' phosphorus haze in Saturn's atmosphere from \citet{Noy1981} and the retrieved constant refractive index from \citet{Fletcher2023}. \nt{More specifically, we follow the methodology of \citet{Sromovsky2020} where the real index of white phosphorus (P$_4$) is utilized for the entire wavelength range and the imaginary indices from \citet{Noy1981} are reduced by a factor of 10 and utilized from 0.25 to 0.6 \textmu m. From 0.6 to 20 \textmu m the retrieved imaginary refractive index of the second cloud layer in \citet{Fletcher2023} is utilized.} 

\nt{For more information on aerosols and their references (such as synthetic vs natural samples, how lab measurements were taken, etc.), see the `Aerosol-Database-Readme.txt' on Zenodo (\S \ref{sec:Links}) or \href{https://poseidon-retrievals.readthedocs.io/en/latest/_static/Aerosol-Database-Readme.txt}{\texttt{POSEIDON}'s documentation}, or  \href{https://poseidon-retrievals.readthedocs.io/en/latest/content/opacity_database.html}{\texttt{POSEIDON}'s Opacity Database}.}

\setlength{\extrarowheight}{1.2pt}
\startlongtable
\begin{deluxetable*}{lllll}
\colnumbers
\tablecolumns{5}
\tablewidth{0pt}
\tabletypesize{\footnotesize}
\tablecaption{Aerosol Refractive Index Data \label{table:aerosol_species}}
\tablehead{Species & Database  & Refractive Index & Notes & (Min, Max) \textmu m}
\startdata
\textbf{Super Hot} \\
\hspace{2pt} CaAl$_{12}$O$_{19}$ & WS15 & \citet{Mutschke2002}$^\mathrm{D}$ & Hibonite (crystalline, E$\parallel$c) & (2, 30)\\
\hspace{2pt} Al$_2$O$_3$ & WS15 & \citet{Koike1995}  & $\gamma$ Corundum (crystalline)  & (0.34, 30)\\
\hspace{2pt} Al$_2$O$_3$ & KH18 & \citet{Koike1995} & $\gamma$ Corundum (crystalline) & (0.2, 30)\\
& & \citet{Begemann1997}$^\mathrm{D}$  & \hspace{2pt} Corundum (amorphous, porous)& \\
\hspace{2pt} CaTiO$_3$ & WS15 & \citet{Posch2003}$^\mathrm{D}$ & Perovskite (crystalline) & (2, 30)\\
\hspace{2pt} CaTiO$_3$ & KH18 & \citet{Posch2003}$^\mathrm{D}$ & Perovskite (crystalline) & (0.2, 30)\\
& & \citet{Ueda1998} & & \\
\hline
\textbf{M-L Dwarf} \\
\hspace{2pt}TiO$_2$ & KH18 & \citet{Zeidler2011}$^\mathrm{D}$ & Anatase (crystalline) & (0.2, 30)\\
&& \citet{Posch2003}$^\mathrm{D}$ &\\
&& \citet{Siefke2016} &\\
\hspace{2pt}TiO$_2$ & gCMCRT & \citet{Zeidler2011}$^\mathrm{D}$ &  Rutile (crystalline, E$\parallel$a,b) & (0.47, 30)\\ 
& & Ribarsky in \citet{Palik1985}$^\mathrm{C}$ & & \\
\hspace{2pt}TiC & KH18 & \citet{Koide1990} & Titanium Carbide (crystalline)  &  (0.2, 30)\\
&&  \citet{Henning2001} & \\
\hspace{2pt}VO & gCMCRT & \citet{Wan2019} & VO$_2$ as a VO analog (crystalline) & (0.3, 30)\\
\hspace{2pt}Nano-Diamond & N/A & \citet{Mutschke2004} & Meteoritic C (crystalline) & (0.2, 30)\\
\hline
\textbf{Iron} \\
\hspace{2pt}Fe & KH18 & Lynch \& Hunter in \citet{Palik1991}$^\mathrm{C}$ & $\alpha$ Fe (crystalline) & (0.2, 30)\\
\hspace{2pt}FeO & WS15 & \citet{Begemann1995} & W{\"u}stite (crystalline) & (0.21, 30)\\
\hspace{2pt}FeS & KH18 & \citet{Pollack1994}$^\mathrm{C}$ & Troilite (Crystalline) & (0.2, 30) \\
&&\citet{Henning1997}$^\mathrm{D}$  &\\
\hspace{2pt}Fe$_2$O$_3$ & WS15 & \citet{Triaud2005}$^\mathrm{D}$ (unpublished) & Hematite ($\alpha$ Fe$_2$O$_3$) & (0.2, 30)\\
\hspace{2pt}& &  & (crystalline, E$\parallel$a,b) & (0.2, 30)\\
\hspace{2pt}FeSiO$_3$ & WS15 & \citet{Day1981} &  Ferrosilite (Iron-rich Pyroxene) & (8.22, 30) \\
\hspace{2pt} & &  & (amorphous) & \\
\hspace{2pt}Fe$_2$SiO$_4$ & KH18 & \citet{Fabian2001}$^\mathrm{D}$ & Fayalite (Iron-rich Olivine) & (0.40, 30)\\
\hspace{2pt} & &  & (crystalline) & \\
\hline
\textbf{Magnesium} \\
\hspace{2pt}MgO & KH18 & Roessler \& Huffman in \citet{Palik1991}$^\mathrm{C}$ & Periclase (crystalline) & (0.2, 30) \\
\hspace{2pt}Mg$_{0.8}$Fe$_{1.2}$SiO$_4$ & WS15 & \citet{Henning2005}$^\mathrm{C}$ & Forsterite (Fe-rich) & (0.21, 30)\\
\hspace{2pt} & & & (amorphous glass) & \\
\hspace{2pt} & & \citet{Dorschner1995}$^\mathrm{D}$ & \\
\hspace{2pt}Mg$_{1.72}$Fe$_{0.21}$SiO$_4$ & WS15 & \citet{Zeidler2011}$^\mathrm{D}$ & Forsterite (Fe-poor) & (0.2, 30)\\
\hspace{2pt} & & & (crystalline, E$\parallel$c) & \\
\hspace{2pt}Mg$_2$SiO$_4$ & B21 & \citet{ScottDuley1996}$^\mathrm{C}$ & Forsterite (amorphous) & (0.27, 30)  \\
\hspace{2pt} & & \citet{DraineLee1984} & \hspace{2pt} `Astronomical' Silicate & \\
\hspace{2pt} & & \citet{NitsanShankland1976} & \hspace{2pt} Forsterite (crystalline, E$\parallel$a, E$\parallel$c) & \\
\hspace{2pt}Mg$_2$SiO$_4$ & KH18 & \citet{Jager2003}$^\mathrm{D}$ & Forsterite (amorphous sol gel) & (0.2, 30) \\
\hspace{2pt}Mg$_2$SiO$_4$ & gCMCRT & \citet{Suto2006} & Forsterite (crystalline, E$\parallel$a, E$\parallel$b) & (0.2, 30) \\
\hspace{2pt}MgFeSiO4 & KH18 & \citet{Dorschner1995}$^\mathrm{D}$ & Olivine (amorphous glass) & (0.2, 30)\\
\hspace{2pt}Mg$_{0.8}$Fe$_{1.2}$SiO$_4$ & KH18 & \citet{Dorschner1995}$^\mathrm{D}$ & Olivine (amorphous glass) & (0.2, 30) \\
\hspace{2pt}MgSiO$_3$ & WS15 & \citet{Egan1975} & Enstatite (crystalline) & (0.2, 30)\\
& & \citet{Dorschner1995}$^\mathrm{D}$ & \hspace{2pt} Enstatite (amorphous glass) & \\
\hspace{2pt}MgSiO$_3$ & B21 & \citet{ScottDuley1996} & Enstatite (amorphous) & (0.27, 30) \\
\hspace{2pt} & & \citet{DraineLee1984} & \hspace{2pt} `Astronomical' Silicate & \\
\hspace{2pt} & & \citet{NitsanShankland1976} & \hspace{2pt} Forsterite (crystalline, E$\parallel$a, E$\parallel$c) & \\
\hspace{2pt}MgSiO$_3$ & KH18 & \citet{Dorschner1995}$^\mathrm{D}$ & Enstatite (amorphous glass) & (0.2, 30) \\
\hspace{2pt}MgSiO$_3$ & KH18 & \citet{Jager2003}$^\mathrm{D}$ & Enstatite (amorphous sol gel)& (0.2, 30)\\
\hspace{2pt}MgSiO$_3$ & B21 & \cite{Jaeger1998}$^\mathrm{D}$ &  Ortho-Enstatite (crystalline) & (0.27, 30) \\
\hspace{2pt}Mg$_{0.4}$Fe$_{0.6}$SiO$_3$ & KH18 & \citet{Dorschner1995}$^\mathrm{D}$ & Pyroxene (amorphous glass) & (0.2, 30)\\
\hspace{2pt}Mg$_{0.5}$Fe$_{0.5}$SiO$_3$ & KH18 & \citet{Dorschner1995}$^\mathrm{D}$ & Pyroxene (amorphous glass) & (0.2, 30)\\
\hspace{2pt}Mg$_{0.8}$Fe$_{0.2}$SiO$_3$ & KH18 & \citet{Dorschner1995}$^\mathrm{D}$ & Pyroxene (amorphous glass) & (0.2, 30)\\
\hspace{2pt}MgAl$_2$O$_4$ & WS15 & \citet{Fabian2001}$^\mathrm{D}$ & Spinel (annealed, crystalline) & (1.69, 30)\\
\hline
\textbf{Silica} \\
\hspace{2pt}SiC & KH18 & \citet{LaorDraine1993}$^\mathrm{C}$ & $\alpha$ SiC (crystalline) & (0.2, 30)\\
\hspace{2pt} & & \citet{Bohren1983} & & \\
\hspace{2pt} & & Philipp \& Taft in \citet{SiCBook} & & \\
\hspace{2pt}SiO & KH18 & Philipp in \citet{Palik1985}$^\mathrm{C}$ & Amorphous (glass) & (0.2, 30)\\
\hspace{2pt} & & \citet{Wetzel2013} & & \\
\hspace{2pt}SiO$_2$ & WS15 & Philipp in \citet{Palik1985}$^\mathrm{C}$ & $\alpha$ Quartz (crystalline) & (0.2, 30)\\
& & \citet{Zeidler2013}$^\mathrm{D}$ & \hspace{2pt} $\beta$ Quartz (crystalline, 928K, E$\parallel$c) & \\
\hspace{2pt}SiO$_2$  & N/A & \citet{Herbin2023} & $\alpha$ Quartz (crystalline) & (0.25, 15.37) \\
\hspace{2pt}SiO$_2$ & N/A & Philipp in \citet{Palik1985}$^\mathrm{C}$ & $\alpha$ Quartz (crystalline) & (0.2,30) \\
\hspace{2pt}SiO$_2$ & KH18 & \citet{Henning1997}$^\mathrm{D}$ & Amorphous Silica (glass) & (0.2, 30)\\
&& Philipp in \citet{Palik1985}$^\mathrm{C}$ & \hspace{2pt} $\alpha$ Quartz (crystalline)\\
\hspace{2pt}SiO$_2$ & N/A & Philipp in \citet{Palik1985}$^\mathrm{C}$ & Amorphous Silica (glass) & (0.2,30) \\
\hline
\textbf{T-Y Dwarf} \\
\hspace{2pt}Cr & KH18  & Lynch \& Hunter in \citet{Palik1991}$^\mathrm{C}$ & Crystalline & (0.2, 30)\\
&& \citet{Raki1998OpticalPO}&\\
\hspace{2pt}MnS & WS15 & \citet{Huffman1967} & Lab Data (crystalline) & (0.2, 13)\\
\hspace{2pt}MnS & KH18 & \citet{Huffman1967} & Extrapolated (crystalline) & (0.2, 30) \\
& & \citet{Montaner1979} & \hspace{2pt} Na$_2$S (15K) & \\
\hspace{2pt}MnS & WS15 & \citet{Huffman1967} & Lab Data + Extrapolated & (0.2, 30) \\
& KH18 & \citet{Montaner1979} & \hspace{2pt} Na$_2$S (15K) & \\
\hspace{2pt}Na$_2$S & WS15 & \citet{Morley2012} & Crystalline & (0.2, 30) \\
& & \citet{Montaner1979} & \hspace{2pt} 15K & \\
& & \citet{Khachai2009} & & \\
\hspace{2pt}ZnS & WS15 & \citet{Querry1987}& Zinc blende (crystalline) & (0.22, 30)\\
\hspace{2pt}NaCl & WS15 & Eldrige \& Palik in \citet{Palik1985}$^\mathrm{C}$ & Halite (crystalline) & (0.2, 30)\\
\hspace{2pt}KCl & WS15 & Palik in \citet{Palik1985}$^\mathrm{C}$ & Sylvite (crystalline) & (0.2, 30)\\
\hline
\textbf{Ices} \\
\hspace{2pt}ADP (NH$_4$H$_2$PO$_4$) & N/A & \citet{Zernike1965} & Crystalline & (0.2, 19.99) \\
& & \citet{Querry1974} & Liquid & \\
\hspace{2pt}H$_2$O & WS15 & \citet{HaleQuerry1973}$^\mathrm{C}$ & Liquid & (0.2, 30)\\
\hspace{2pt}H$_2$O & WS15 & \citet{Warren1984}$^\mathrm{C}$ & Ice 1h (266.15K) & (0.2, 30)\\
\hspace{2pt}NH$_4$SH & N/A & \citet{Howett2007} (pers. comm.) & Crystalline ($\sim$160K) & (0.5, 30) \\
\hspace{2pt}NH$_3$ & optool &  \citet{Martonchik1984}$^\mathrm{C}$ & Crystalline (77-88K) & (0.2, 30)\\
\hspace{2pt}CH$_4$ & WS15 & \citet{Martonchik1994}$^\mathrm{C}$ & Liquid (111K) & (0.2, 30)\\
\hspace{2pt}CH$_4$ & WS15 & \citet{Martonchik1994}$^\mathrm{C}$ & Crystalline (90K) & (0.2, 30)\\
\hspace{2pt}Ice Tholins & N/A & \citet{Khare1993} & C$_2$H$_6$-H$_2$O Photolysis & (0.2, 30)\\
\hspace{2pt} & & & (amorphous, 77K) & \\
\hline
\textbf{Soots and Hazes} \\
\hspace{2pt}C & KH18 & \citet{Draine2003} & Graphite (crystalline) & (0.2, 30)\\ 
& & \citet{Draine2003b} &  & \\ 
\hspace{2pt}ExoHaze (300K) & N/A & \citet{He2023}  & H$_2$O-CH$_4$-N$_2$-CO$_2$-He & (0.4, 28.6) \\
\hspace{2pt} & & & Photolysis (amorphous) & \\
\hspace{2pt}ExoHaze (400K) & N/A & \citet{He2023} & H$_2$O-CH$_4$-N$_2$-CO$_2$-H$_2$-He  &
(0.4, 28.6) \\
\hspace{2pt} & & & Photolysis (amorphous) & \\
\hspace{2pt}Flame Soot & gCMCRT & \citet{Lavvas2017}$^\mathrm{C}$ & Amorphous & (0.2, 30)\\
\hspace{2pt}Hexene (C$_6$H$_{12}$) & WS15 & \citet{Anderson2000} & Liquid & (2, 25)\\
\hspace{2pt}H$_2$SO$_4$ & N/A & \citet{Palmer1975} & Sulfuric Acid (liquid, 300K) & (0.36, 24.98)\\
\hspace{2pt}S$_8$ & gCMCRT & Fuller, Downing, \& Querry in & $\alpha$-Sulfur (crystalline) & (0.2, 30)\\
\hspace{2pt} & & \citet{Palik1998}$^\mathrm{C}$ &  & \\
\hspace{2pt}Saturn Phosphorus Haze  & N/A & \citet{Noy1981} & P$_2$H$_2$ Proxy (amorphous) & (0.25, 20) \\
& & \citet{Sromovsky2020}& &\\
& & \citet{Fletcher2023}& &\\
\hspace{2pt}Soot 6mm & N/A & \citet{Chang1990} & Amorphous & (0.2, 28.4) \\
\hspace{2pt}Tholin & WS15 & \citet{Khare1984} & N$_2$-CH$_4$ Photolysis (amorphous) & (0.2, 30)\\
& & \citet{Ramirez2002} & & \\
\hspace{2pt}Tholin CO 1 & N/A & \citet{Corrales2023} & N$_2$-CH$_4$-CO$_2$ Photolysis & (0.2, 9.99) \\
\hspace{2pt} & &  & (amorphous) & \\
\hspace{2pt}Tholin CO 0.625 & N/A & \citet{Corrales2023}& N$_2$-CH$_4$-CO$_2$ Photolysis & (0.2, 9.99)\\
\hspace{2pt} & &  & (amorphous) &\\
\enddata
\tablecomments{Aerosol species included in the precomputed, open-source aerosol datbase. Categories: Super-Hot \citep[e.g.,][]{Wakeford2017}, M-L Dwarf \citep[e.g.,][]{Burrows1999,Lodders2002,Lee2016}, Iron, Magnesium, Silica \citep[e.g.,][]{Sudarsky2003,Visscher2010,Lee2016}, T-Y Dwarf \citep[e.g.,][]{Morley2012}, Ices \citep[e.g.,][]{Morley2014, Fletcher2023}, Soots/Hazes \citep[e.g.,][]{Gao2017,Lavvas2017, He2023}. Species name (1), refractive index database reference (2) (\citet{WakefordSing2015} (WS15, \url{https://stellarplanet.org/science/condensates/}), \citet{KitzmannHeng2018} (KH18, \url{https://github.com/NewStrangeWorlds/LX-MIE/tree/master/compilation}), \citet{Burningham2021} (B21), \citet{Dominik2021} (optool, \url{https://github.com/cdominik/optool/tree/master/lnk_data}), \citet{Lee2022} (gCMCRT, \url{https://github.com/ELeeAstro/gCMCRT/tree/main/data/nk_tables}), N/A refers to this work) where refractive index files can be found on the POSEIDON github (\url{https://github.com/MartianColonist/POSEIDON} under POSEIDON\slash reference data) or \S \ref{sec:Links}, refractive index reference (3), notes (common name, crystalline or amorphous, specific direction/polarization/temperature if one was used) (4). The absolute minimum and maximum wavelength of the database is 0.20 and 30 \textmu m. The minimum and maximum wavelengths of individual species are listed in (5), where we do not extrapolate past the bounds of the refractive index data. Note that refractive indices are interpolated between their minimum and maximum wavelength, and therefore might not be accurate where there are gaps in the refractive index data (see `aerosol\_database.pdf' on Zenodo (\S \ref{sec:Links}) or \href{https://poseidon-retrievals.readthedocs.io/en/latest/content/opacity_database.html}{\texttt{POSEIDON}'s Opacity Database} to see refractive indices vs interpolated indices). Optical properties are precomputed for particle sizes 1e-3 to 10 \textmu m. \nt{$^\mathrm{D}$ refers to refractive indices that can be found on the Database of Optical Constants for Cosmic Dust (DOCCD, \url{https://www.astro.uni-jena.de/Laboratory/OCDB/index.html}). $^\mathrm{C}$ refers to references that compile refractive index data (sometimes to supplement their own lab data). For more information on aerosols and refractive index citations (such as synthetic vs natural samples, how lab measurements were taken, crystal shape, etc.), see the `Aerosol-Database-Readme.txt' on Zenodo (\S \ref{sec:Links}) or \href{https://poseidon-retrievals.readthedocs.io/en/latest/_static/Aerosol-Database-Readme.txt}{\texttt{POSEIDON}'s documentation}.}  We recommend that for colder ices and hydrocarbons, users should refer to the cosmic ice laboratory (\url{https://science.gsfc.nasa.gov/691/cosmicice/constants.html}) \nt{or the optical constants database (\url{https://ocdb.smce.nasa.gov/})} and add species to the precomputed database following the `Making an Aerosol Database' tutorial in \href{https://poseidon-retrievals.readthedocs.io/en/latest/content/forward_model_tutorials.html}{Forward Model Tutorials}}
\end{deluxetable*}
\clearpage

\subsection{Aerosol Models}\label{sec:cloud-models}

The impact of aerosols on exoplanet spectra also depends crucially on their spatial distribution within an atmosphere. We have extended \texttt{POSEIDON} to consider a variety of aerosol spatial distributions. Here, we first summarize the original aerosol parameterization in \texttt{POSEIDON},  before describing the new spatial distribution parameterizations for Mie scattering aerosols.

\subsubsection{Opaque Cloud Deck + Power law Haze}\label{sec:cloud-models_deck_haze}

Before the launch of JWST, exoplanet transmission spectra at short wavelengths showed abundant evidence of aerosols. At short wavelengths (bluewards of 0.5 \textmu m), this manifested as scattering in the form of `enhanced' Rayleigh slopes. While in the infrared, aerosols muted H$_2$O bands at 1.8 \textmu m \citep[e.g.,][]{Wakeford2019}. There was also evidence for patchy clouds on the terminator \citep[e.g.,][]{Barstow2020}.  

\texttt{POSEIDON} has generally employed an opaque cloud deck combined with a power law `haze' for transmission spectra aerosol models \citep[e.g.,][]{MacDonaldMadhusudhan2017, Alderson2022}. This model places a surface / cloud deck at the cloud pressure $P_{cloud}$, with the atmosphere assumed completely opaque for deeper layers with $P > P_{cloud}$. Cloud decks such as this mute out and flatten spectra by increasing a planet’s size in all wavelengths. The layers above the cloud are described by a generic haze power law given by
\begin{equation}
\kappa_{cloud}(\lambda,P) = n_{tot} \; a \; \sigma_0 \; (\lambda/\lambda_0)^\gamma \;\;\;\;; P < P_{cloud}
\end{equation} 
where $\lambda_0$ = 350\,nm (reference wavelength), $\sigma_0$ is the Rayleigh scattering cross section at 350\,nm, $a$ is the Rayleigh enhancement factor, and $\gamma$ is the scattering slope. 
The user can optionally allow for patchy clouds with this model, either with a cloud fraction parameter ($\bar{\phi_{cloud}}$; \citealt{MacDonald2017}) or with multiple parameters to describe the spatial location and extend of a 3D `Iceberg' cloud \citep{MacDonald2022}. This cloud deck + haze model has been successful in detecting the presence of aerosols in planetary atmospheres in the optical and near-infrared without assuming the identity of said aerosol \citep[e.g.,][]{Pinhas2019,Alderson2022,Taylor2023b}. However, the enhanced precision and mid-infrared capabilities of JWST motivate more sophisticated aerosol models.

\subsubsection{Mie Scattering `Fuzzy Deck' }\label{sec:cloud-models_fuzzy}

\begin{figure*}[ht!]
     \centering
     \includegraphics[width=0.80\textwidth]{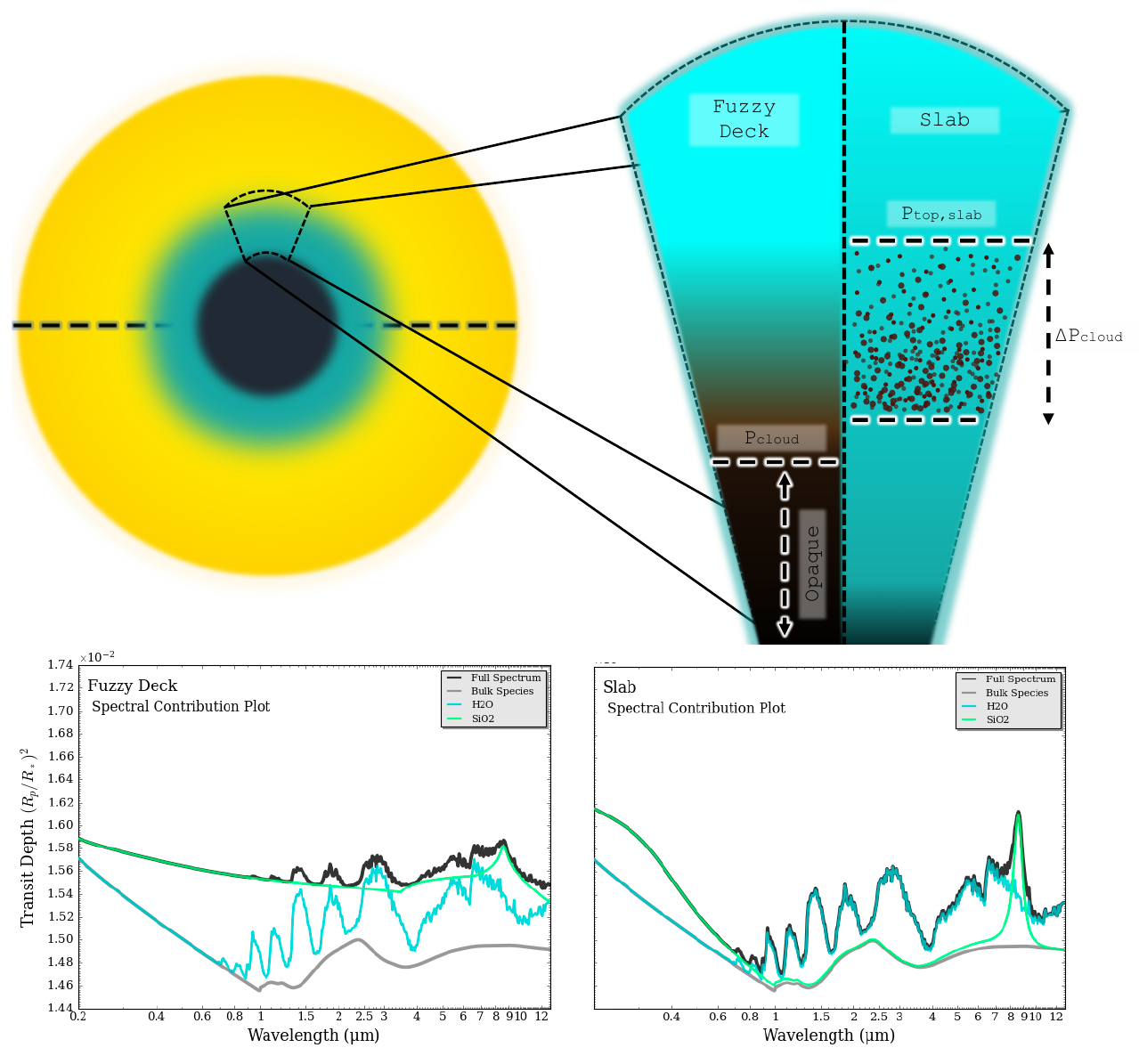}
     \caption{Top: Graphic displaying the two main Mie aerosol models introduced to \texttt{POSEIDON}. The Fuzzy Deck is modeled with an opaque cloud deck with a `fuzzy' cloud top of aerosols. The Slab is modeled with a cloud top pressure, cloud width, and constant mixing ratio for aerosols in the cloud. Both are parameterized with mean particle radius. Bottom: Example transmission spectra utilizing the two aerosol models are shown. There are a myriad of other aerosol models that are built from these two models : opaque deck + slab(s), fuzzy deck + slab(s), uniform mixing ratio with pressure (uniform X), and opaque deck + uniform X. For more a  more in-depth look at aerosol models and their effect on spectra, refer to the `Clouds in Transmission Spectroscopy' tutorial in \href{https://poseidon-retrievals.readthedocs.io/en/latest/content/forward_model_tutorials.html}{Forward Model Tutorials}.}
     \label{fig:cloud-models}
\end{figure*}

We have added two main categories of aerosol spatial models to \texttt{POSEIDON} for specific Mie scattering aerosol species (Figure \ref{fig:cloud-models}). The first is the `Fuzzy Deck' model, adapted directly from \texttt{PLATON} \citep{Zhang2019}. This model is similar to the cloud deck + haze model in that it includes an opaque cloud deck at $P_{top, \, deck}$. However, above the cloud deck, this aerosol model prescribes the number density of aerosols as
\begin{equation}
n = n_0 \; exp \left(\frac{-h}{f_H \, H_{gas}}\right)
\end{equation} 
where $n_0$ is the maximum number density of the aerosol species at the cloud deck, $h$ is the altitude above the cloud deck, $H_{gas}$ is the atmospheric scale height, and $f_H$ is a free parameter that weighs how fast the number density decreases with height (i.e., the `fuzziness'). Alongside these spatial distribution parameters, the Fuzzy Deck also requires a mean particle radius, $r_m$, and the specific aerosol composition. Similar aerosol models have been used in brown dwarf codes such as \texttt{Brewster} \citep[e.g.,][]{Burningham2021} to model opaque iron decks in the deep atmosphere. The Fuzzy Deck model produces similar aerosol distributions to the models of \citet{AckermanMarley2001} and \citet{Pont2013}, where larger particles sediment and form an opaque cloud deck while smaller particles can remain lofted above the deck. One important feature of the Fuzzy deck model is that the aerosol volume mixing ratio is not constant with altitude (due to $f_H \neq 1$). In particular, we note that for $f_H > 1$ the aerosol number density decreases slower than the background gas, which causes the aerosol volume mixing ratio to increase with altitude. We test allowing $f_H$ to exceed one in \S~\ref{sec:different_cloud_models}.

\subsubsection{Mie Scattering `Slab' Model}\label{sec:cloud-models_slab}

The second new aerosol model in \texttt{POSEIDON} is the `Slab’ model. This model is more agnostic towards aerosol formation mechanisms and does not include any opaque regions. The slab model is described by a slab top pressure, $P_{top, \, slab}$, a vertical slab width, $\Delta P$, and the (constant) aerosol mixing ratio within the slab. As with the Fuzzy Deck, the mean particle radius, $r_m$, and the aerosol composition are also required for the Mie scattering calculation. Since the slab model does not assume any aerosol formation mechanism, the retrieved aerosol identity and properties can be used to compare against both bottom-up equilibrium and top-down disequilibrium aerosol formation \citep[e.g.,][]{AckermanMarley2001, Helling2006}. We show a graphical representation of the Fuzzy Deck and Slab aerosol models, alongside their effect on transmission spectra, in Figure \ref{fig:cloud-models}.

\subsubsection{Mie Scattering `Hybrid' Models }\label{sec:cloud-models_hybrid}

We have additionally introduced a myriad of `Hybrid' aerosol models that are a combination of the ones above. Namely, one can define multiple slabs, an opaque deck + slab(s), a fuzzy deck + slab(s), a uniform with pressure Mie scattering cloud (i.e., defined by two parameters: a constant mixing ratio and the mean particle radius), and, finally, an opaque deck + uniform with pressure cloud. We also have implemented a 1+1D implementation of patchy clouds for each of these models, where $f_{cloud}$ defines the fraction of cloud coverage \citep[e.g.,][]{Marley2010, Morley2014}. Patchy cloud models calculate spectra as a linear combination of the spectrum from a clear atmosphere and a cloudy atmosphere, weighted by (1-$f_{cloud}$) and $f_{cloud}$, respectively.\footnote{For more information on the aerosol models, and how they imprint onto transmission spectra, see `Clouds in Transmission Spectroscopy' tutorial in \href{https://poseidon-retrievals.readthedocs.io/en/latest/content/forward_model_tutorials.html}{Forward Model Tutorials}.}

\subsubsection{Retrieval Aerosol Model Comparison}\label{sec:aerosol_model_comparison}

\nt{We now compare our new aerosol models directly to aerosol models found in other atmospheric retrieval codes. The main difference between our aerosols models and those implemented in other codes is that we chose to parameterize the number density/volume mixing ratio of aerosols with respect to pressure, whereas other codes typically parameterize optical depth. Another difference is that while our aerosol optical properties ($\sigma_{\mathrm{ext,eff}}$, $g$, $\omega$) are precomputed from wavelength-dependent refractive index data with respect to mean particle radius $r_m$, other retrieval codes will parameterize aerosol optical properties with the asymmetry parameter and single scattering albedo being free parameters \citep{ Marley2014, Mukherjee2021, Taylor2021} or derive optical properties from freely fit wavelength-dependent imaginary refractive index spectrum \citep{Irwin2015}.}

\nt{Our `Fuzzy Deck' model is adapted directly from \texttt{PLATON} \citep{Zhang2019} and defines the number density of aerosols above an opaque deck. This model is similar to the `deck' model found in \texttt{BREWSTER} \citep{Burningham2017} that is defined to become optically thick ($\tau \sim 1$ at 1 \textmu m) at some pressure with a decay height that parameterizes how the optical depth of the cloud falls off with decreasing pressure.} 

\nt{Our slab model defines the vertical extent of aerosols in pressure space and the uniform mixing ratio of aerosols in the slab. This model is similar to the `slab' model found in \texttt{BREWSTER} \citep{Burningham2017} that is defined by the vertical extent of aerosols in pressure space, and the total optical depth at 1 \textmu m. Both the `deck' and `slab' aerosol models in \texttt{BREWSTER} compute wavelength dependent optical-properties from refractive index data of aerosols. Our slab model is also similar to the box cloud model and single cloud profile model (case 1 and 2) found in \citet{Mukherjee2021}. Their box cloud model defines the vertical extent of aerosols in pressure space with a constant optical depth, whereas their single cloud profile model is defined with a cloud base pressure and an optical depth that falls off with respect to a freely retrieved scale height (similar to our `Fuzzy Deck' aerosol number density parameterization above the opaque deck). The optical depth, asymmetry parameter, and single scattering albedo in their models are free parameters that are retrieved. \texttt{ExoRel$^\Re$} \citep{Damiano2020} implements a slab-like model that incorporates feedback on the gas-phase abundances. Specifically, their model defines the uniform volume mixing ratio of an aerosol precursor gas below the cloud, the pressure of the cloud top, the vertical extent of the cloud, and the condensation ratio that defines how much of the aerosol-precursor gas forms into aerosols in the cloud layer. From this, the density and particle size of aerosols within the cloud layer are computed, and from that wavelength-dependent optical properties.}

\nt{Our uniform mixing ratio model defines a constant-with-pressure volume mixing ratio of an aerosol throughout the entire model atmosphere. In \citet{Taylor2021}, a uniform mixing ratio gray opacity cloud with a parameterized single scattering albedo is implemented into the \texttt{NEMESIS} retrieval framework.} 

\nt{Our hybrid opaque deck + slab model defines an opaque deck extending from some pressure in addition to a compositionally-specific aerosol slab. This is similar to the 2-cloud model found in \citet{Marley2014}, \citet{Lupu2016}, and \citet{Nayak2017}, which defines a bottom cloud layer that extends to the bottom of the model atmosphere with a constant single scattering albedo and an upper cloud slab with a constant single scattering albedo, asymmetry factor, and total optical depth. In this model, the bottom cloud layer is optically thick and acts as a reflective surface contrasting with our infinitely-absorptive deck. A similar model is utilized in \citet{Feng2018}, where their model has been adapted for terrestrial emission+reflection spectroscopy with a reflective surface (in lieu of a reflective bottom cloud layer) and patchy water clouds. The inclusion of surface albedos into \texttt{POSEIDON} will be the focus of future work. Our hydrbid multiple slab model is similar to the double cloud model (cases 3 and 4) in \citet{Mukherjee2021}.}

\nt{\texttt{petitRADTRANS} has implemented an adaptation the \citet{AckermanMarley2001} cloud model, which balances the upward turbluent mixing of aerosols and vapor (parameterized by $K_{zz}$) and the downward transport of aerosols through sedimentation (parameterized by $f_{sed}$). This model solves for both the vertical extent and particle sizes (as a function of pressure) of aerosols. Our fuzzy deck and opaque deck + slab models, which assume a constant particle size and instead vary number density or volume mixing ratio, can mimic the \citet{AckermanMarley2001} cloud model in cases where large particles sediment to form an optically thick region and small particles stay lofted.}


\subsection{Radiative Transfer Updates} \label{sec:rad_model_improvements}

Here, we present our implementation of new radiative transfer functionality in \texttt{POSEIDON} for modeling spectra of secondary eclipses or directly imaged objects. We describe, in turn, the extension of thermal emission to consider multiple scattering and a newly added reflected starlight prescription. Both emission and reflection can be combined in forward models and retrievals.

\subsubsection{Emission Spectra with Multiple Scattering}\label{sec:MS}

\texttt{POSEIDON}'s initial implementation of emission spectra radiative transfer, described in \citet{Coulombe2023}, used a simple single-stream approach with no directional scattering. In this prescription, the optical depth attenuates intensity across an atmospheric layer by assuming that any scattered light is lost to the beam. However, the presence of strongly scattering clouds in substellar objects necessitates a more complex radiative transfer model that considers directional scattering.

Our multiple scattering prescription adapts and closely follows the techniques used in the open-source radiative transfer code \texttt{PICASO} \citep{Batalha2019, Mukherjee2023}. \texttt{PICASO} uses the two-stream approach to multiple scattering radiative transfer from  \citet{Toon1989} and \citet{Cahoy2010}, which calculates the net upwards and downwards flux in each atmospheric layer and returns the resultant emergent flux from the top of the atmosphere. 
In order to link \texttt{POSEIDON}'s opacity and atmosphere modules with \texttt{PICASO}'s radiative transfer functions, we require the overall asymmetry parameter and single scattering albedo in each atmospheric layer. For radiative transfer purposes, these optical properties must consider the relative balance of both Rayleigh and Mie scattering. The weighted single scattering albedo in each layer is given by
\begin{equation}
\bar{\omega} = \frac{\Delta\tau_{Ray}}{\Delta\tau_{tot}} \; \omega_{Ray} + \frac{\Delta\tau_{Mie}}{\Delta\tau_{tot}} \; \omega_{Mie}
\end{equation}
where $\Delta\tau_{tot}$ is the total optical depth across the layer (with contributions from Rayleigh scattering, chemical absorption, and Mie extinction), $\Delta\tau_{Ray}$ is the layer optical depth from Rayleigh scattering alone, $\Delta\tau_{Mie}$ is the layer Mie scattering extinction optical depth, $\omega_{Ray}$ = 1 (since Rayleigh scattering is a pure scattering process), and $\omega_{Mie}$ is the Mie scattering single scattering albedo (see \S~\ref{sec:Mie-scattering}). The weighted asymmetry parameter is similarly given by
\begin{align}
\bar{g} = \frac{\omega_{Ray} \; \Delta\tau_{Ray} \; g_{Ray} + \omega_{Mie} \; \Delta\tau_{Mie} \; g_{Mie}}{\omega_{Ray} \; \Delta\tau_{Ray} + \omega_{Mie} \; \Delta\tau_{Mie}} 
\end{align}
where $g_{Ray}$ = 0 since Rayleigh particles are symmetric scatterers. The relative balance of back-scattering, forward-scattering, and multiple-scattering in both the upward and downward directions is encoded in the weighted asymmetry parameter and single-scattering albedo. We then invert the altitude order of these variables (since \texttt{POSEIDON} uses layer `0' at the bottom of the atmosphere, the opposite convention to \texttt{PICASO}) and plug these variables into \texttt{PICASO}'s implementation of \citet{Toon1989}'s radiative transfer equations.\footnote{More details on our implementation are included in the `\texttt{POSEIDON} Preamble' in the emission\_Toon() function found in \texttt{POSEIDON}'s \texttt{emission.py}. How thermal scattering imprints on emission spectra can be seen in the `Thermal Scattering' and `A Y-Dwarf with Scattering Clouds' tutorials in \href{https://poseidon-retrievals.readthedocs.io/en/latest/content/forward_model_tutorials.html}{Forward Model Tutorials}.}

This emission scattering implementation has several advantages. First, it foregoes the need for ray tracing and makes a more complete prescription of scattering feasible for atmospheric retrievals. Second, the inclusion of forward scattering not only accurately describes large particle Mie scattering but also includes the small amount of light that is forward scattered by Rayleigh scattering. The inclusion of scattering is of utmost importance for colder planetary atmospheres with strongly scattering aerosols, such as Y-dwarfs. Furthermore, scattering can also allow for large particles to imprint spectral features, even for isothermal atmospheres, as discussed in \citet{Taylor2021} and explored further in \S~\ref{sec:forward_model_explore}.

\subsubsection{Reflection Spectra}\label{sec:reflection}

Reflected light from aerosols has been shown in the solar system to be an important source of observed flux. Historically, reflected light has only been considered an important source of flux in wavelengths $\leq$ 1 \textmu m. However, recent JWST observations of Saturn \citep{Fletcher2023} have shown that reflected light from aerosols can be an important component up to 5 \textmu m. Compared to the solar system planets, close-in hot exoplanets are expected to have dark albedos due to alkali (Na + K) absorption \citep[e.g.,][]{Marley1999,Sudarsky2000,Cahoy2010}. If the observed planet-star flux ratio of a hot gas giant is brighter than expected from theory, a back-scattering cloud or a cloud that cuts off the absorption wings of alkali species can provide an explanation \citep[e.g.,][]{Evans2013, Barstow2014}. In order to model and retrieve reflected light jointly with thermal emission, and therefore gain more information on the dayside clouds of hot Jupiters, we have added reflection spectra radiative transfer into \texttt{POSEIDON}.

Reflected light can be modeled using a multiple scattering implementation similar to that described above. The main difference between the emission scattering prescriptions described in \S~\ref{sec:MS} and reflection is the computation of separate forward and back-scattering asymmetry parameters directly from $g$, and the inclusion of the $\delta$-eddington correction. For our implementation, we do not include any Raman scattering, use the 5 Gauss angles x 1 Chebychev angle integration scheme, and we adopt the modeling recommendations from Appendix A of \citet{Batalha2019} for direct and multiple scattering.\footnote{More details on our implementation are included in the `\texttt{POSEIDON} Preamble' in the reflection\_Toon() function found in \texttt{POSEIDON}'s \texttt{emission.py}. How reflection imprints on emission spectra can be seen in the `Reflection in Hot Jupiters' tutorial in \href{https://poseidon-retrievals.readthedocs.io/en/latest/content/forward_model_tutorials.html}{Forward Model Tutorials}.} The reflection forward model solves for a wavelength-dependent geometric albedo, $A_g(\lambda)$, that is subsequently converted to the observed reflected flux via 
\begin{equation}
F_{p, \, reflected, \, obs}(\lambda) = A_g(\lambda) \; \left(\frac{R_p(\lambda)}{a_p}\right)^2 \;  \left(\frac{R_s}{d}\right)^2 \; F_s(\lambda)
\end{equation}
\noindent where $a_p$ is the planet's orbital distance from the star, $R_p(\lambda)$ is the wavelength-dependent photosphere radius computed at $\tau=2/3$, $R_s$ is the radius of the star, $d$ is the distance to the system, and $F_s(\lambda)$ is the stellar flux. The reflected flux contribution can be added directly to the emission flux to determine the total observed flux. 

We note that, since three-dimensional GCMs have found that the effects of scattering in hot Jupiters can be ignored past 5\,\textmu m \citep[e.g.,][]{Lee2017}, we allow the option to only compute reflection up to 5 \textmu m. This expedites retrievals that jointly fit reflection and emission at the same time. We therefore use this option in our HD~189733~b retrievals in \S~\ref{sec:hd189-emission-retrievals}.

\subsection{Other Improvements}\label{sec:other-improvements}

In addition to adding aerosols and scattering to \texttt{POSEIDON}, we have also added three new P-T profile parameterizations, contribution visuals, and modified alkali cross sections.

\subsubsection{P-T Profiles}\label{sec:pt_profiles}

The first new P-T profile is the double-gray Guillot profile \citep{Guillot2010}. The Guillot P-T profile is given by 
\begin{align}
T^4 &= \frac{3T_{int}^4}{4} \left(\frac{2}{3} + \tau\right) \notag \\ & \hspace{2pt} + f\frac{3T_{irr}^4}{4}\left(\frac{2}{3} + \frac{1}{\gamma\sqrt{3}} + \left(\frac{\gamma}{\sqrt{3}}-\frac{1}{\gamma\sqrt{3}}\right)e^{-\gamma\tau\sqrt{3}}\right) \; .
\label{eq:guillot}
\end{align}
Our implementation of the Guillot profile, following \texttt{petitRadtrans} \citep{Molliere2019}, uses four free parameters: $\gamma$, $T_{int}$, $T_{eq}$, and $\kappa_{IR}$. The latter two parameters are used to calculate the irradiation temperature, $T_{irr} = \sqrt{2}T_{eq}$, and the `infrared optical depth', $\tau = P \kappa_{IR}/g$ (note that this $\tau$ is simply a coordinate with the dimensions of optical depth, but it is not related to the atmospheric optical depth of the model). 
The redistribution factor, $f$, is fixed to 0.5 for hot Jupiter daysides and 0.25 for hot Jupiter terminators or directly imaged substellar objects. 

The second profile, which we refer to as the `Line' profile, is an adaptation of the Guillot profile assuming two independent downwelling visible streams and one upwelling thermal stream \citep{Line2013}. The main feature of the Line profile is that it allows for thermal inversions, whereas the Guillot profile does not. The Line profile is given by 
\begin{equation}
T^4 = \frac{3T_{int}^4}{4}\left(\frac{2}{3} + \tau\right) + \frac{3T_{irr}^4}{4}\left(1-\alpha\right)\xi_{\gamma}(\tau) + \frac{3T_{irr}^4}{4}\alpha\xi_{\gamma2}(\tau) \; .
\label{eq:line}
\end{equation}
Our implementation of the Line profile, following \texttt{Platon} \citep{Zhang2020}, has six free parameters: $\gamma$, $\gamma_2$, $\alpha$, $T_{int}$, $\beta$, and $\kappa_{IR}$. The irradiation temperature is here given by $T_{irr} = \beta T_{eq}$, where $T_{eq}$ is instead fixed to the measured equilibrium temperature (instead of fit, as in the Guillot profile above). The $\beta$ parameter serves as a  `catch-all' term for heat redistribution, geometric arguments, emissivity, albedo, and errors in the equilibrium temperature. The $\xi$ term is given by Eq. 14 in \citet{Line2013}. We note that the Line profile reduces to become equivalent to the Guillot profile when $\alpha \sim$ 1. 

The third new P-T profile is the non-parametric, `knot' P-T profile introduced by \citet{Pelletier2021}. The knot profile is defined by a set of temperature `knots', spaced uniformly in log pressure, that are interpolated to produce the P-T profile. This model includes a smoothing parameter, $\sigma_s$, which penalizes the second derivative of the P-T profile by applying an evidence penalty (see Eq. 11 in \citet{Pelletier2021}). Given that this profile makes no physical assumptions, it can be used to flexibly probe the temperature structure that is best able to fit a spectrum.

Finally, we also adapted the `slope' P-T profile parameterization \citep{PietteMadhu2021}, which was introduced into \texttt{POSEIDON} in Gressier et al. (2024). The original P-T profile, as defined in \citet{PietteMadhu2021}, fits for a photospheric temperature at 3.16 bars and a fixed number of $\Delta$T parameters between specific pressure anchor points (e.g., $\Delta T$ from 10 to 1\,mbar). The implementation in \texttt{POSEIDON} now allows the user to define the pressure of the photosphere and the pressure edges where the $\Delta T$ parameters are defined (i.e., the number of $\Delta T$ parameters is a user choice). This parameterization is much better suited than other P-T profiles at fitting a deep temperature adiabat with a decreasing temperature with height. The slope profile explicitly assumes monotonically decreasing temperature with altitude and thus does not allow for thermal inversions (i.e., the $\Delta T > 0$). We deploy all these new P-T profiles in \S~\ref{sec:hd189-emission-retrievals} and \ref{er_pt_profiles}.\footnote{For more details on P-T profiles in \texttt{POSEIDON}, see the `Pressure-Temperature Profiles' tutorial in \href{https://poseidon-retrievals.readthedocs.io/en/latest/content/forward_model_tutorials.html}{Forward Model Tutorials}}

\subsubsection{Contribution Visuals}

Contribution visuals are an important tool for interpreting forward models and retrieved spectra. We differentiate here between two types of contribution visuals: (i) the pressure contribution function highlights the atmospheric layers driving the formation of spectral features and (ii) spectral decomposition plots highlight the impact of a single chemical species on a spectrum.

Our implementation of the pressure contribution function for the i$^{th}$ atmospheric layer is given by 
\begin{equation}
C_{i, \lambda} = \delta_{\lambda} - \delta_{\lambda} (\kappa_{i, \, \lambda} = 0)
\end{equation}
where $\delta_{\lambda}$ is either the transit or eclipse depth and $\delta_{\lambda} (\kappa_{i, \, \lambda} = 0)$ is the equivalent spectrum but with the opacity in the i$^{th}$ layer set to 0. The pressure contribution function is a measure of how important a single pressure layer is to the resultant spectrum, and hence which pressure layers are being probed at a specific wavelength. We also allow users to calculate the pressure contribution for a specific chemical species or cloud, such that only the opacity for that species/cloud in the i$^{th}$ is set to 0. We recommend using log normalization per wavelength for all pressure contributions for visualization purposes.

The pressure contribution function can also be integrated over a specific wavelength range to produce a `photometric' pressure contribution function. The photometric contribution of the i$^{th}$ layer is simply $\int_{\lambda_1}^{\lambda_2} C_{i, \lambda} d\lambda$.
%
%
We recommend wavelength bins of size $\Delta \lambda = 1$\,\textmu m, to show which pressure layers contribute most over a given wavelength range. For visualization purposes, the photometric contribution is normalized to range from 0 to 1.

The second type of contribution visual is spectral decomposition plots. The spectral contribution of a single chemical species, $q$, to a spectrum is given by 
%
%
$\delta_{\lambda} (\kappa_{j\neq q} = 0)$ (i.e., the spectrum with every opacity source disabled except for species $q$ --- a `put-one-in' visual), where $\delta_{\lambda}$ can be a transmission or emission spectrum. An important aspect of this method is that the model atmosphere is initialized with all the chemical species present, such that the mean molecular weight (and hence the scale height) remains the same. Maintaining the background mean molecular weight for spectral decomposition is especially important for high-metallicity atmospheres. We note that we generally display emission spectra decompositions in terms of the brightness temperature, defined as
\begin{equation}
T_{bright}(\lambda) = \left(\frac{hc}{k_b \lambda}\right) \left(\ln\left[1 + \frac{2 h c^2}{(F_{p, \lambda}/\pi) \lambda^5}\right] \right)^{-1}
\end{equation}
where $F_{p, \lambda}$ is the emergent flux of the atmosphere. 

Example contribution plots, for both pressure contributions and spectral decomposition, can be seen in Figures~\ref{fig:cloud-models}, \ref{fig:transmission-MNS-Contribution}, and \ref{fig:emission-contribution}. Further contribution examples can be found in the tutorial notebooks in \texttt{POSEIDON}'s documentation\footnote{For more details on contribution visuals for transmission and emission spectra, see the `Transmission Spectra Model Visuals' and `Emission Spectra Model Visuals' tutorials in in \href{https://poseidon-retrievals.readthedocs.io/en/latest/content/forward_model_tutorials.html}{Forward Model Tutorials}}. 

\subsubsection{Opacity Database Updates}

\texttt{POSEIDON}'s opacity database for gas-phase chemical species is currently undergoing a major update to use the latest high-temperature line lists. The new opacity database will be released to coincide with \texttt{POSEIDON} v1.2\footnote{\href{https://poseidon-retrievals.readthedocs.io/en/latest/content/opacity_database.html}{\texttt{POSEIDON}'s Opacity Database}}. This study uses the existing opacity database for molecules, but we found during the retrieval investigations in \S~\ref{sec:hd189-transmission-retrievals} that an update to the alkali line wings was necessary to adequately fit HD~189733b's secondary eclipse spectrum.

\begin{figure}[b!]
     \centering
        \includegraphics[width=0.95\linewidth]{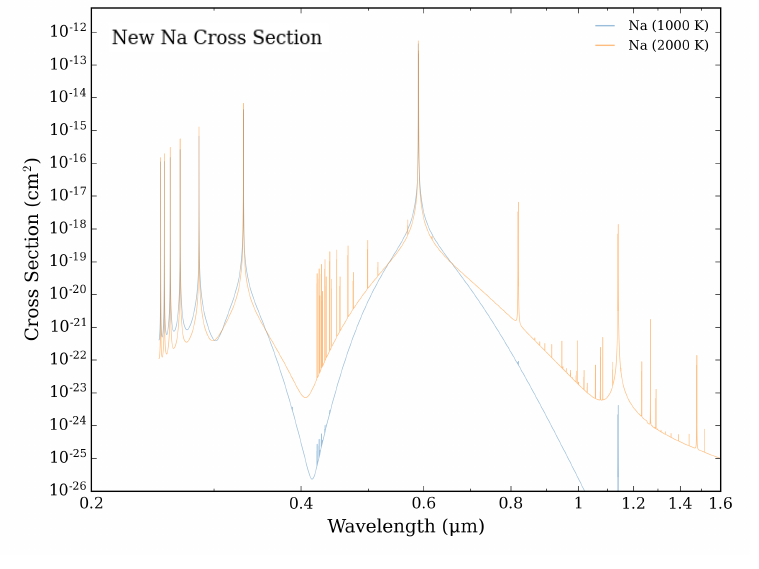}
     \caption{Updated Na cross section in \texttt{POSEIDON}. The retrievals in this paper use a sub-Voigt line wing prescription for all lines, based on the wing fit from \citet{Baudino2015} for the Na and K resonance doublets.}
     \label{fig:New-Na-cross-sections}
\end{figure}

We updated the alkali cross sections (namely Na and K) by modeling the wings for all lines with a sub-Voigt profile, after \citet{Baudino2015}, as shown in Figure~\ref{fig:New-Na-cross-sections}. Previously, \texttt{POSEIDON} used this sub-Voigt line wing shape only for the Na and K resonance doublet while other lines used Voigt profiles (with a cutoff at the minimum of 500 Voigt half width half maxima or 30\,cm$^{-1}$ from the line core). This alkali opacity update was motivated by our forward model hot Jupiter albedo spectra being far brighter at wavelengths shorter than 0.4\,$\mu$m, where Na has several strong lines, compared to the same model in \citet{Batalha2019}. By replacing the sharp Voigt line wing cutoffs with broader sub-Voigt wings, our alkali absorption sufficiently darkens the albedo spectra to be consistent with both theoretical expectations \citep[e.g.,][]{Cahoy2010} and HD~189733b's observed albedo spectrum (see \S~\ref{sec:hd189-transmission-retrievals}).

\section{Aerosol Model Validation on HD~189733~b - Transmission}\label{sec:hd189-transmission-retrievals}

\begin{figure*}[ht]
     \centering
         \includegraphics[width=1.0\linewidth]{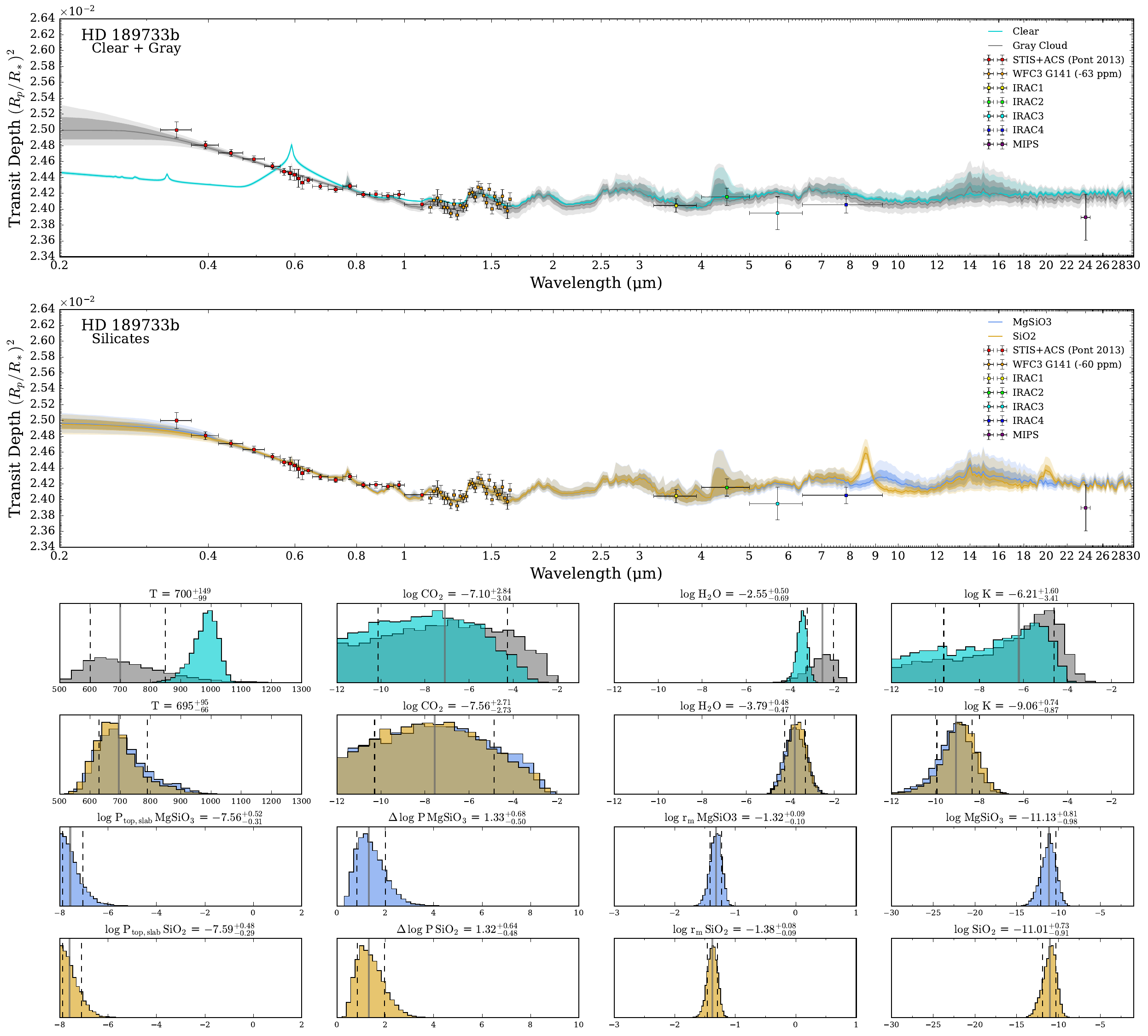}
     \caption{Atmospheric retrievals on \textit{HST} and \textit{Spitzer} transmission data of HD~189733~b, with retrieved offsets applied to the WFC3 data. Retrieved spectra display median retrieved spectra (solid lines), with 1$\sigma$ and 2$\sigma$ confidence intervals (dark and light shaded regions). Top: Retrieval results for a clear atmosphere (aqua) and an atmosphere with a parameterized gray cloud and haze (gray). Middle: Retrieval results for silicate aerosol species, namely enstatite (MgSiO$_3$, blue) and crystalline quartz (SiO$_2$, yellow). Note the silicate absorption features at 9.6 \textmu m (MgSiO$_3$) and 8.5 \textmu m (SiO$_2$). Bottom panels: retrieved posteriors for atmospheric isothermal temperature, gas-phase volume mixing ratios of CO$_2$, H$_2$O, and K. Retrievals with a parameterized cloud retrieve a super-solar H$_2$O abundance and do not constrain K, while retrievals with Mie scattering aerosols retrieve sub-solar H$_2$O and K abundances. Subsequent rows display retrieved posteriors for the slab-top pressure in log space $\log$\,$P_{\rm{top,slab}}$, width of the slab in log-pressure space $\Delta\log$\,$P$, mean particle size $\log$\,$r_{\rm{m}}$, and constant aerosol mixing ratio in the slab. Aerosols are needed to fit the 0.3-1 \textmu m slope. Both silicate aerosol species retrieve a high altitude, thin, low density ($\sim$ -11) slab composed of sub-micron particles ($\sim$ 0.1 \textmu m).}
     \label{fig:transit-clear-gray-silicates}
\end{figure*}

\begin{figure*}[ht]
     \centering
         \includegraphics[width=1.0\linewidth]{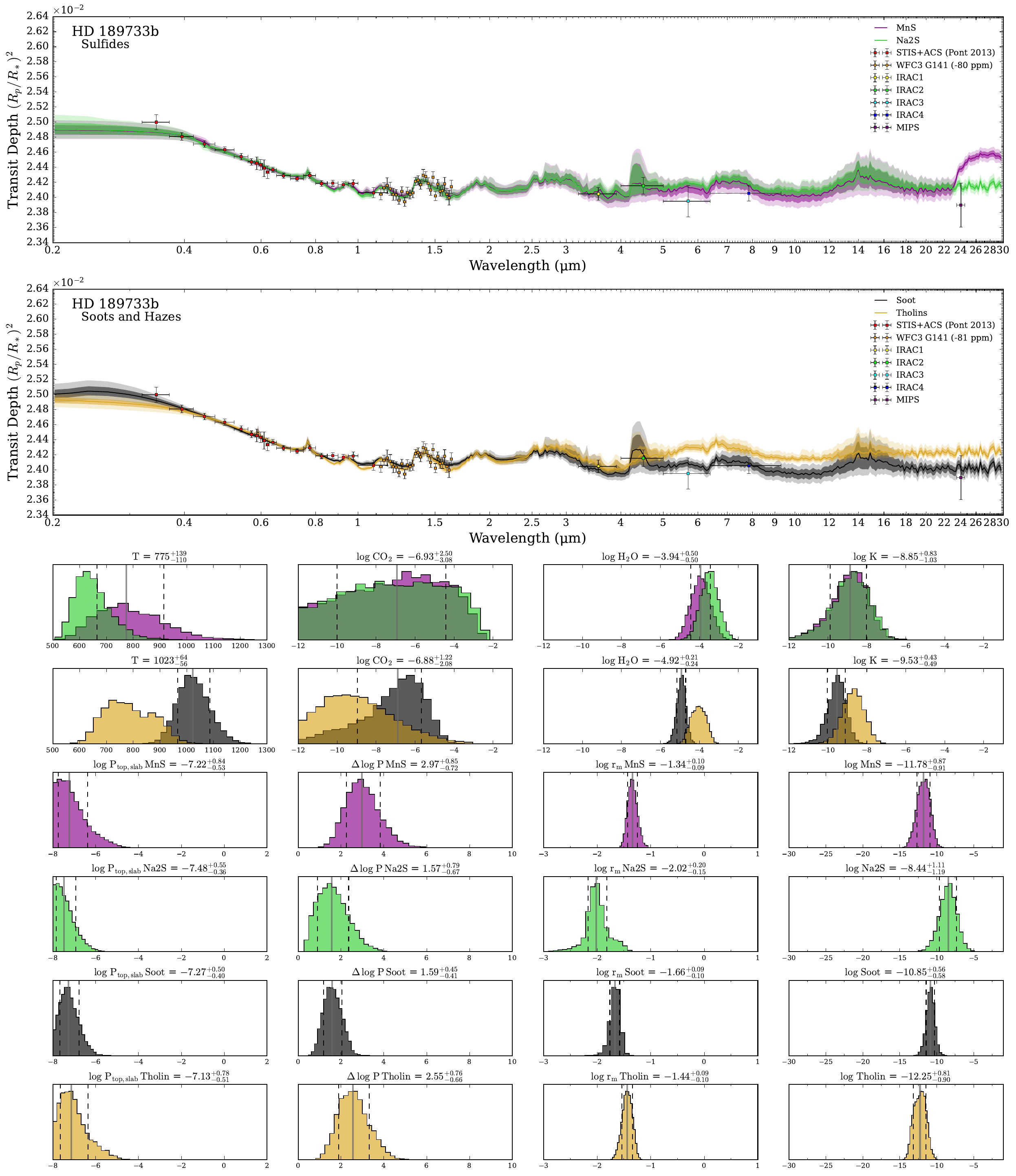}
     \caption{Same as Figure \ref{fig:emission-clear-gray}, but highlighting the retrieval results when including sulfide, soot, and haze aerosols. Top: Retrieval results for sulfide aerosol species, namely manganese sulfide (MnS, purple) and sodium sulfide (Na$_2$S, green). Middle: Retrieval results for disequilibrium aerosols, namely flame soot (Soot, black) and Titan tholins (Tholins, gold). MnS has an absorption feature in the longest wavelengths, 23-30 \textmu m. We note that the longer wavelength data used is from a sulfide extrapolation, not lab data. Na2S has no signifigant absorption in the 0.2-30 \textmu m range. Soot is featureless save a small feature in the shortest wavelengths (0.2-0.3 \textmu m). Tholins have a large feature from 5-8 \textmu m with contiuum opacity from 9-30 \textmu m. All four cloud species retrieve properties consistent with silicates (high altitude, thin, low density slab composed of sub-micron sized particles).}
     \label{fig:transit-sulfides-soots}
\end{figure*}

We benchmark the new aerosol capabilities introduced above on available \textit{HST} and \textit{Spitzer} data of the well studied hot Jupiter HD~189733~b. HD~189733~b is a hot Jupiter with an orbital period of 2.2 days. HD~189733~b has a mass, radius, semi-major axis, and equilibrium temperature of $1.13 \, \rm{M}_{\rm{Jup}}$, $1.13 \, \rm{R}_{\rm{Jup}}$, $0.03142 \, AU$, and $1200 \, \rm{K}$, respectively \citep{Southworth2010, Sing2016, Stassun2017}. This target has been studied in much detail pre-JWST due to its relative closeness (20pc), and its tidally-locked orbit and permanent day-side irradiation inducing very interesting conditions at the day-night terminator that transmission spectra probes. In particular, the terminator is at a temperature where many silicate and alkali species are expected to condense, making the transmission spectra of HD~189733~b a perfect test-bed for our new aerosol features \cite[e.g.,][]{Showman2002,Knutson2007, Kataria2016, Lee2016}. .



\subsection{Datasets and Offsets}\label{sec:datasets}

For our analysis of the transmission spectrum, we utilized the \nt{passband-}averaged \textit{HST}/STIS G430L, G750M and ACS G800L data from \citet{Pont2013}, \textit{HST}/WFC3 G141 from \citet{McCullough2014}, all four \textit{Spitzer} IRAC \nt{channels} and MIPS from \citet{Pont2013}. \nt{The passband-averaged data from \citet{Pont2013} was generated by averaging combined instrumental datasets across arbitrary passbands.} Transmission data used in our analysis can be found in Table \ref{table:transmission_data} and is consistent with data utilized in \citet{Zhang2020}.

It has been shown in extant work that data offsets are preferred when fitting multiple transmission datasets due to instrumental, limb darkening, and data reduction effects \citep[e.g.,][]{Zhang2020,MacDonald2019}.  We fit for one offset in the transmission dataset, namely allowing the \textit{HST}/WFC3 data to have free offsets between -100 to 100 ppm, consistent with the retrieval performed in \citet{Zhang2020}.

\begin{deluxetable}{llll}
\colnumbers
\tablecolumns{4}
\tablecaption{Fixed Constants\label{table:fixed_constants}}
\tablehead{ & Variable & Value & Reference}
\startdata
 & M$_{\rm{p}}$ [M$_{\rm{J}}$] & 1.129 & \citet{Stassun2017} \\
Planetary & d [pc] &  19.7638 & $\cdots$ \\
 & a$_{\rm{p}}$ [AU] & 0.03142 & \citet{Southworth2010}\\
 & T$_{\rm{eq}}$ [K] & 1200 & \citet{Sing2016} \\
\hline
 & R$_{\rm{s}}$ [R$_\odot$] & 0.78 & \citet{Stassun2017}) \\
Stellar& T$_{\rm{s}}$ [K] & 5014 & \citet{Stassun2017} \\
 & log $g_{\rm{s}}$ [log10 cm/s$^2$] & 4.58 & \citet{Addison2019} \\
\enddata
\tablecomments{Planetary and stellar constants (1), variable name and unit (2), value assumed in this work (3), and pertinent reference (4). Stellar spectrum loaded in with the PHOENIX grid, which assumes solar metallicity.}
\end{deluxetable}

\subsection{Retrieval Configuration}\label{sec:transmission_config}

We conduct free-chemistry retrievals with and without the inclusion of a parameterized gray cloud and compositionally specific aerosols on the transmission spectrum of HD~189733~b. Planetary and stellar constants can be found in Table \ref{table:fixed_constants}. We assume an isothermal pressure–temperature profile for each model, which has been shown to be sufficient in most transmission geometries \citep[e.g.,][]{MacDonald2020}. We initialize the atmosphere from 100 to 1e-8 bars with 1 bar being the reference pressure. We assume a H$_2$-He dominated atmosphere, with a fixed H$_2$-He fraction (with He/H$_2$ = 0.17), and freely retrieve the gas phase mixing ratios of alkalis (Na, K), major carbon bearing species (CH$_4$, CO, CO$_2$), H$_2$O, HCN, and NH$_3$. Retrievals include continuum absorption from H$_2$ and He collision-induced absorption \nt{(H$_2$-H$_2$, H$_2$-He)} and H$_2$ Rayleigh scattering. \nt{H$_2$-CH$_4$, H$_2$-CO$_2$, CO$_2$-CH$_4$, and CO$_2$-CO$_2$ collision-induced absorption is also included.} The specific line lists and CIA parameterizations used are detailed in the appendix of \citet{MacDonald2022}, with particular references specified in Table \ref{tab:line-lists}.  For compositionally specific aerosols, we include the slab aerosol model parameterized by the mean particle size, $r_{\rm{m}}$, the cloud-top pressure, $P_{\rm{top,slab}}$, the width of the cloud in log pressure space, $\Delta\log$\,$P$, and the constant log mixing ratio of the aerosol in the cloud. The effective aerosol extinction cross section is pulled from the precomputed database with mean particle sizes ranging from 0.001--10\,\textmu m and wavelengths spanning 0.2--30\,\textmu m at $R=1000$. We retrieve with six distinct aerosol species: two silicate species, crystalline silicon dioxide (SiO$_2$ (crystalline)), enstatite (MgSiO$_3$); two sulfides species, manganese sulfide (MnS (Morley 2012)), sodium sulfide (Na$_2$S); and two soots/hazes: flame soot, and titan Tholins. \nt{See Appendix \S \ref{sec:appendex_refractive_indices} for specific details on refractive index data utilized.} Model transmission spectra are computed at a spectral resolution of R = 10,000 from 0.2-30.0 \textmu m. Priors for all transmission models can be seen in Table \ref{table:transmission_results}. Retrievals sample the parameter space using \texttt{MultiNest} \citep{Feroz2008} with 1,000 live points and a stopping criteria of a difference of 0.5 evidence between all live points. Results are presented in \S~\ref{transmission:results}.

In \S~\ref{sec:different_cloud_models}, we explore different aerosol models and patchy clouds, choosing one cloud species (MgSiO$_3$), with priors for those models listed in Table \ref{table:transmission_cloud_model_results}. In \S~\ref{sec:aerosol_showcase}, we only run retrievals on the \textit{HST} STIS and WFC3 data, with an offset on the WFC3 data, and a chemical inventory of just H$_2$O and K with cloud species ranging from the super-hot category to ices (Table \ref{table:aerosol_species}), in order to showcase the infrared absorption and scattering slopes of multiple species. We run these exploratory retrievals at a spectral resolution of R = 5,000 from 0.2-30.0 \textmu m with 500 live points for expediency.

\subsection{Results} \label{transmission:results}

As a point of comparison with the new aerosol features, we conducted retrievals with a clear atmosphere and an atmosphere with a parameterized gray aerosol model. Results for this and silicate species can be seen in Figure \ref{fig:transit-clear-gray-silicates}. Results for the sulfide clouds and the soots/hazes can be seen in Figure \ref{fig:transit-sulfides-soots}. All retrieved posteriors for each model can be found in Table \ref{table:transmission_results}. We find that aerosols are needed to fit the transmission spectrum of HD~189733~b, in line with many extant studies \citep[e.g.,][]{Pont2013, Zhang2020}. Notably, without aerosols the clear retrieval is unable to fit for the slope found shortwards of 1 \textmu m. The gray aerosol model retrieved an opaque cloud deck extending up to $\sim$ 0.1 bars, with a steep parameterized scattering slope. The low retrieved isothermal temperature ($\sim$ 700K) for the gray aerosol model compared to the equilibrium temperature (1200K, \citet{Sing2016}) is likely a byproduct of assuming a one-dimensional atmosphere \citep{MacDonald2020}. The model with gray aerosols detects a super-solar water abundance ($\sim$ -2.5) and does not detect carbon dioxide, sodium, and potassium\footnote{\nt{Solar abundance mixing ratios we are comparing to are log Na = -5.78, log K = -6.93, and log H$_2$O = -3.31 \citep{Asplund2021}. \citet{Barstow2014} use 5 ppmv Na (log Na = -5.3) as approximately solar. Utilizing the chemical equilibrium grids found in \citet{Marley2021}, log CO$_2$ is approximately -7 at 10mbar, 1x solar metallicity (C/O ratios ranging from 0.25x to 1.25x solar, where solar C/O $\approx$ 0.9 \citep{Teske2014}.)}}. The retrieved offset for the \textit{HST}/WFC3 is $\sim$ -60 ppm (note that a positive offset means the data is shifted `down' on the y-axis), consistent with the retrieved offset found in \citet{Zhang2020}. This offset results in the final \textit{HST}/STIS+ACS data point around 1.1 \textmu m becoming the first point in the first water absorption feature centered at 1.15 \textmu m.

Silicate aerosols are able to fit the transmission spectrum with a low density ($\sim$ -11) high altitude slab spanning about a bar from the top of the initialized atmosphere. Both silicate species retrieved sub-micron particle sizes ($\sim$ 0.01 \textmu m) and imprint aerosol absorption features in the infrared between 8-10 \textmu m. The models agree with the gray model in both isothermal temperature, unconstrained carbon dioxide abundances, and dataset offset, but disagree slightly in water and potassium abundances. Notably, the inclusion of specific silicate clouds result in retrieved water abundances that are an order of magnitude lower ($\sim$ -3.7) and a constrained potassium abundances ($\sim$ -9). The inclusion of compositionally specific Mie scattering aerosols allows for potassium to be detected and constrained, and retrieves a sub-solar water abundance in lieu of a super-solar one.

Sulfide aerosols are also able to fit the transmission spectrum with slight deviations in retrieved aerosol properties. MnS fits for a low density ($\sim$ -12) high altitude slab spanning about three bars from the top of the initialized atmosphere composed of sub-micron particle sizes ($\sim$ 0.01 \textmu m). MnS fits for a slightly hotter atmosphere ($\sim$ 800K). Na$_2$S fits for a slightly higher density ($\sim$ -8) high altitude slab spanning about one-two bars from the top of the initialized atmosphere composed of much smaller particle sizes ($\sim$ 0.001 \textmu m). MnS imprints a feature from 23-30 \textmu m, whereas Na$_2$S imprints no observable feature. Both sulfide species retrieve gas properties consistent with silicate aerosol retrievals. 

Photochemically produced tholins are able to fit the data with aerosol properties nearly identical to MnS, but has difficulty fitting \textit{Spitzer} points due to additional continuum opacity at longer, infrared wavelengths. Tholin has an aerosol absorption feature from 3-5 \textmu m, explaining the lack of a carbon-dioxide feature and retrieved low ($\sim$ -9) abundance. Flame soot retrieved aerosol properties nearly identical to other aerosols, but has notable differences in other retrieved properties. The retrieved isothermal temperature for soot is hotter than the other retrievals ($\sim$ 1000K), fits for a pronounced carbon dioxide peak at 4.5 \textmu m ($\sim$ -7), and smaller water ($\sim$ -5) and potassium ($\sim$ -9.5) abundances. Flame soot is unique in that it has a feature in the shortest wavelengths (0.2-0.3 \textmu m). 

\nt{To further visualize the opacity contributions of each aerosol model an optical depth plot (versus wavelength and pressure) can be seen in Appendix Figure \ref{fig:optical-depth}.} Statistical analysis for the clear vs gray model, and the compositionally specific aerosol models can be found in Table \ref{table:transmission_stats}. A gray aerosol parameterization is preferred over the clear model by 19.1 $\sigma$. MnS is the preferred aerosol species (by 2-5 $\sigma$ over other species), but we note that long wavelength MnS refractive indices are extrapolated from other laboratory data. A spectral contribution plot for the MnS retrieval can be seen in Appendix Figure \ref{fig:transmission-MNS-Contribution}. All aerosol models prefer a high altitude, thin, low density cloud composed of sub-micron sized particles. We compare our results with extant retrieval results in \S~\ref{sec:Dicussion}.

\subsection{Different Aerosol Models}\label{sec:different_cloud_models}

\begin{figure*}[ht]
     \centering
         \includegraphics[width=1.0\textwidth]{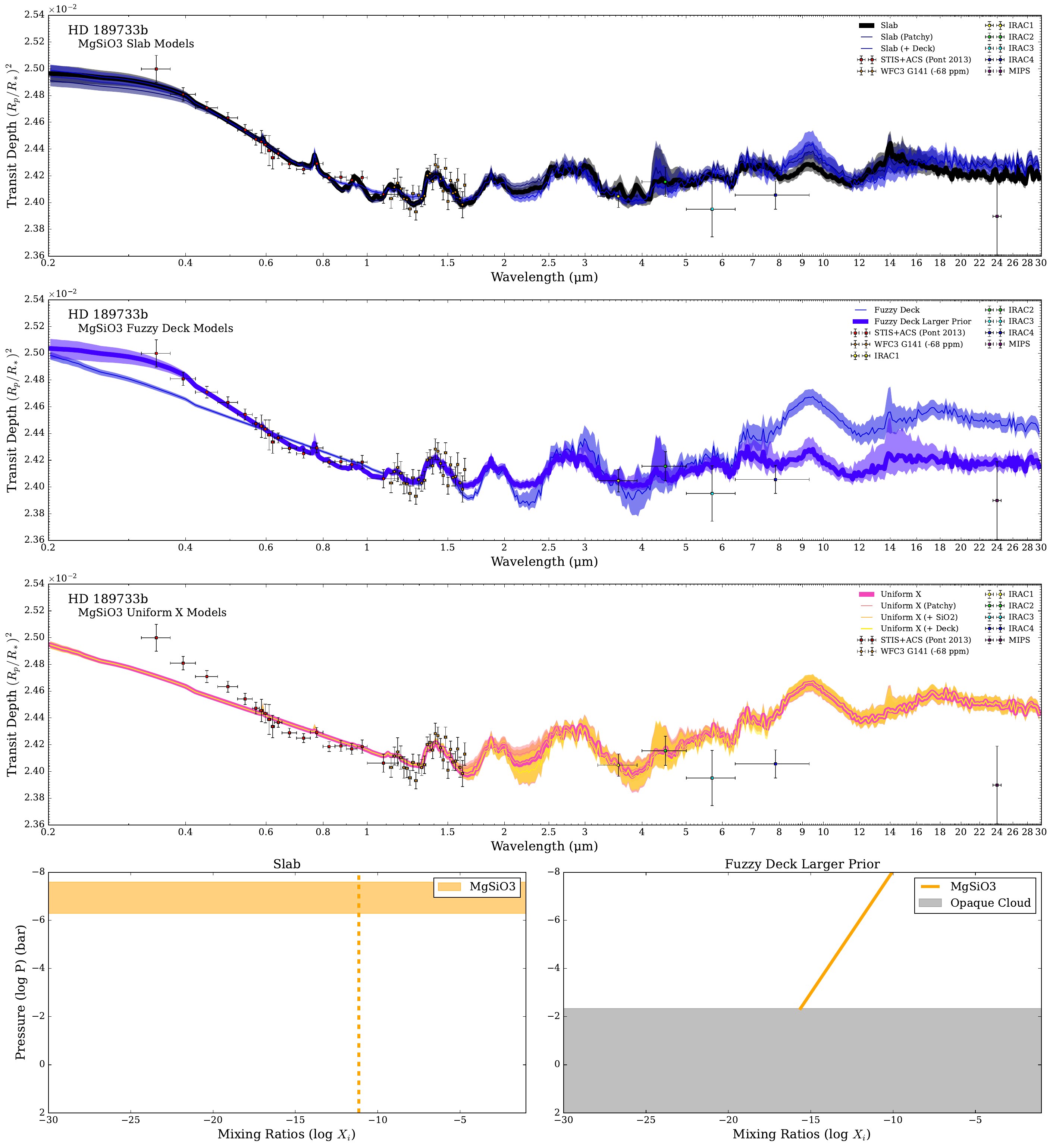}
     \caption{Top: Resulting transmission spectra from atmospheric retrievals assuming different aerosol models \nt{(median spectra and 1-sigma regions; from top to bottom: slab models, fuzzy deck models, and uniform X models)}. We find that the fuzzy deck larger prior (purple, thick line) results in the best fit that can fit the short wavelength slope without imparting a strong silicate feature in the longest infrared wavelengths. The modified fuzzy deck is statistically consistent with the slab model (black, thick line). All uniform mixing ratios retrieve nearly identical spectra that cannot fit the slope (pink, thick line). Bottom: Panels displaying retrieved aerosol properties for the slab and modified fuzzy deck models. (Left) Shaded region represents vertical extent of slab in log pressure space, with the dotted vertical line representing the mixing ratio of the aerosol in that region. A low density, high altitude, thin slab was retrieved. (Right) Gray shaded region represents vertical extent of opaque deck in log pressure space, with the solid line representing the mixing ratio of the aerosol above the deck. The fuzzy deck larger prior retrieves an opaque deck at 0.01 bars, with an aerosol volume mixing ratio that rises from -16 to -10 to the top of the atmosphere. A full list of retrieved parameters can be seen in Table \ref{table:transmission_cloud_model_results}.}
     \label{fig:transit-cloud-models}
\end{figure*}

In the previous section, we ran retrievals with solely the slab model. We now explore the full range of aerosol parameterizations in order to investigate how the assumed aerosol model affects the resultant retrieved spectrum. The retrieved aerosol properties from testing different aerosol models are presented in Table \ref{table:transmission_cloud_model_results} and Figure \ref{fig:transit-cloud-models}, with Bayesian and frequentist statistical measures presented in Table \ref{table:transmission_stats}. We first tested three variations of the `slab' model: the nominal slab model discussed in \S~\ref{transmission:results}, a patchy slab, and a slab with an opaque deck. Out of the three slab models, the nominal slab was preferred statistically. This indicates that the present observational HST and \textit{Spitzer} data do not necessitate patchy aerosol models nor the addition of an opaque deck with the slab. We then tested two variations of the `fuzzy deck' model. The first variation does not allow the fractional scale height $f_H$ to be larger than one, while the second variation does (as discussed in \S~\ref{sec:different_cloud_models}). Out of the two fuzzy deck models, the fuzzy deck with the larger prior was preferred. The retrieved fractional scale height for the larger prior model was $\sim$ 60, resulting in a log number density that does not fall off substantially with pressure (the log number density falls from 6.96 at the cloud deck top at $\sim$ 0.001 bars to 6.86 at 1e-8 bars) resulting in the volume mixing ratio rising from -15.62 at the cloud deck top to -10.1 at the top of the atmosphere (see Figure \ref{fig:transit-cloud-models}, bottom right panel). We tested four uniform mixing ratio models: a uniform mixing ratio of MgSiO$_3$, a patchy version, a version with both MgSiO$_3$ and SiO$_2$, and a uniform mixing ratio of MgSiO$_3$ with an opaque deck. We find that across the board assuming a uniform mixing ratio of aerosols produces a poor fit that cannot fit the scattering slope. 

The `Fuzzy Deck Larger Prior' model with the larger prior on $f_H$ was the most preferred statistically out of all the aerosol models tested, but has a reduced chi square similar to the basic slab model discussed in \S~\ref{transmission:results}. These two most statistically preferred aerosol models are displayed in the bottom panels of Figure \ref{fig:cloud-models}, where both models have a similar aerosol mixing ratio in the lowest pressures. The retrieved spectra for these two are also similar (purple and black), where it is revealed that both models are preferred due to being able to fit the slope without imparting a strong silicate feature in infrared wavelengths.  

\citet{Pont2013} postulate that aerosols in the atmosphere of HD~189733~b are not confined to a single high altitude, but instead extend over most of the atmosphere. The model they utilized varied particle size by balancing vertical wind speed and gravity, allowing for larger particles to sediment lower in the atmosphere and smaller particles to remain lofted high in the atmosphere, and was invoked so that upper atmospheric smaller particles could cause the observed amplified scattering slope and that lower atmospheric large particles would act as an opaque deck to explain the observed flatter infrared data. Our current models do not account for physically motivated particle size distributions and instead opt for parameterizations with constant mean particle size. However, we plan to implement aerosol models with changing particle size in future work.

\subsection{Aerosol Showcase}\label{sec:aerosol_showcase}

\begin{figure*}[ht]
     \centering
     \includegraphics[width=1.0\linewidth]{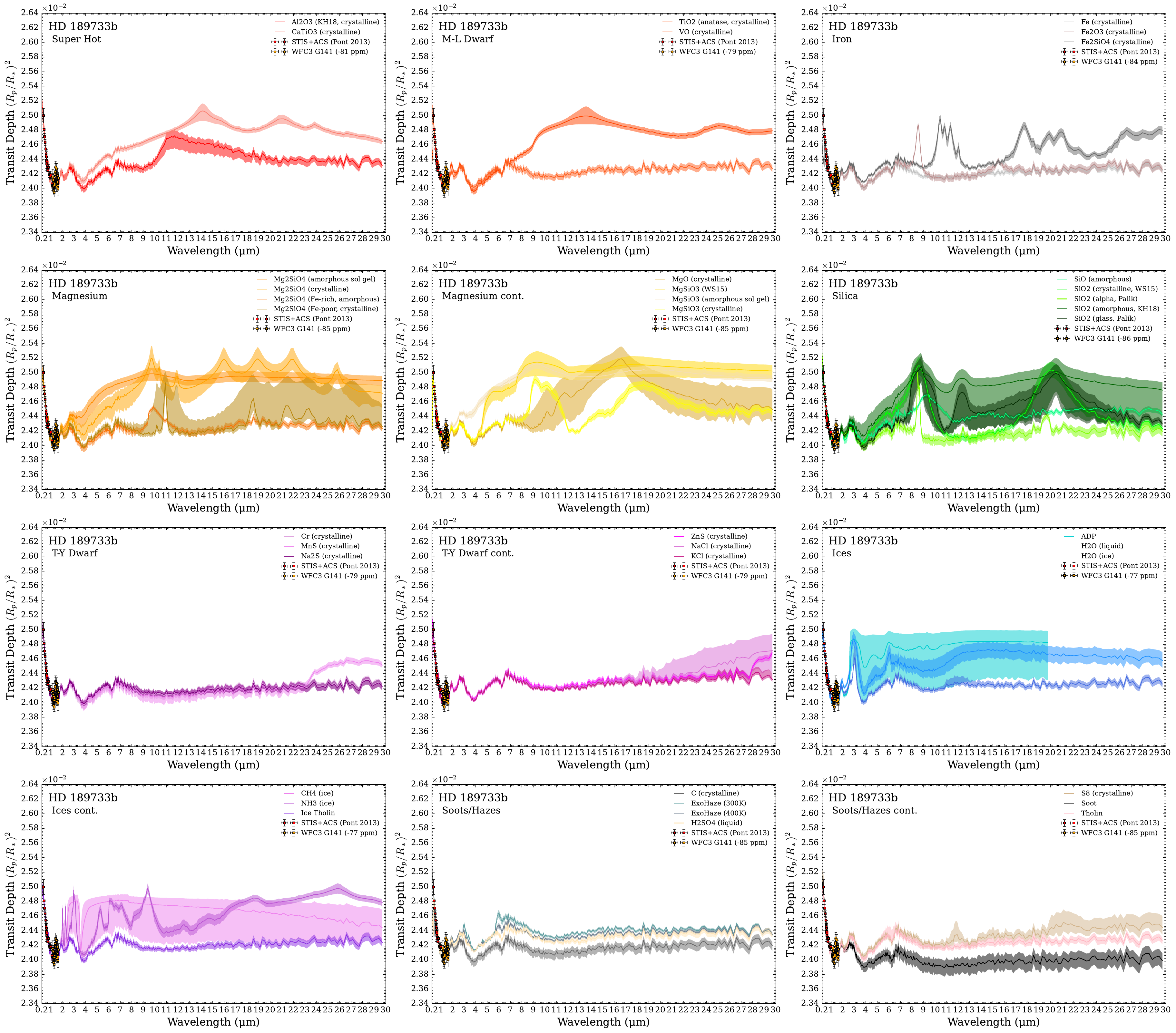}
     \caption{Showcasing each aerosol in the pre-computed database by performing retrievals fitting for only the short wavelength slope and WFC3 data and extending to longer wavelengths to show aerosol identifying absorption features. Plots display 1-sigma regions and are ordered from most hot aerosols (super hot) to coldest (ices), with the last two panels showcasing photochemical products. Most aerosols have infrared absorption features, especially aerosol species found in hot Jupiters (super-Hot to silica class) and ices. Many clouds found in T-Y dwarf temperature regimes, as well as many soots and hazes, do not have features readily available via current space-based telescopes. We note that from the slope alone, TiO$_2$ (anatase) provides the best fit \nt{(see Appendix Figure \ref{fig:transmission-TiO2}). Aerosol opacity plots are available for every aerosol in Table \ref{table:aerosol_species}, see the `aerosol\_database.pdf' on Zenodo (\S \ref{sec:Links}) or \href{https://poseidon-retrievals.readthedocs.io/en/latest/content/opacity_database.html}{\texttt{POSEIDON}'s Opacity Database}.}}
     \label{fig:transit-aerosol-showcase}
\end{figure*}

In this section, we showcase aerosols and their absorption features included in Table \ref{table:aerosol_species} by performing retrievals on only the short wavelength HST transmission data of HD~189733~b and extending to 30 \textmu m. The retrieved spectra of these retrievals can be found in Figure \ref{fig:transit-aerosol-showcase}. We conducted these tests to show that regardless of species, the scattering slope can be fit. This is not entirely unsurprising, since the scattering properties of Mie scattering aerosols is more dependent on particle size than aerosol identity \citep[e.g.,][]{WakefordSing2015}. It is to additionally show that many species have identifying absorption features in the infrared. Without infrared data to anchor the feature amplitude, the features become exaggerated which might be nonphysical, but do provide good reference point to the general shape the absorption features are expected to make. We also note that many species do not have significant absorption features in wavelengths accessible by current space-based observatories (0.2 - 30 \textmu m), such as many T-Y dwarf cloud species and Soot/Haze species found in Table \ref{table:aerosol_species} and shown Figure \ref{fig:transit-aerosol-showcase}. Additionally, some aerosols only imprint spectral features for specific particle sizes. Aerosols like Ice Tholins in Figure \ref{fig:transit-aerosol-showcase} have aerosol absorption features when particles are small (0.0001-0.001 \textmu m), but have no identifiable features for for the particle size needed to fit the scattering slope. Figure \ref{fig:transit-aerosol-showcase} expresses the need for caution when applying specific species in order to fit data. For example, ferrous oxide (Fe$_2$O$_3$) has an absorption feature at 8.6 \textmu m, similar SiO$_2$, but theory predicts that a majority of iron will homogeneously condense as an Fe cloud and that ferrous oxide is only expected to form in highly oxidizing atmospheres \citep[e.g.,][]{Visscher2010}. We note that by fitting the \textit{HST} data alone, TiO$_2$ (anatase) had the best evidence out of all the species displayed in Figure \ref{fig:transit-aerosol-showcase} due to being able to fit the slope near perfectly (see Appendix Figure \ref{fig:transmission-TiO2}). This result is interesting, since \citet{Vahidinia2014} and \citet{Pinhas2017} found that the use of a single aerosol species (MgSiO$_3$) could not fit the steepness of the slope shortwards of 0.55 \textmu m, or that stellar variability might need to induced to fit the steepness of the curve. TiO$_2$ holds significance as being the most commonly invoked `seed' particle in `top-down' aerosol formation schemes \citep[e.g.,][]{Lee2016,Lee2018}. 

\section{Aerosol Model Validation on HD~189733~b - Emission + Reflection}\label{sec:hd189-emission-retrievals}

\subsection{Datasets}

We utilized six data sets for our analysis of the emission spectrum. Namely the two-channel \textit{HST}/STIS G430L emission data from \citet{Evans2013} that is attributed to reflected light, \textit{HST}/WFC3 G141 from \citet{Crouzet2014}, all four \textit{Spitzer} IRAC \nt{channels}, IRS, and MIPS from \citet{Kilapatrick2020,Charbonneau2008,Agol2010}. Emission data used in our analysis can be found in Table \ref{table:emission_data} and is consistent with data utilized in \citet{Zhang2020} with the exception of the albedo points, which \citet{Zhang2020} did not fit for. Emission datasets are much less likely to suffer from effects that cause the need to fit for an offset \citep[e.g.,][]{Alexoudi2020}. However, HD~189733~A is known to be variable \citep[][]{Knutson2012} and could require offsets due to a changing stellar flux. We tested for offsets and found that free offsets applied to the emission data retrieved a blackbody spectrum with no absorption features, in tension with previous studies \citep[e.g.,][]{Sing2016, Zhang2020}. Therefore, we do not fit for an \textit{HST}/WFC3 offset, deviating from the retrieval set up of \citet{Zhang2020}.

\subsection{Retrieval Configuration}

We perform retrievals with just the emission spectrum (1-30 \textmu m) and retrievals with the reflected light and emission spectrum (0.2-30 \textmu m) jointly fit to glean what information can be gained by including the two datasets together. We assume the Dayside Guillot pressure-temperature profile introduced in \S~\ref{sec:pt_profiles} (where f = 0.5), since \citet{Zhang2020}'s retrieved Line P-T profile reduced down to the simpler Guillot profile. We initialize the atmosphere from 100 to 1e-6 bars with 1 bar being the reference pressure. All models utilized have \nt{thermal} multiple scattering (\S~\ref{sec:MS}) turned on, with reflection models having both \nt{thermal} multiple scattering and reflection (\S~\ref{sec:reflection}) turned on \nt{(note that `scattering’ in \texttt{POSEIDON}’s modules refers specifically to the multiple scattering of thermal flux from a substellar object while `reflection’ refers to the multiple scattering of incident stellar flux)}. For expediency, we only compute reflection up to 5 \textmu m. We once again assume a H$_2$-He dominated atmosphere with a fixed H$_2$-He fraction and freely retrieve the same eight species as transmission (Na, K, CH$_4$, CO, CO$_2$, H$_2$O, HCN, and NH$_3$). In emission and emission + reflection we choose to define our wavelength grid from 0.2 to 1 \textmu m linearly spaced with 1000 points, and then 1 to 30 \textmu m at a spectral resolution of R = 10,000. We chose to utilize 1000 linearly spaced points in the shortest wavelengths in lieu of the R = 10,000 spectral resolution due to the lower resolution of the \textit{HST}/STIS G430L data points in combination with  the excess of wavelength points in the shortest wavelengths when utilizing spectral resolution, which slowed down reflection computations substantially. We ran tests with a higher resolution wavelength grid and found no difference in our results. In transmission, we allowed both Na and K to retrieve as two separate species. However, in emission we bind the alkalies together as one parameter (Na+K) where the abundance of potassium is one order of magnitude lower than sodium \nt{, a method utilized in other retrieval schemes \citep{Kreidberg2015}. This choice is motivated by the SNR and spectral resolution of the emission spectra in the STIS dataset \citep{Barstow2014} as well as the prediction that gas-phase alkalis are in thermochemical equilibrium (and therefore approximately an order of magnitude apart) on the dayside of HD~189733~b where KCl and Na$_2$S aerosols are not predicted to form \citep{Kataria2016,Asplund2021}.} Retrievals are run with 1,000 live points.

We conduct free-chemistry retrievals with a clear atmosphere and an atmosphere with a parameterized gray aerosol model. We then chose to focus on retrievals with enstatite (MgSiO$_3$), the most common dayside condensate invoked in extant literature fitting HD~189733~b's STIS emission spectrum \citep[e.g.,][]{Barstow2014}. Fitting for emission and jointly fitting reflection and emission, we test out four aerosol models: the slab model described in \S~\ref{sec:transmission_config}, a patchy slab model, a uniform mixing ratio model parameterized by just the mean particle size, $r_{\rm{m}}$ and the constant log mixing ratio of the aerosol, and a patchy uniform mixing ratio model. Our aim is to explore aerosol models that allow our results to be consistent with both \citet{Barstow2014} and \citet{Zhang2020}. Priors for our emission and emission + reflection retrievals can be seen in Table \ref{table:emission_results}. In particular we utilize a tight radius prior (1.11,1.13 R$_J$), motivated by the abundance of extant transit measurement of HD~189733~b and confident reported planetary radius \citep[e.g.,][]{Sing2016}. We note for future users that our tight radius prior was imperative for emission retrievals, and that when we ran retrievals with uniform, large radius priors our retrieved water abundances and thermal structure were nonphysical and unconstrained.

Similar to transmission, we run a suite of additional retrievals that explore other aerosol species and thermal structure parameterizations to explore how they affect retrieved spectra. Using the best fit enstatite aerosol model as a basis, we conduct emission + reflection retrievals with alternative condensate species, namely crystalline silicon dioxide (SiO$_2$) and magnesiums sulfide (MnS, Morley). We then explore our choice of P-T profile by running retrievals with the Line, Madhu, Pelletier, and Slope profiles with a clear model with priors in Table \ref{table:pt_results}. We run these exploratory retrievals at a spectral resolution of R = 5,000 from 1 - 30.0 \textmu m (0.2 to 1 \textmu m linearly spaced with 1000 points) with 500 live points. Finally, we explore forward models with different combinations of aerosol abundance, cloud fraction, particle size, and species to investigate how the albedo and aerosol features affect resultant spectra. 

\subsection{Results}

\begin{figure*}[ht]
     \centering
         \includegraphics[width=1.0\linewidth]{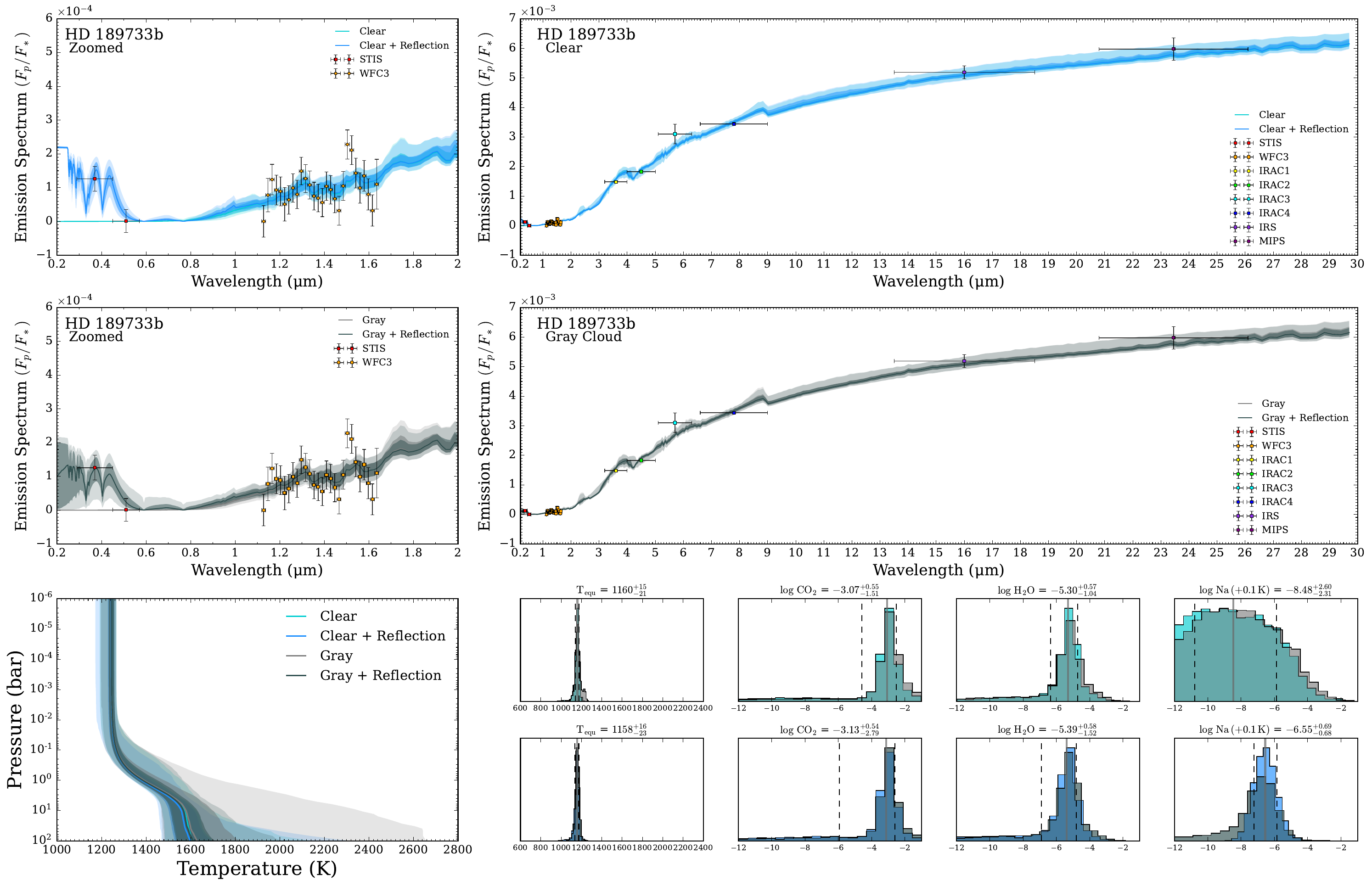}
     \caption{Atmospheric retrievals on \textit{HST} and \textit{Spitzer} emission data of HD~189733~b. Retrieved spectra display median retrieved spectra (solid lines), with 1$\sigma$ and 2$\sigma$ confidence intervals (dark and light shaded regions). Top: Retrieval results for a clear atmosphere, displaying results for both emission and emission + reflection retrievals. All retrievals include thermal multiple scattering. Left panel is zoomed on \textit{HST} data and shows that reflection is necessary to fit the \textit{STIS} data. Right panel shows full retrieved emission spectra. Middle: Retrieval results for a gray aerosol model with an opaque cloud and parameterized haze. The retrieval does not prefer an opaque cloud deck, but does utilize the absorptive haze to shape the reflection spectrum. Bottom panels: retrieved atmospheric thermal structures (left) and posteriors for equilibrium temperature and gas-phase volume mixing ratios of CO$_2$, H$_2$O, and Na + (0.1 K). Top row represents retrievals with just emission, bottom row represents retrievals with emission + reflection. A clear atmosphere is sufficient to fit the data, with no preference for an absorptive cloud deck or haze. Retrieved posteriors are consistent between emission and emission + reflection retrievals with the exception of alkalis, which are constrained due to the short wavelength reflection spectra. A full list of retrieved parameters can be seen in Table \ref{table:emission_results}.}
     \label{fig:emission-clear-gray}
\end{figure*}

\begin{figure*}[ht]
     \centering
         \includegraphics[width=1.0\linewidth]{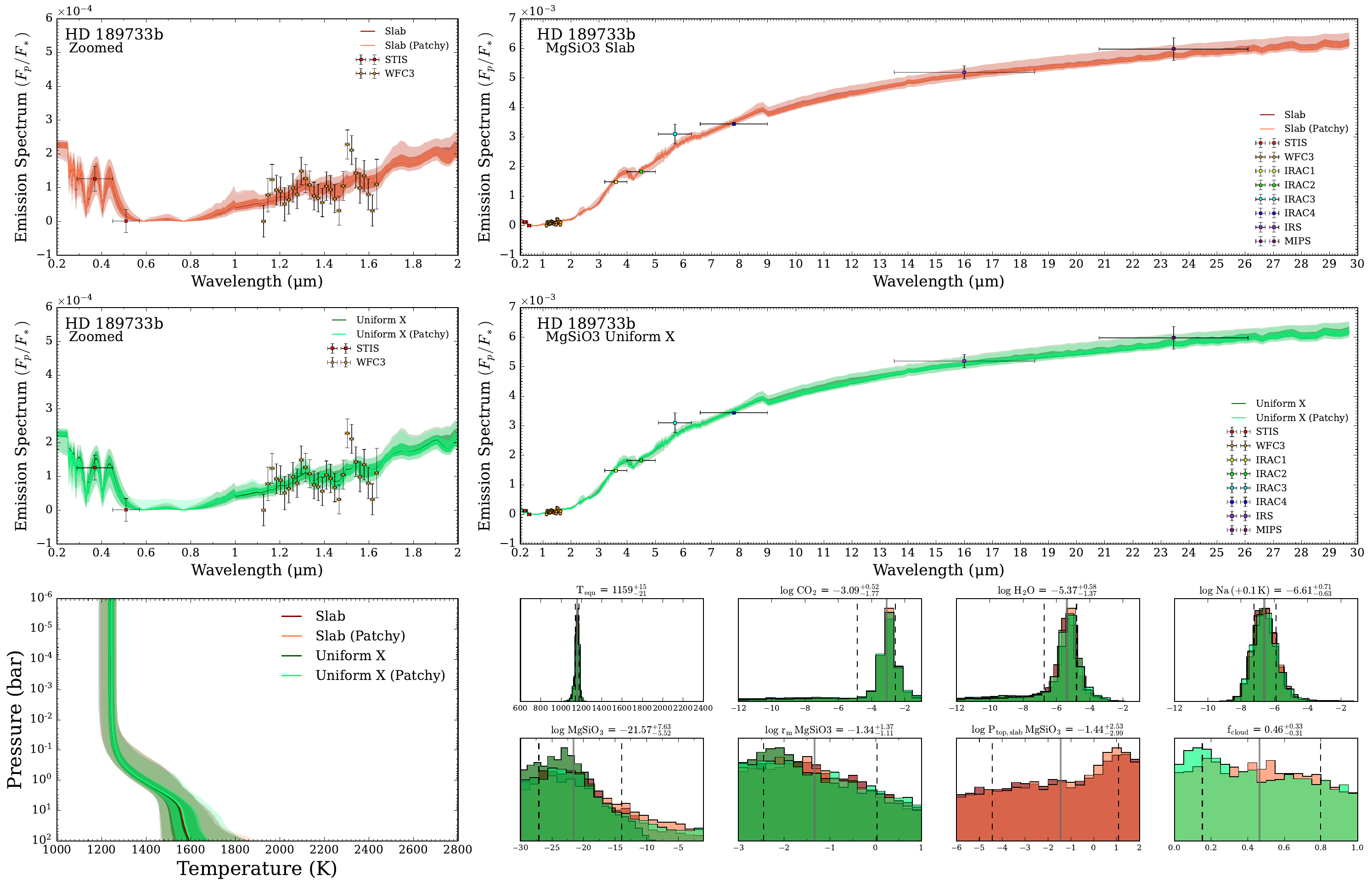}
     \caption{Same as Figure \ref{fig:emission-clear-gray}, but highlighting the emission + reflection retrieval results for enstatite (MgSiO$_3$) clouds. Top: Retrieval results for the slab model, with and without fractional cloud coverage as a free parameter. Middle: Retrieval results for the uniform mixing ratio model, with and without fractional cloud coverage as a free parameter. Bottom: Retrieved temperature structure of each model. Top row of posteriors is consistent with the results displayed in Figure \ref{fig:emission-clear-gray}. Bottom row shows retrieved aerosol properties, in particular the mean particle size $\log$\,$r_{\rm{m}}$, and constant aerosol mixing ratio in the cloud for all four clouds models, the slab-top pressure $\log$\,$P_{\rm{top,slab}}$ for the two slab models, and the fractional cloud coverage percentage $f_{cloud}$ for the patchy aerosol models. We find that all retrievals fit for a minimal-density aerosol abundance, consistent with the aerosol-free clear model. A full list of retrieved parameters can be seen in Table \ref{table:emission_results}.}
     \label{fig:er-mgsio3}
\end{figure*}

We conducted emission retrievals with a clear atmosphere and an atmosphere with a parameterized gray aerosol model, similar to the analysis we performed in \S~\ref{sec:hd189-transmission-retrievals}. Results for this can be seen in Figure \ref{fig:emission-clear-gray}. Retrievals with the inclusion of MgSiO$_3$ clouds are highlighted in Figure \ref{fig:er-mgsio3}. All retrieved posteriors for each model can be found in Table \ref{table:emission_results}. We find that aerosols are not needed to fit the emission, nor the joint emission + reflection spectrum. A clear atmosphere is preferred over the gray aerosol model by 2.1$\sigma$. The clear retrieval retrieves an equilibrium temperature (in the context of the Dayside Guillot formulation, Equation \ref{eq:guillot}) within one sigma of the true equilibrium temperature (1200K), a sub-solar water abundance ($\sim$ -5) to fit the\textit{HST}/WFC3 G141 data, and a high carbon dioxide abundance ($\sim$ -3) to fit \textit{Spitzer} IRAC Ch2. The emission-only retrieval fits for an unconstrained alkali abundance, whereas the emission + reflection retrieval retrieved a \nt{slightly sub-}solar alkali abundance (Na $\sim$ -6.55, K $\sim$ -7.55). A clear atmosphere being able to fit the albedo spectra is consistent with \citet{Barstow2014}'s retrieval analysis. The retrieved P-T profile is unconstrained below 10 bars, has a negative temperature gradient from 10 (1600K) to 0.1 bars (1200K), and then is isothermal at the equilibrium temperature for the remainder of the atmosphere. The inclusion of reflection does not alter any of the retrieved values except for the retrieved alkali abundance, which is constrained by the STIS emission spectrum (see Table \ref{table:emission_results}, Clear results). 

The gray aerosol model places the opaque deck at 10 bars, below the infrared photosphere at 100 mbar (1e-1 bar), in both the emission and emission+reflection retrievals. When including reflection, the retrieval explores a wide range of absorptive, power-law hazes in order to shape the STIS emission spectrum in the shortest wavelengths, as highlighted in the 1 and 2 sigma shaded regions in Figure \ref{fig:emission-clear-gray} (Gray cloud, zoomed). All MgSiO$_3$ aerosol models fit for a minimal-density (cloud abundance $\leq$ -20) cloud, indicating that the retrieval does not favor a cloud when fitting the joint reflection + emission spectrum. We further explore how the resultant spectra changes with the inclusion of compositionally specific clouds in \S~\ref{sec:forward_model_explore}. The retrieved gas species and thermal structure properties are consistent within 1 sigma for all clear and cloudy emission and emission + reflection retrievals (see Table \ref{table:emission_results}). Statistical analysis for the clear vs gray cloud, and the MgSiO$_3$ models can be found in Table \ref{table:emission_stats} (emission statistics) and Table \ref{table:reflection_stats} (emission + reflection statistics). The spectral, pressure, and photometric contribution functions for the clear retrieval can be seen in Appendix Figure \ref{fig:emission-contribution}. The pressure contribution plot reveals that a bulk of the observed spectrum is being formed in the 10 to 0.1 bar region where the temperature gradient is located.

\subsection{SiO$_2$ and MnS}\label{er_cloud_species}

We utilized the best-fit MgSiO$_3$ aerosol model (patchy slab) in order to test the sensitivity of our results to cloud species in the emission + reflection retrieval. Priors and retrieved cloud properties can be seen in Table \ref{table:emission_cloud_model_results}, with statistics found in Table \ref{table:reflection_stats}. We find that, similar to our MgSiO$_3$ model that both aerosol models retrieve a minimal-density cloud.

\subsection{P-T Profiles}\label{er_pt_profiles}

\begin{figure*}[ht]
     \centering
         \includegraphics[width=1.0\linewidth]{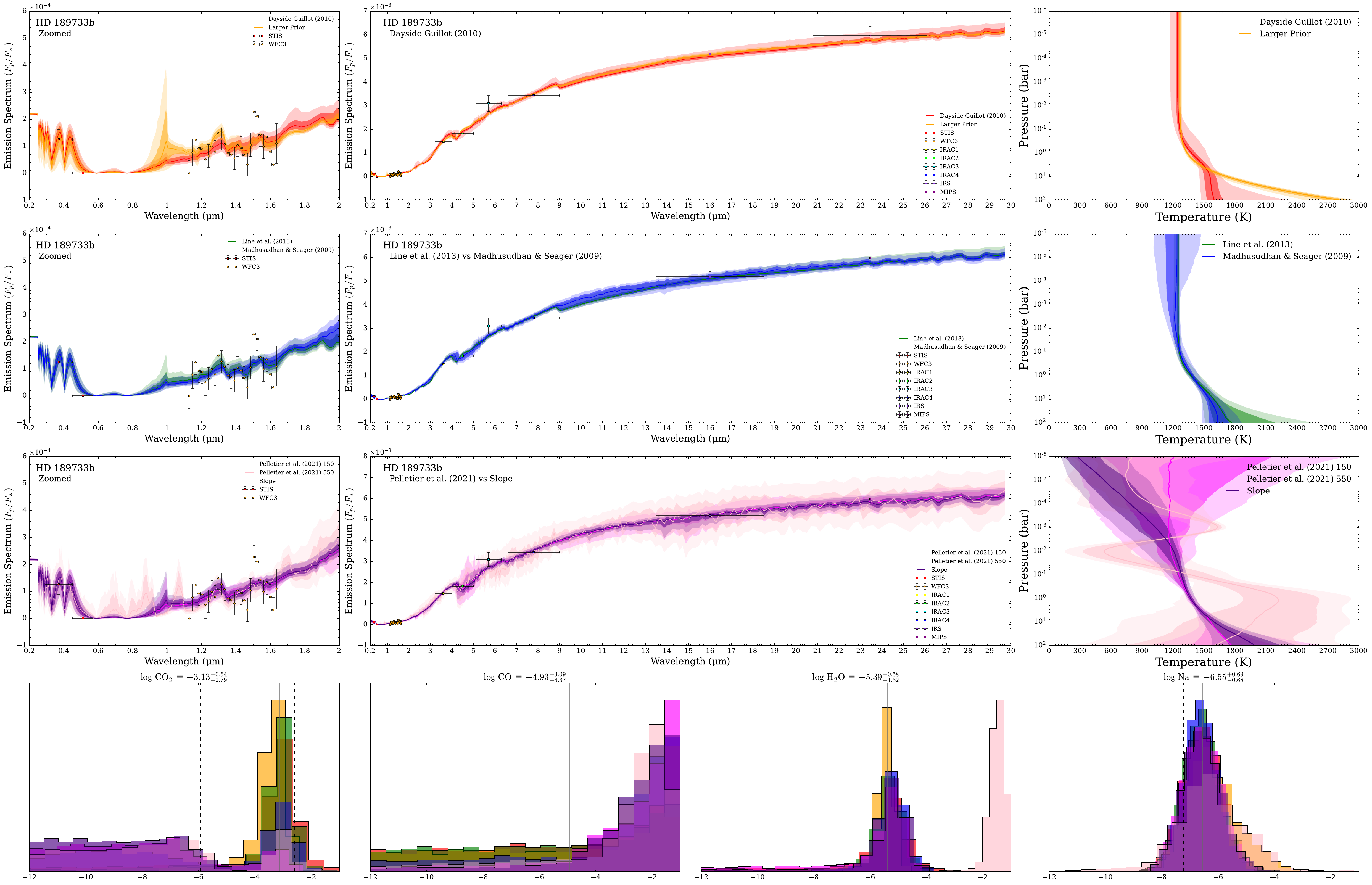}
     \caption{ Emission + reflection retrieval results for cloud-free atmospheres assuming different P-T profiles. Top: Showcasing the results for the Dayside Guillot profile (which was featured in Figure \ref{fig:emission-clear-gray} and \ref{fig:er-mgsio3}) and the same profile with a larger prior on the infrared opacity (log $\kappa_{IR}$) and gamma ($\gamma$). With a larger prior, the P-T profile fits for a deep, steep adiabat. Middle: Showcasing results for two six-parameter P-T profiles, specifically the \citet{Line2013} and \citet{Madhusudhan2009} parameterizations. Bottom: Showcasing results for two flexible P-T profiles, specifically the slope profile (which assumes a negative temperature gradient) from \citet{PietteMadhu2021} and the knot-fitting profile (which assumes a penalty ($\sigma_s$) on the second-derivative) from \citet{Pelletier2021}. We showcase both a strict penalty on the second derivative (150) as well a relaxed penalty (550) to showcase the effect of the penalization. All P-T profiles, with the exception of the nonphysical Pelletier 550 profile, share a negative temperature gradient from 10 (1600K) to 0.1 bars (1200K) that is driving the water and carbon-dioxide absorption features in the spectra. The retrieved parameters for gas phase species are consistent, with the exception of a bimodal carbon-dioxide abundance in the Madhu, Slope, and Pelletier profiles which can allow for upper atmospheric temperature gradients. Retrieved P-T parameters can be seen in Table \ref{table:emission_cloud_model_results}.}
     \label{fig:er-pt-profile}
\end{figure*}

In the previous sections, we ran retrievals with solely the Dayside Guillot thermal profile. We now explore a wide-range of thermal structure parameterizations in order to investigate how the assumed thermal profile affects resultant retrieved spectra. The priors and retrieved results for testing different P-T profiles are presented in Table \ref{table:pt_results}, with statistics in Table \ref{table:reflection_stats}. Retrieved spectra and thermal structures are highlighted in Figure \ref{fig:er-pt-profile}. 

We first tested the Dayside Guillot P-T profile with larger priors on the infrared opacity (log $\kappa_{IR}$) and ratio of the visible and infrared opacity (log $\gamma$). We conducted this exercise due to some of our retrievals converging on the edge of the log $\kappa_{IR}$ prior, namely converging near -5. We find that with a larger prior, the retrieved P-T profile has a steep adiabat from 3000K to 1200K spanning 100 to 0.1 bars. This deep adiabat results in a increase in flux from the planet between 0.95 and 1.3 \textmu m. A comparison between the base and larger prior Dayside Guillot retrievals can be found in the top panel of Figure \ref{fig:er-pt-profile}.

We then test the six-parameter \citet{Line2013} (Equation \ref{eq:line}) and \citet{Madhusudhan2009} P-T profiles (middle panel) and the more flexible slope \citep{PietteMadhu2021} and Pelletier \citep{Pelletier2021} P-T profiles (bottom panel). All four parameterizations indicate a sensitivity to a commonly retrieved negative temperature gradient from 10 (1600K) to 0.1 bars (1200K) that is driving both the water and carbon dioxide absorption features, similar to the Guillot profile. The two six-parameter profiles retrieve nearly identical P-T profiles with the exception of the Line profile fitting for a slightly hotter interior, causing the same increase in flux at 0.95 and 1.3 \textmu m found in the larger prior Dayside Guillot profile. The slope P-T parmeterization fits for a deep adiabat up to 10 bars, and after the commonly retrieved temperature gradient, continues to have a steep negative temperature gradient from 0.1 bars to the top of the atmosphere. The slope P-T profile is defined to have a decrease in temperature with decreasing pressure, and the emission signal is not sensitive to those pressure regions, which is further indicated by the large 1 and 2 sigma regions in shallow pressures  We tested the Pelletier P-T profile with two $\sigma_s$ values, where $\sigma_s$ is the evidence penalization on the second derivative. We find that a value of $\sigma_s$ = 150 is informative in discovering which portions of the P-T profile can be constrained, with tight constraints in the 10 to 0.1 bar region and large 1 and 2 sigma regions in the P-T profile indicating which regions of the atmosphere the retrieval is not sensitive to at a larger degree than P-T profiles with strict assumptions (namely the deepest and shallowest pressures). In the Pelletier 150 spectra, the large 1-2 sigma region between IRAC Ch1 and Ch2 is due to this profile allowing for positive and negative temperature gradients in the upper atmosphere, causing both CO$_2$ and CO absorption and emission features in these wavelengths. A larger $\sigma_s$ = 550 retrieves a nonphysical P-T profile with many sharp temperature inversions. We include this retrieved profile as a reference for future users.

The Line profile was the most preferred model (Table \ref{table:reflection_stats}), with an evidence similar to the base Dayside Guillot P-T profile. \citet{Zhang2020} found a similar result, in that in their retrievals the Line profile reduced down to the Guillot profile. The retrieved water and alkali abundance is consistent across all models (with the exception of the Pelletier 550 profile, which fits for a high water abundance). We find that the Madhusudhan \& Seager, Pelletier, and Slope profiles fit for a bimodal carbon dioxide abundance due to the additional flexibility of those models to have upper atmospheric temperature gradients. We find that retrieved carbon monoxide abundances are pushing up against the prior in all retrievals. We took the best fit retrieval and ran forward models with and without high CO abundances and found that models with higher CO abundance had a slight dip in the emission spectrum around 5 \textmu m (between IRAC Ch2 and Ch3). This slight CO feature is displayed as a slightly larger 1-2 $\sigma$ region at these wavelengths in every plot. Forward models with high CO abundances also had brighter peaks in their emission spectra. Though this seems counter-intuitive, the retrieval is utilizing CO in order to increase the mean molecular weight of the atmosphere, thereby decreasing the scale height and allowing flux to escape from deeper in the atmosphere. \nt{This effect can be seen in Appendix Figure \ref{fig:No-CO}.}


\subsection{Forward Model Exploration}\label{sec:forward_model_explore}

\begin{figure*}[ht]
     \centering
         \includegraphics[width=1.0\linewidth]{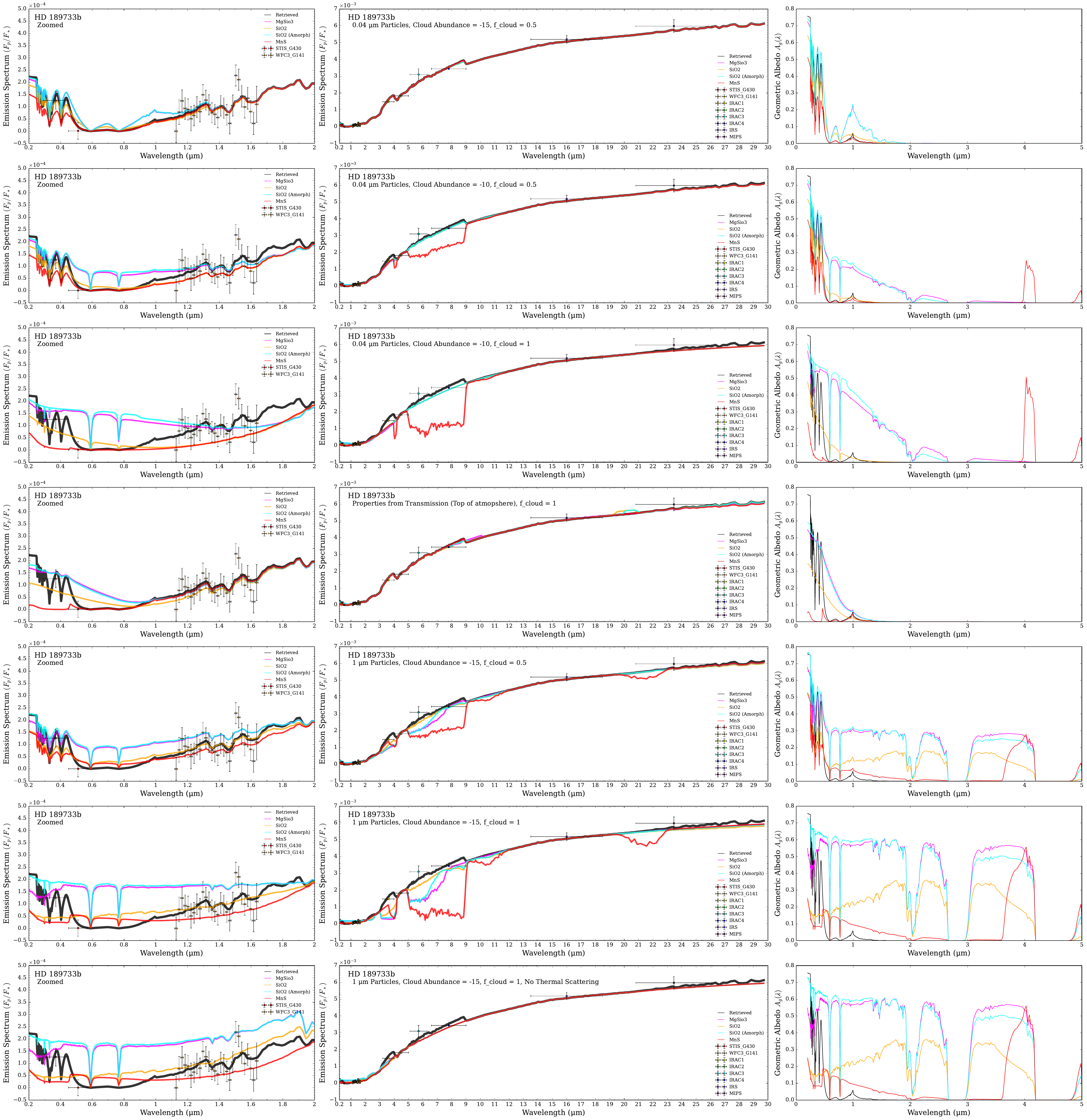}
     \caption{Figure displaying the effect of aerosol properties and species on the resultant emission and reflection spectra. For all models, we test enstatite (MgSiO$_3$), quartz (SiO$_2$, both crysalline and amorphous), and manganese sulfide (MnS). Unless specified, the cloud top pressure is at 0.1 bars and extends to 100 bars, is composed of sub-micron particle sizes, and has a fractional cloud coverage of 0.5. Each row displayed a zoom-in on the HST data, the full spectra, and the geometric albedo up to to 5 \textmu m. First two rows display the effect of increasing the aerosol abundance to -15 and -10, with a general increase in reflected light, especially for MgSiO$_3$ and amorphous SiO$_2$. The third row displays the effect of increasing the cloud fraction to 1, which washes out the water absorption features and amplifies the reflected light. The fourth row utilizes the results from our transmission retrievals and initialized a high altitude, low density slab which results in a smooth reflection signal with no alkali absorption features. The fifth and sixth rows display the effect of particle size by utilizing 1 \textmu m sized particles, with half and full cloud coverage. Larger particles dominate the reflection spectra, and impart strong scattering features in longer wavelengths, even when the atmospheric thermal structure contributing to those wavelengths is isothermal. In row six we display that thermal scattering imparts the features in long wavelengths by turning it off.}
     \label{fig:er-pt-profile}
\end{figure*}

Because our retrievals did not necessitate a cloud to fit the joint emission+reflection spectrum, we took the patchy MgSiO$_3$ slab model's retrieved properties (thermal structure, chemical abundances, and aerosol properties) and explored the effect that changing the aerosol properties and species had on the resultant spectra. For all forward models, unless specified, we kept the slab spanning the retrieved 0.1 bars to 100 bars. We tested these models with MgSiO$_3$, crystalline SiO$_2$, amorphous SiO$_2$, and MnS as the aerosol species. Resultant spectra are displayed in columns 1 and 2 of Figure \ref{fig:er-pt-profile}, with the third column displaying the geometric albedo up to 5 \textmu m. We first explored increasing the aerosol abundance to a mixing ratio of -15, and -10. For the -15 aerosol abundance model we tested half cloud coverage, and for the -10 cloud abundance model we tested both half and full cloud coverage. We find that increasing the aerosol abundance dramatically increases the geometric albedo of models with MgSiO$_3$ and amorphous SiO$_2$ slabs. Both crystalline SiO$_2$ and MnS are much more absorptive in UV and visible wavelengths, but do also have a slight increase in albedo. There is an expectation for patchy clouds on the dayside of hot Jupiters \citep[e.g.,][]{Lee2016}, so we also tested how patchy clouds affect spectra. The half-cloud coverage models still have water absorption features in their spectra, while the full cloud coverage models block flux from escaping the layers with the negative temperature gradient, and therefore blocks all water absorption features. This indicates that combining reflection spectra from clouds with infrared emission features could be a potential test for patchy clouds. In these models, we note that in longer wavelengths MnS imparts a strong feature between IRAC Ch1 and Ch3 due to its scattering properties. \nt{This is because MnS particles have a single scattering albedo $\sim$ 1 in infrared wavelengths, meaning that their interaction with thermal flux is 100\% dominated by scattering. The scattering properties of an aerosol is dictated by its asymmetry parameter, which will impart `scattering features’ due to thermal multiple scattering processes (see the `MnS\_Mor' entry in the `aerosol\_database.pdf' on Zenodo (\S \ref{sec:Links}) or \href{https://poseidon-retrievals.readthedocs.io/en/latest/content/opacity_database.html}{\texttt{POSEIDON}'s Opacity Database} for plots of the single scattering albedo and asymmetry parameter).}


We then modeled the emission spectra with the retrieved MgSiO$_3$ high altitude, low density slab properties from the transmission analysis done in \S~\ref{sec:hd189-transmission-retrievals}. We find that a high altitude cloud imparts a smooth albedo spectra without any alkali absorption features. With high altitude clouds, there isn't much of a column of alkali gas above the clouds to absorb photons. These models still can fit the water absorption features. 

Finally, we tested 1 \textmu m particle sizes. We find that this larger particle size (compared to the retrieved particle size found in the transmission retrievals, but relevant for the wavelength range we are probing with Mie scattering) greatly increase the albedo and causes the entirety of the \textit{HST} spectrum to be dominated to the reflected component (with water features imprinting themselves in the reflection spectrum in lieu of the emission spectrum). Cloud fraction here has a similar effect, where full cloud coverage results in a flat albedo spectrum in short wavelengths with alkali and water absorption features imprinted on it, similar to the class 5 EGP models in \citet{Sudarsky2000}. Patchy clouds allow for water absorption features in \textit{HST}-WFC3 wavelengths to imprint partially on the spectrum. Larger particles also imprint strong features in longer wavelengths due to the multiple scattering treatment. Even though these longer wavelengths are probing higher up in the atmosphere where the thermal structure is isothermal in the retrieved Dayside Guillot profile, large particles such as these can still impart features \citep[e.g.,][]{Taylor2021}. We turned off thermal multiple scattering in the final panel in Figure \ref{fig:er-pt-profile} to demonstrate this phenomena. \nt{Higher resolution emission spectra from JWST could reveal absorption features in the \textit{Spitzer} data wavelength range, which will necessitate the thermal gradient needed to produce aerosol absorption features (from sub-micron sized aerosols) in lieu of scattering features (from micron sized aerosols).}

\section{Discussion}\label{sec:Dicussion}

HD~189733b's transmission spectrum has been the subject of over 10 years of extensive retrieval studies. \citet{Lee2014} ran retrievals with \texttt{NEMESIS} on the 0.3 - 1.0 \textmu m \textit{HST} NICMOS spectra and found that a uniform with pressure aerosol layer of particles less than 0.1 \textmu m in size can fit the transmission dataset, regardless of aerosol species. We retrieve a similar mean particle size, but with thin, high altitude aerosol slabs being preferred over a uniform with pressure model. \citet{Pinhas2019} used the \texttt{AURA} retrieval code to run free retrievals on the \textit{HST} STIS, ACS, WFC3, and \textit{Spitzer} IRAC Ch1 and IRAC Ch2 datasets, excluding IRAC Ch3, IRAC Ch4, and MIPS. Their retrieval used a patchy aerosol model with an opaque cloud deck and power-law haze. They found a sub-solar water abundance of log H$_2$O $\sim$ -5, a value that is lower than our retrieved values of between -3 and -4 (with the exception of our flame soot retrieval, which retrieved a substantially lower water abundance of $\sim$ -5). \citet{Zhang2020} used the same dataset examined in this work and used \texttt{Platon} to run chemical equilibrium retrievals where the emission and transmission spectra were jointly retrieved. Their chemical equilibrium water abundance at 10 mbar on the limb was reported to be super-solar ($\sim$ -2.5). Our free retrieval results are more in line with the chemical equilibrium retrievals of \texttt{Platon}, hinting that the inclusion of IRAC Ch3, IRAC Ch4, and MIPS notably affects the water abundance. \citet{Zhang2020} retrieve a haze parameterized by mean particle size, maximum number density, and fractional scale height. They additionally fit for a wavelength-dependent refractive imaginary index, keeping the real portion constant, therefore making their retrieval agnostic of lab refractive index data. They retrieve a collection of arbitrarily small particles ($<$ 0.001 \textmu m) with no lower limit, with a strong inverse degeneracy with number density. We find that utilizing lab data to generate aerosol properties puts a tighter constraint on the mean particle size, with a constrained sub-micron size from our suite of transmission retrievals. The \texttt{Platon} retrieved fractional scale height indicates that haze particles are more abundant in the upper atmosphere than the lower, similar to both our slab and fuzzy deck (larger prior) retrievals showcased in Figure~\ref{fig:cloud-models}.

\nt{More recently, \texttt{CHIMERA} \citep{Mai2019} was utilized to conduct a free-chemistry retrieval analysis on the JWST NIRCam transmission spectrum of HD~189733~b (2.4-5 \textmu m) \citep{Fu2024}. They detect gas-phase CO$_2$ (-6.24), CO (-4.13), H$_2$S (-4.50), H$_2$O (-3.69), and an upper limit on CH$_4$ (-7.9). Our retrieved H$_2$O abundance (for all transmission models except Gray and Soot) and our CO$_2$ abundance for the Soot retrieval are consistent within 1 sigma of their results. We did not detect CO nor CH$_4$ in our transmission retrievals due to the lower spectral resolution of the \textit{Spitzer} data. \texttt{CHIMERA} retrievals included inhomogeneous clouds and hazes and found that their properties were unconstrained due to NIRCam's wavelength range being relatively insensitive to aerosols \citep{WakefordSing2015}. Tight constraints on the predicted main reservoirs of oxygen (H$_2$O), carbon (CO), and sulfur (H$_2$S) at HD~189733~b's equilibrium temperature, and constraints on carbon-bearing species CO$_2$ and CH$_4$, demonstrates JWST's ability to measure planetary atmospheric metallicities and carbon-to-oxygen ratios that determine what aerosols can form.}

Our retrieval results also offer insights into the dayside atmosphere of HD~189733~b from secondary eclipse retrievals. We first consider a comparison with retrievals assuming only thermal emission. \citet{Zhang2020} ran chemical equilibrium retrievals with \texttt{Platon} and find that a 10x solar metallicty and C/O of 0.69 provide the best fit to the data. Their retrieved chemical equilibrium model predicts a high CO abundance, a marginally low CO$_2$ abundance, and a water mixing ratio of $\sim$ -2.5 (at 10 mbar). Our dayside emission + reflection (clear) retrieval indicates a dayside sub-solar abundance of water ($\sim$ -5, 0.001x solar) and high CO$_2$ abundance ($\sim$ -3, 100x solar)\footnote{\nt{Utilizing the chemical equilibrium grids found in \citet{Marley2021}, log CO$_2$ is approximately -3 at 10mbar, 100x solar metallicity (C/O ratios ranging from 0.25x to 1.25x solar, where solar C/O $\approx$ 0.9 \citep{Teske2014}.)}}. The discrepancy in retrieved water abundances can come from the strict requirements that equilibrium chemistry places on abundances, or that \citet{Zhang2020}  retrieved an offset on the WFC3 emission dataset. In future work, we will investigate the effect that implementing equilibrium and disequilibrium chemistry grids into our retrieval has on our results. Our retrieved thermal structure is consistent with the retrieved P-T profile of \citet{Zhang2020} (see their Figure 10, our models range from 100 to 1e-6 bars, or 1e7 to 0.1 Pa). 

Our joint emission + reflection retrievals over a wide wavelength range offer new insights into HD~189733\,b's dayside atmosphere. \citet{Barstow2014} performed retrievals on the STIS secondary eclipse spectrum of HD~189733~b and found that a range of clear and cloudy models could fit the data. In particular, they found that a Na volume mixing ratio of 50 ppmv ($\sim$ \nt{-4.3}, 10x solar) was needed to explain the STIS emission spectrum without a cloud. We instead find that a Na mixing ratio of -6.5 (hence a K mixing ratio of -7.5) is sufficient to fit the data. However, we also note that \citet{Barstow2014} utilized the six channel, lower precision version of the dataset where the fifth data point has a negative $F_{p} / F_{s}$. Our retrieved dayside alkali abundance is slightly enhanced by an order of magnitude when compared to our terminator abundances. The slightly enhanced alkali species abundance on the dayside could be explained by salt clouds on the terminator evaporating on the dayside \citep[e.g.,][]{Parmentier2016, Lee2016}. However, with the current data quality, our joint emission + reflection retrievals do not favor a cloud. Therefore, we conclude that the current secondary eclipse data of HD~189733~b is much less sensitive to clouds than the transmission spectrum and will require higher-quality data (i.e., JWST) to identify the dayside cloud composition and properties.

\nt{Our joint emission + reflection retrievals do not include TiO and VO in their gaseous forms, which have strong absorption short wavelengths (0.3-1 \textmu m). \citet{Barstow2014} explored the effects of including TiO and VO in their retrieval analysis of the STIS secondary eclipse spectrum and found that their inclusion alongside Na improves the fit to the spectrum, but note that the solution is incredibly degenerate. We decided to exclude these species from our retrieval setup since both gasses are predicted to have condensed out and subside below the photosphere of HD~189733~b \citep{Burrows1999, Lodders2002, Fortney2010} and that there has been no evidence of TiO and VO absorption in the transmission spectra of HD~189733~b \citep{Pont2013}. Further studies have shown that even in hotter planetary atmospheres (i.e., HD~209458~b), TiO and VO can become cold-trapped below the photosphere \citep{Spiegel2009} or on the nightside \citep{Parmentier2013}. \citet{Parmentier2016} note that the inclusion of TiO and VO in atmospheres of hotter planets (i.e., Kepler-76~b, HAT-P-7~b) provides a better fit to their apparent geometric albedo. In future retrieval studies with reflection, gaseous TiO and VO will be included when fitting spectra of these hotter planets.}

\section{Conclusion}\label{sec:Conclusion}

We present new formulations for the exploration of aerosols into the open-source atmospheric retrieval code \texttt{POSEIDON} that are compatible with transmission, emission, and reflection spectra. We additionally present an open-source, precomputed aerosol database that is easy to add new aerosols to and to use. We benchmark these new capabilities on available \textit{HST} and \textit{Spitzer} data of the well studied hot Jupiter HD~189733~b. Aerosols are needed to explain the transmission spectrum of HD~189733~b, best explained by a high-altitude, low density, thin slab composed of sub-micron sized particles. We find that manganese sulfide (MnS) produces the best fit to the data, in line with equilibrium aerosol predictions found in \citet{Kataria2016}, but note that HD~189733~b's current transmission data in the mid-infrared (5-25 \textmu m) is limited by precision. We find a sub-solar abundance of water and sub-solar abundance of potassium in HD~189733~b's atmosphere. Notably, the inclusion of Mie-scattering aerosols in lieu of a parametric gray cloud + power law haze changes the retrieved water abundance from super- to sub-solar and constrains the potassium abundance. We also tested a variety of aerosol models and found that an opaque deck with an increasing aerosol volume mixing ratio above it also provide a good fit to the data. Fitting for only the \textit{HST} STIS and WFC3 data, we find that TiO$_2$ (anatase) provides the best fit to the slope's intrinsic shape, a species that holds significance as the most commonly invoked seed particle in disequilibrium aerosol modeling \citep[e.g.,][]{HellingFomins2013, Lee2016}.

We then co-fit the combined reflection and emission spectrum of HD~189733~b, which provides better constraints on alkalli abundance and aerosol properties than emission or reflection alone. We find that a cloudless, clear atmosphere provides a reasonable match to current observations. We find a dayside sub-solar water abundance, enhanced carbon dioxide abundance, and \nt{slightly sub-}solar alkali abundance fit the joint spectra. We then explore a variety of P-T profiles and find that all the profiles agree on a common negative temperature gradient from 10 (1600K) to 0.1 (1200K) bars and fit for similar chemical abundances. We also conduct a forward model exploration of the emission and reflection spectrum by changing different cloud properties. We find that joint albedo and emission spectra can provide a method by which to test for dayside patchy clouds, namely that forward models with patchy reflective clouds imprint both alkali reflection features and water emission features in the observed dayside spectrum. We also find that multiple scattering is important to consider, as it can imprint spectral features even in isothermal regions of the atmosphere. Looking forward, JWST will provide more precise data with a broader wavelength coverage for hot Jupiters like HD~189733~b that will be well suited to exploring aerosol properties using our new capabilities in \texttt{POSEIDON}. Retrieved aerosol properties will, in turn, open a rich multi-dimensional picture of cloud formation processes in exoplanetary atmospheres.

\subsection{Links}\label{sec:Links}


\nt{The links in the paper point primarily to the \texttt{POSEIDON} github and readthedocs which house the most current version of the code and documentation, and are subject to change with time. The zenodo repository for this paper specifically houses the version of the code (and aerosols) released subsequently with this paper. The zenodo repository for this paper (\href{https://doi.org/10.5281/zenodo.13755771}{Zenodo (doi: 1375577)}) contains 1) a pdf with the interpolated refractive indices and precomputed optical properties of every aerosol in the aerosol database, 2) a aerosol database readme file with detailed information on each aerosol in the precomputed aerosol database, 3) the precomputed aerosol database inputs and refractive indices, 4) the precomputed aerosol database, 5) retrieval scripts and results for the HD~189733~b retrievals presented in this paper, 6) a stable version of the \texttt{POSEIDON} V1.2 code used to run these retrievals and make the figures, as well as tutorial notebooks for each new feature added in this paper (aerosols in transmission and emission geometries, thermal multiple scattering, reflection, contribution visuals, new pressure-temperature profiles, retrievals with multiple offsets, and how to add aerosols to the precomputed aerosol database).} 

\nt{The \href{https://github.com/MartianColonist/POSEIDON}{\texttt{POSEIDON} github} houses the most current version of \texttt{POSEIDON}. \texttt{POSEIDON} V1.2 was released alongside this paper. An installation of \texttt{POSEIDON} comes with the precomputed aerosol database, refractive indices, the aerosol database readme file, and tutorial notebooks showcasing each new feature added in this paper.}

\nt{The \href{https://poseidon-retrievals.readthedocs.io/en/latest/}{\texttt{POSEIDON} readthedocs} houses the most current version of \texttt{POSEIDON} documentation. \texttt{POSEIDON} V1.2 was released alongside this paper which updated the readthedocs with new tutorial notebooks, and an opacity database page (\href{https://poseidon-retrievals.readthedocs.io/en/latest/content/opacity_database.html}{\texttt{POSEIDON}'s Opacity Database}) with gas-phase and aerosol opacity previews and references. }

\section{Acknowledgements}

\nt{We would like to express our gratitude to a multitude of working groups for providing insight and guidance to this project. This work stands on the hard work of the PICASO team who integrated multiple scattering into their radiative transfer scheme. We specifically thank Dr. Natasha Batalha and Sagnick Mukherjee for their continued guidance on helping us integrate multiple scattering into our code. We thank the BDNYC working group for helpful discussions on how to improve POSEIDON for brown dwarf retrievals, and providing a workspace to develop the code. We thank the JWST-TST-DREAMS team for their continued feedback and support when first developing this code to retrieve quartz clouds on WASP-17~b. We thank the anonymous referee for a very constructive and thorough report of our manuscript that significantly improved the quality of this work. We thank Cornell University's support of this project by funding an in-person poster presentations at the 243rd meeting of the American Astronomical Society and an oral presentation at ERES IX. E.M. thanks Dr. Sarah Moran, Dr. Hannah Wakeford, Dr. Ben Burningham, Dr. Daniel Kitzmann, Dr. Michael Radke, and Dr. Carly Howett for help in uncovering the secrets of aerosols. E.M. acknowledges that this material is based upon work supported by the National Science Foundation Graduate Research Fellowship under Grant No.\ 2139899. R.J.M. is supported by NASA through the NASA Hubble Fellowship grant HST-HF2-51513.001, awarded by the Space Telescope Science Institute, which is operated by the Association of Universities for Research in Astronomy, Inc., for NASA, under contract NAS 5\-26555.}


\appendix 

\begin{figure*}[t]
     \centering
         \includegraphics[width=1.0\linewidth]{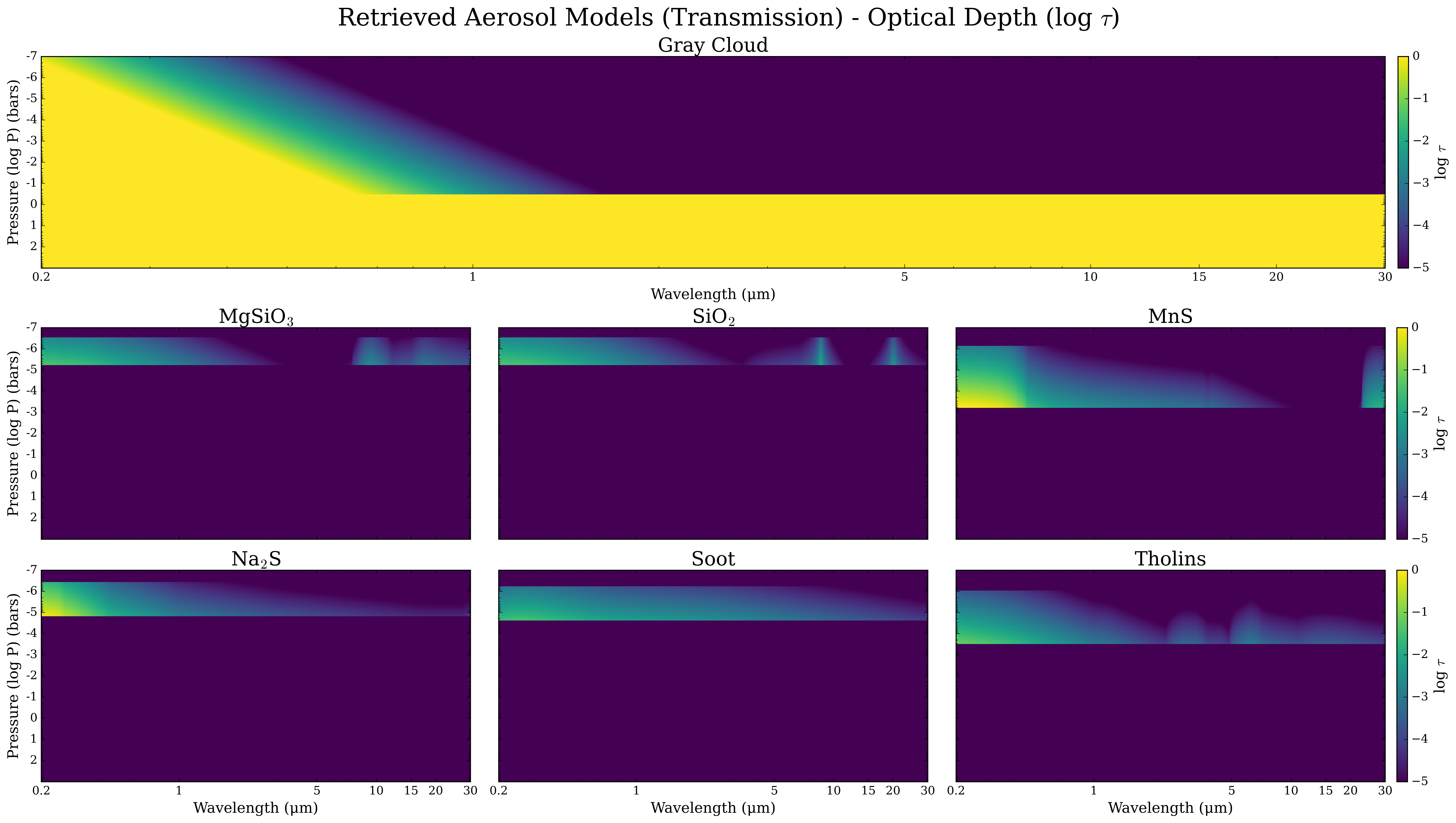}
     \caption{Log optical depth (log $\tau$) of each aerosol model explored in the transmission suite of retrievals (\S \ref{sec:hd189-transmission-retrievals}) overlaid on a wavelength (\textmu m) vs log pressure (bar) plot. Forward models were generated with the median retrieved aerosol properties found in Table \ref{table:transmission_results}. The gray cloud model (top panel) displays the infinite opacity deck and parameterized haze. Each aerosol model (bottom panels) displays the retrieved pressure region spanned by the aerosol slab, as well as opacity contributions in the short wavelength scattering region (0.2-1 \textmu m) and infrared absorption (1-30 \textmu m). Color bar is normalized to $\tau$ = 1 and $\tau$ = 1e-5.}
     \label{fig:optical-depth}
\end{figure*}

\begin{figure*}[t]
     \centering
         \includegraphics[width=1.0\linewidth]{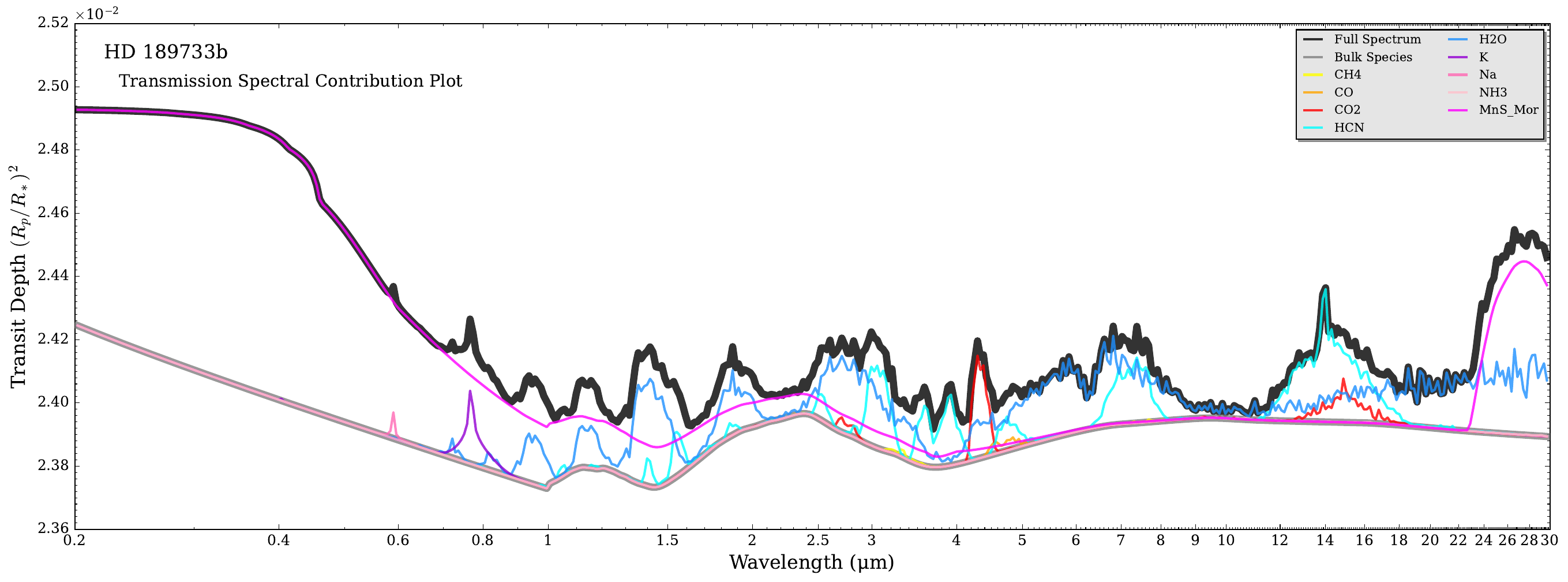}
     \caption{Spectral contribution of highest evidence transmission retrieval, the MnS spectrum (black) displayed in Figure \ref{fig:transit-sulfides-soots}, Sulfides. H$_2$ + He continuum opacity (gray) provides a baseline to the spectrum. MnS (magenta) dominates the opacity 0.2--0.6 \textmu m via Mie scattering, and has an absorption feature at the longest wavelengths, 23-30 \textmu m. Na (hot pink) has a weak absorption feature at 0.60 \textmu m. K (purple) has an absorption feature at 0.76 \textmu m. H$_2$O (blue) dominates the spectrum in a series of absorption features in the infrared. H$_2$S (cyan) broadens a few water features, and has a significant feature at 14 \textmu m. CO$_2$ (red) has an absorption feature at 4.5 \textmu m. CO (orange), CH$_4$ (yellow), and NH$_3$ (pink) do not contribute to the spectrum.}
     \label{fig:transmission-MNS-Contribution}
\end{figure*}

\begin{figure*}[t]
     \centering
         \includegraphics[width=1.0\linewidth]{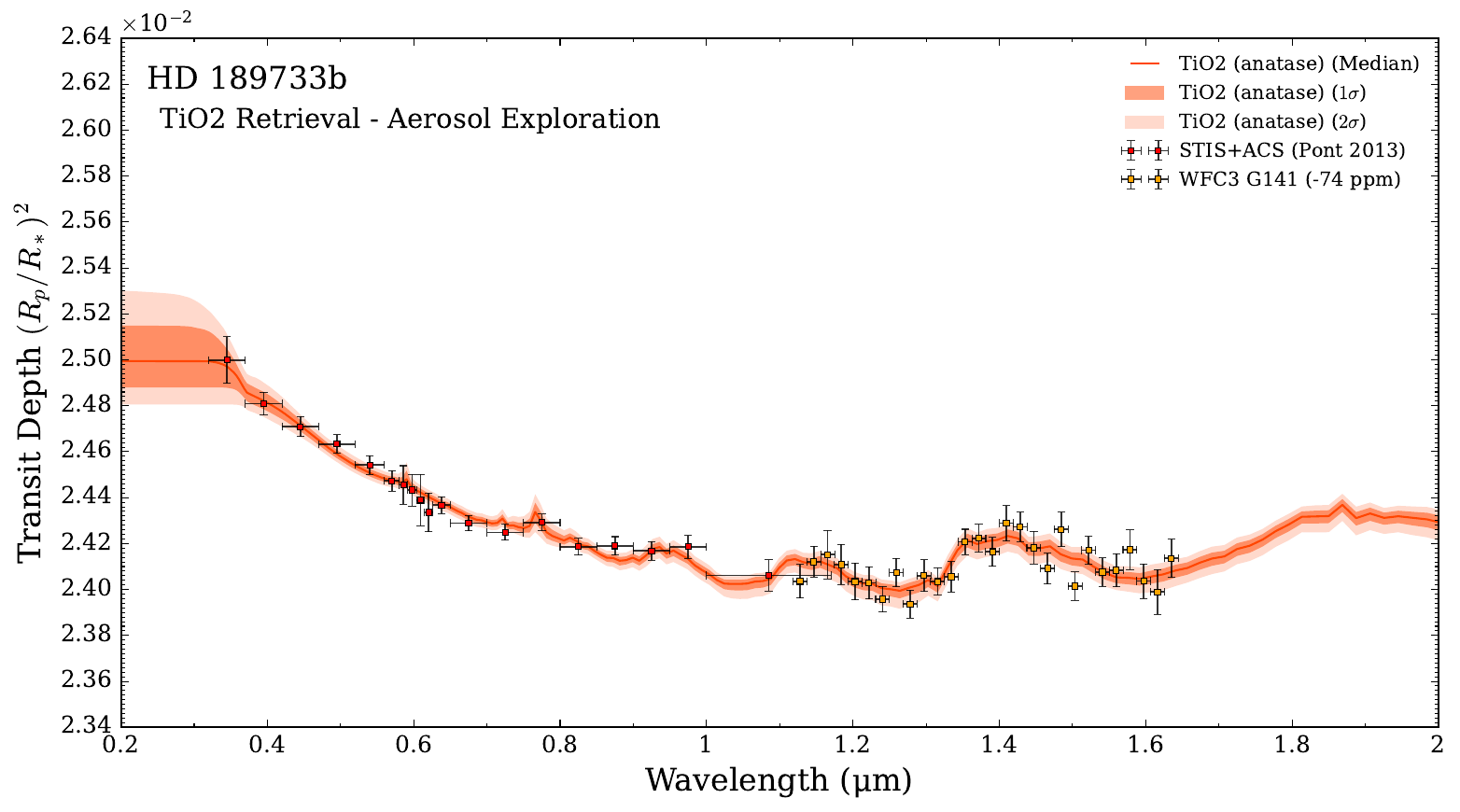}
     \caption{Zoom in on the highest evidence aerosol (TiO$_2$) from \S~\ref{sec:aerosol_showcase}, which only fit the \textit{HST} transmission data. TiO$_2$ is able to near perfectly fit the slope. TiO$_2$ hold significance due to being commonly invoked as a `seed' particle by which other volatiles condense on to.}
     \label{fig:transmission-TiO2}
\end{figure*}

\begin{figure*}[t]
     \centering
         \includegraphics[width=1.0\linewidth]{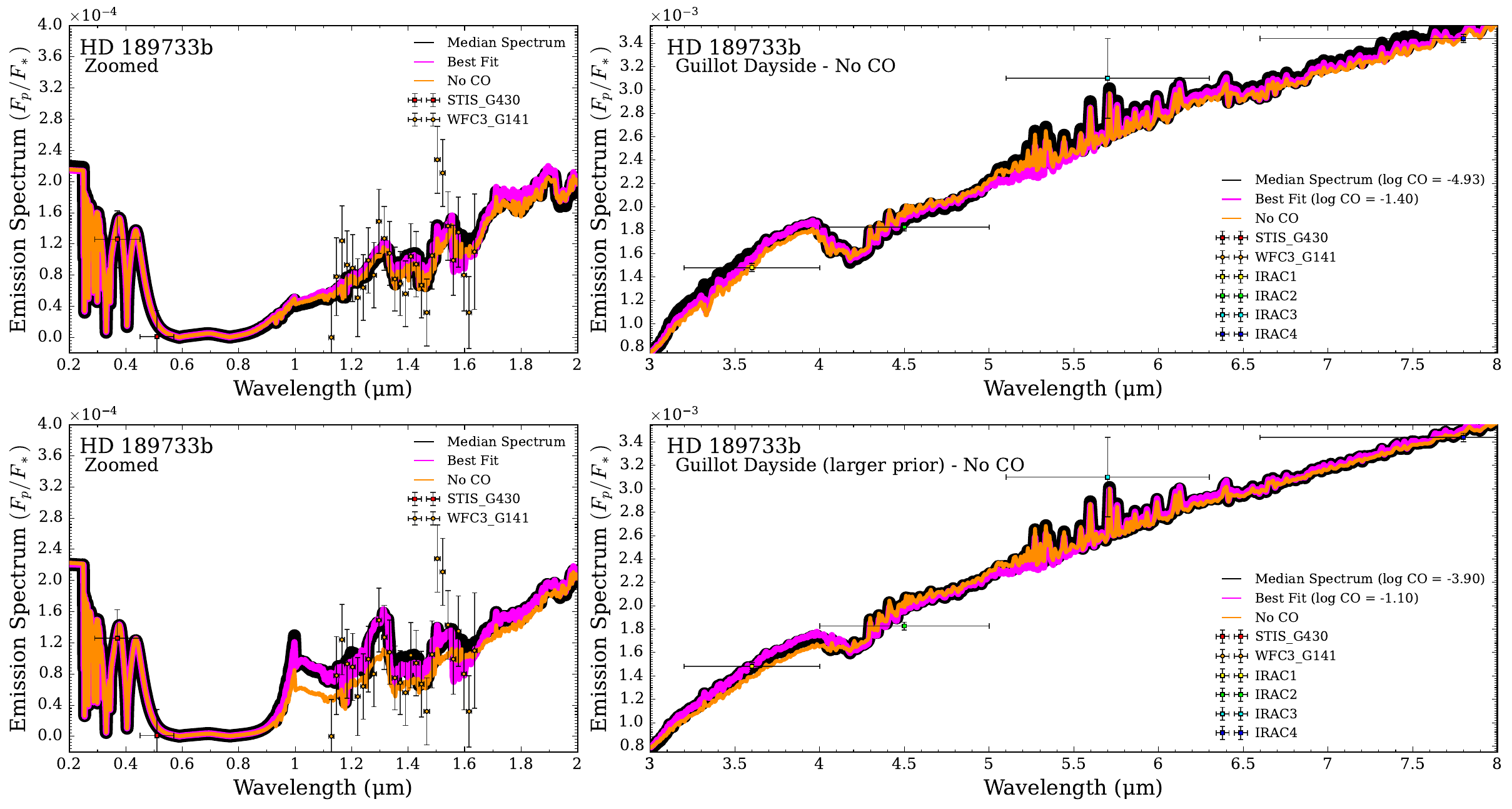}
     \caption{Figure displaying the effect of removing CO from the median (black) and best-fit (magenta) retrieved atmospheric properties of the Guillot dayside (top) and Guillot dayside (larger prior) (bottom) retrievals (no CO, dark orange). The best-fit retrieval for both cases has high CO abundances. CO imparts only a slight feature in the data (around 5 \textmu m where it widens the CO$_2$ feature at 4.5 \textmu m). The retrieval is utilizing CO to increase the mean molecular weight of the atmosphere, decreasing the scale height, and allowing flux to escape from deeper in the atmosphere. This effect can be seen especially in the 0.8 - 2 \textmu m region of the Dayside Guillot (larger prior) retrieval. Spectra are plotted at R = 500.}
     \label{fig:No-CO}
\end{figure*}

\begin{figure*}[t]
     \centering
         \includegraphics[width=1.0\linewidth]{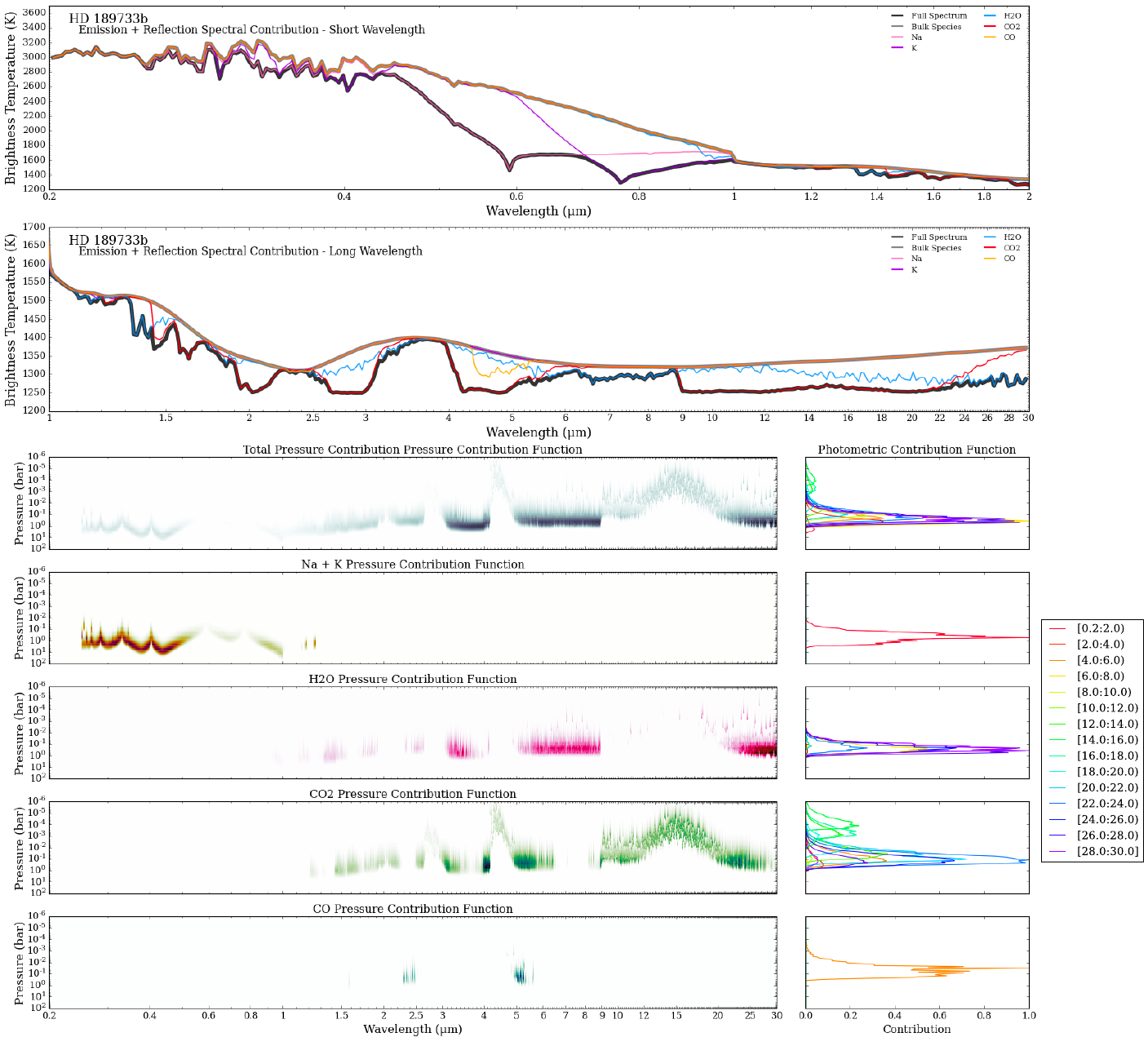}
     \caption{Spectral and pressure contribution of the clear emission + reflection retrieval displayed in Figure \ref{fig:emission-clear-gray}. Top: Spectral contribution of detected species, with the emission spectrum broadcasted into brightness temperature. Bottom: Pressure and photometric contribution function for the total spectrum, as well as specific chemical species. Each pressure contribution function (left) is normalized with darker regions representing the pressure regions where the opacity of that species is contributing to the spectrum. The photometric contribution (right) shows which wavelengths are the most important for which pressure regions and species. The pressure and photometric contribution functions show that a bulk of the spectrum is forming in the 10 to 0.1 bar region, where the temperature gradient is.}
     \label{fig:emission-contribution}
\end{figure*}

\nt{Section \ref{sec:appendex_refractive_indices} makes special note of the refractive indices utilized in our suite of HD~189733~b transmission and emission retrievals by giving details (such as synthetic vs natural samples, how lab measurements were taken, polymorphs, polarizations, and how databases compiled the sources) on the original sources of the refractive indices. Table \ref{table:transmission_data} tabulates the archival \textit{HST} and \textit{Spitzer} transmission data of HD~189733~b. Table \ref{table:emission_data} tabulates the archival \textit{HST} and \textit{Spitzer} emission data of HD~189733~b. Table \ref{table:transmission_results} tabulates the priors and retrieved median and 1$\sigma$ results for the transmission models discussed in \S \ref{sec:hd189-transmission-retrievals}. Table \ref{table:transmission_cloud_model_results} tabulates the priors and the retrieved median and 1$\sigma$ results for the transmission aerosol model exploration discussed in \S \ref{sec:different_cloud_models}. Figure \ref{fig:optical-depth} displays the aerosol optical depth (versus pressure and wavelength) of each of the retrieved aerosol models in \S \ref{sec:hd189-transmission-retrievals}. Figure \ref{fig:transmission-MNS-Contribution} displays the spectral contribution function for the highest evidence transmission retrieval with a compositionally specific aerosol (model with MnS). Figure \ref{fig:transmission-TiO2} displays a zoom-in on the highest evidence aerosol (TiO$_2$) from the aerosol exploration retrievals in \S \ref{sec:aerosol_showcase}. Table \ref{table:transmission_stats} tabulates the evidence, $\chi$, $\chi_{red}^2$, and $\sigma$ significance of the transmission retrievals. Table \ref{table:emission_results} tabulates the priors and retrieved median and 1$\sigma$ results for the emission and emission + reflection models discussed in \S \ref{sec:hd189-emission-retrievals}. Figure \ref{fig:No-CO} displays the effect of removing CO from the median and best-fit retrieved atmospheric properties from the clear emission + reflection retrievals. Figure \ref{fig:emission-contribution} displays the spectral contribution (in brightness temperature), total pressure contribution and specific chemical species pressure contribution (as well as photometric contributions), of the clear emission + reflection retrieval displayed in Figure \ref{fig:emission-clear-gray}. Table \ref{table:emission_cloud_model_results} tabulates the retrieved aerosol properties for the emission + reflection aerosol species exploration in \S \ref{er_cloud_species}. Table \ref{table:pt_results} tabulates the retrieved pressure-temperature properties for the emission + reflection P-T profile exploration in \S \ref{er_pt_profiles}. Table \ref{table:emission_stats} tabulates the evidence, $\chi$, $\chi_{red}^2$, and $\sigma$ significance of the emission retrievals while Table \ref{table:reflection_stats} tabulates them for emission + reflection retrievals in \S \ref{sec:hd189-emission-retrievals}. Table \ref{table:chemical_inventory} displays the chemical inventory used in our retrievals with relevant opacity references.}

\section{Refractive Index Information}\label{sec:appendex_refractive_indices}

\nt{Here, we make special note of the refractive indices used in our suite of HD~189733~b transmission and emission retrievals. Similar details for every aerosol in Table \ref{table:aerosol_species} are found in the `Aerosol-Database-Readme.txt' on Zenodo (\S \ref{sec:Links}) or \href{https://poseidon-retrievals.readthedocs.io/en/latest/_static/Aerosol-Database-Readme.txt}{\texttt{POSEIDON}'s documentation}, or  \href{https://poseidon-retrievals.readthedocs.io/en/latest/content/opacity_database.html}{\texttt{POSEIDON}'s Opacity Database}.}

\nt{The MgSiO$_3$ refractive indices from the \citet{WakefordSing2015} database include the short wavelength data of \citet{Egan1975} and longer wavelength data of \citet{Dorschner1995}. \citet{Egan1975} focused on UV measurements of many silicate species and used a sample of natural, brown enstatite from a suppler in India that was originally noted in \citet{HuffmanStapp1971}. Given that this was a natural sample, it is assumed that it was crystalline. Forsterite is an anisotropic, biaxial crystal, which means its refractive indices depend on how light is polarized with respect to its polarization axes. \citet{Egan1975} mention no polarization, so it is assumed that they were averaged over a surface. \citet{Dorschner1995} explored many different amorphous glass pyroxene and olivines that were produced by quenching a melt (the preparation of samples is detailed in \citet{Jager1994}). Amorphous aerosols interact with light isotropically. Refractive indices from \citet{Dorschner1995} are publicly available on the \texttt{Database of Optical Constants for Cosmic Dust} (\url{https://www.astro.uni-jena.de/Laboratory/OCDB/amsilicates.html}).}

\nt{The SiO$_2$ refractive indices from the \citet{WakefordSing2015} database include the short wavelength data from Philipp in \citet{Palik1985} (Volume 1, Section 34) and the longer wavelength data of \citet{Zeidler2013}. Philipp in \citet{Palik1985} compile a variety of $\alpha$ quartz (room-temperature crystalline polymorph SiO$_2$) lab data. Since $\alpha$ quartz is an anisotropic, uniaxial crystal, \citet{Palik1985} report both the ordinary (E $\parallel$ a,b) and extraordinary (E $\parallel$ c) ray refractive indices. \citet{WakefordSing2015} record only the ordinary-ray in their entry. \citet{Zeidler2013} explore the temperature and directional dependence of a natural $\alpha$ quartz crystal from Brazil. They capture the phase transition from $\alpha$ to $\beta$ quartz that occurs at 846-847K. \citet{WakefordSing2015} record only the 928K, $\beta$ quartz, extraordinary ray (E $\parallel$ c) indices. \citet{Zeidler2013} refractive indices are publicly available on the \texttt{Database of Optical Constants for Cosmic Dust} (\url{https://www.astro.uni-jena.de/Laboratory/OCDB/crsilicates.html}). We note that the \citet{WakefordSing2015} indices are sparse from 0.15 to 6.66 \textmu m. This does not change our results since the interpolated refractive indices in those regions match the more complete data found in \citet{Palik1985} and that optical properties in this wavelength region are dominated by particle size, not composition. We also note that the original \citet{WakefordSing2015} paper cites \citet{Andersen2006}, which utilized amorphous glass indices from Phillip in \citet{Palik1985} (Volume 1, Section 35). The indices were updated post publication to utilize only crystalline indices.}

\nt{The MnS refractive indices utilized for our retrievals are described in detail in \S \ref{sec:aerosol_database}. \citet{Huffman1967} grew single crystal samples from $\alpha$ MnS (a cubic, isotropic crystal) powder and explore optical properties using three different techniques depending on wavelengths being explored. Na$_2$S, which was used to perform the sulfide extrapolation, references are detailed in the next paragraph. We note that in \citet{KitzmannHeng2018} they mention that either Na$_2$S or ZnS can be used to perform a sulfide extrapolation, but only cite \citet{Montaner1979} which contains only Na$_2$S indices. \citet{KitzmannHeng2018} utilize only the real portion of Na$_2$S and used Kramers-Kronig analysis to compute the imaginary indices (pers. comm). 
}

\nt{The Na$_2$S refractive indices from the \citet{WakefordSing2015} database were compiled originally in \citet{Morley2012}. \citet{Morley2012} utilized the short wavelength data of \citet{Khachai2009} and longer wavelength data of \citet{Montaner1979}. \citet{Khachai2009} computed the optical properties of Na$_2$S from first principles (i.e., completely computational) assuming a cubic, anit-flourite structure. \citet{Montaner1979} took measurements low-temperature (15K) infrared data of crystals grown from powder (details on how crystals were grown can be found in \citet{Moret1977}). We note that Na$_2$S has a sulfide absorption feature past 30 \textmu m, the upper bound of our precomputed database.} 

\nt{The Tholin refractive indices from the \citet{WakefordSing2015} database include data from \citet{Khare1984} and \citet{Ramirez2002}. \citet{Khare1984} exposed a `Titan-like' atmosphere (90\% N$_2$ + 10\% CH$_4$) to electrical discharge for four months to produce a dark-red solid. \citet{Khare1984} note that their 0.2-0.4 \textmu m region was unreliable. \citet{Ramirez2002} took short wavelength data of Tholins produced from a 98\% N$_2$ + 2\% CH$_4$ atmosphere and were used to fill in the 0.2 - 1 \textmu m uncertain data in \citet{Khare1984}.}

\nt{The flame soot refractive indices from the \texttt{gCMCRT} database \citep{Lee2022} were compiled originally in \citet{Lavvas2017} who compiled many different lab sources \citep{Chang1990, Stagg1993, LEE1981, Gavilan2016} and checked for consistency between the real and imaginary indices by using the Kramers-Kronig relation. They derived average real and imaginary indices from their compiled lab sources and then calculated corresponding imaginary and real indices using the Kramers-Kronig relation. The refractive indices used in this study are specifically the set of indices derived from the Kramers-Kronig analysis.}

\nt{Aerosols can have multiple polymorphs and crystalline aerosols can have directional-polarization specific refractive indices. Exploring temperature-dependent polymorphs and their directionality will be featured in future work \citep[][Mullens \& Lewis (in prep)]{Moran2024}. In the meantime, for SiO$_2$ specifically we have compiled three additional SiO$_2$ datasets: airborne $\alpha$ quartz in \citet{Herbin2023}, $\alpha$ quartz compiled in Philipp in \citet{Palik1985} (Volume 1, Section 34), and silica glass compiled in Philipp in \citet{Palik1985} (Volume 1, Section 35)) where Kramers-Kronig analysis (using the open-source \texttt{PyElli} package, \url{https://github.com/PyEllips/pyElli}) was iterated twice on the \citet{Palik1985} datasets to generate a self-consistent set of refractive indices and fill in empty entries. The $\alpha$ quartz ordinary and extraordinary refractive indices were weighted by 2/3 and 1/3 respectively \citep{Mogli2007,Reed2017}.}

\startlongtable
\begin{deluxetable*}{llllll}
\colnumbers
\tablecolumns{6}
\tablewidth{0pt}
\tabletypesize{\footnotesize}
\tablecaption{Transmission Data \label{table:transmission_data}}
\tablehead{Wavelength (\textmu m) & Half-Width (\textmu m) & (R$_p$/R$_s$)$^2$ & Error & Instrument & Source}
\startdata
0.345&	0.025&	0.024999&	0.000101& HST STIS (\textit{G430L,G750M}), ACS (\textit{G800L}) & \citet{Pont2013}, Table 6 \\
0.395&	0.025&	0.024809&	0.00005& HST STIS (\textit{G430L,G750M}), ACS (\textit{G800L}) & \citet{Pont2013}, Table 6 \\
0.445&	0.025&	0.024709&	0.000044& HST STIS (\textit{G430L,G750M}), ACS (\textit{G800L}) & \citet{Pont2013}, Table 6 \\
0.495&	0.025&	0.024633&	0.000041& HST STIS (\textit{G430L,G750M}), ACS (\textit{G800L}) & \citet{Pont2013}, Table 6 \\
0.54&	0.02&	0.024542&	0.000041& HST STIS (\textit{G430L,G750M}), ACS (\textit{G800L}) & \citet{Pont2013}, Table 6 \\
0.57&	0.01&	0.024473&	0.000044& HST STIS (\textit{G430L,G750M}), ACS (\textit{G800L}) & \citet{Pont2013}, Table 6 \\
0.586&	0.006&	0.024455&	0.000084& HST STIS (\textit{G430L,G750M}), ACS (\textit{G800L}) & \citet{Pont2013}, Table 6 \\
0.598&	0.006&	0.024433&	0.000069& HST STIS (\textit{G430L,G750M}), ACS (\textit{G800L}) & \citet{Pont2013}, Table 6 \\
0.6095&	0.0055&	0.024389&	0.000112& HST STIS (\textit{G430L,G750M}), ACS (\textit{G800L}) & \citet{Pont2013}, Table 6 \\
0.6205&	0.0055&	0.024336&	0.000084& HST STIS (\textit{G430L,G750M}), ACS (\textit{G800L}) & \citet{Pont2013}, Table 6 \\
0.638&	0.012&	0.024367&	0.000037& HST STIS (\textit{G430L,G750M}), ACS (\textit{G800L}) & \citet{Pont2013}, Table 6 \\
0.675&	0.025&	0.024289&	0.000034& HST STIS (\textit{G430L,G750M}), ACS (\textit{G800L}) & \citet{Pont2013}, Table 6 \\
0.725&	0.025&	0.024249&	0.000034& HST STIS (\textit{G430L,G750M}), ACS (\textit{G800L}) & \citet{Pont2013}, Table 6 \\
0.775&	0.025&	0.024292&	0.000037& HST STIS (\textit{G430L,G750M}), ACS (\textit{G800L}) & \citet{Pont2013}, Table 6 \\
0.825&	0.025&	0.024186&	0.000037& HST STIS (\textit{G430L,G750M}), ACS (\textit{G800L}) & \citet{Pont2013}, Table 6 \\
0.875&	0.025&	0.02419&	0.00004& HST STIS (\textit{G430L,G750M}), ACS (\textit{G800L}) & \citet{Pont2013}, Table 6 \\
0.925&	0.025&	0.024168&	0.00004& HST STIS (\textit{G430L,G750M}), ACS (\textit{G800L}) & \citet{Pont2013}, Table 6 \\
0.975&	0.025&	0.024186&	0.00005& HST STIS (\textit{G430L,G750M}), ACS (\textit{G800L}) & \citet{Pont2013}, Table 6 \\
1.085&	0.085&	0.024062&	0.000068& HST STIS (\textit{G430L,G750M}), ACS (\textit{G800L}) & \citet{Pont2013}, Table 6 \\
\hline
1.1279&	0.0095&	0.023962&	0.000073 & HST WFC3 (\textit{G141}) & \citet{McCullough2014}, Table 3 \\
1.1467&	0.0095&	0.024047&	0.000067& HST WFC3 (\textit{G141}) & \citet{McCullough2014}, Table 3 \\
1.1655&	0.0095&	0.024078&	0.000105& HST WFC3 (\textit{G141}) & \citet{McCullough2014}, Table 3 \\
1.1843&	0.0095&	0.024035&	0.000087& HST WFC3 (\textit{G141}) & \citet{McCullough2014}, Table 3 \\
1.2031&	0.0095&	0.023961&	0.00008& HST WFC3 (\textit{G141}) & \citet{McCullough2014}, Table 3 \\
1.2218&	0.0095&	0.023955&	0.00007& HST WFC3 (\textit{G141}) & \citet{McCullough2014}, Table 3 \\
1.2406&	0.0095&	0.023884&	0.000056& HST WFC3 (\textit{G141}) & \citet{McCullough2014}, Table 3 \\
1.2594&	0.0095&	0.024&   	0.000062& HST WFC3 (\textit{G141}) & \citet{McCullough2014}, Table 3 \\
1.2782&	0.0095&	0.023863&	0.000061& HST WFC3 (\textit{G141}) & \citet{McCullough2014}, Table 3 \\
1.297&	0.0095&	0.023987&	0.000069& HST WFC3 (\textit{G141}) & \citet{McCullough2014}, Table 3 \\
1.3157&	0.0095&	0.023961&	0.00006& HST WFC3 (\textit{G141}) & \citet{McCullough2014}, Table 3 \\
1.3345&	0.0095&	0.023982&	0.000066& HST WFC3 (\textit{G141}) & \citet{McCullough2014}, Table 3 \\
1.3533&	0.0095&	0.024134&	0.000055& HST WFC3 (\textit{G141}) & \citet{McCullough2014}, Table 3 \\
1.3721&	0.0095&	0.024149&	0.000061& HST WFC3 (\textit{G141}) & \citet{McCullough2014}, Table 3 \\
1.3909&	0.0095&	0.024091&	0.000063& HST WFC3 (\textit{G141}) & \citet{McCullough2014}, Table 3 \\
1.4096&	0.0095&	0.024215&	0.000077& HST WFC3 (\textit{G141}) & \citet{McCullough2014}, Table 3 \\
1.4284&	0.0095&	0.024199&	0.000064& HST WFC3 (\textit{G141}) & \citet{McCullough2014}, Table 3 \\
1.4472&	0.0095&	0.024108&	0.000071& HST WFC3 (\textit{G141}) & \citet{McCullough2014}, Table 3 \\
1.466&	0.0095&	0.024018&	0.000067& HST WFC3 (\textit{G141}) & \citet{McCullough2014}, Table 3 \\
1.4847&	0.0095&	0.024188&	0.000075& HST WFC3 (\textit{G141}) & \citet{McCullough2014}, Table 3 \\
1.5035&	0.0095&	0.023941&	0.000062& HST WFC3 (\textit{G141}) & \citet{McCullough2014}, Table 3 \\
1.5223&	0.0095&	0.024097&	0.000061& HST WFC3 (\textit{G141}) & \citet{McCullough2014}, Table 3 \\
1.5411&	0.0095&	0.024002&	0.000062& HST WFC3 (\textit{G141}) & \citet{McCullough2014}, Table 3 \\
1.5599&	0.0095&	0.02401& 	0.000072& HST WFC3 (\textit{G141}) & \citet{McCullough2014}, Table 3 \\
1.5786&	0.0095&	0.0241&  	0.000087& HST WFC3 (\textit{G141}) & \citet{McCullough2014}, Table 3 \\
1.5974&	0.0095&	0.023963&	0.000075& HST WFC3 (\textit{G141}) & \citet{McCullough2014}, Table 3 \\
1.6162&	0.0095&	0.023916&	0.000098& HST WFC3 (\textit{G141}) & \citet{McCullough2014}, Table 3 \\
1.635&	0.0095&	0.024062&	0.000084& HST WFC3 (\textit{G141}) & \citet{McCullough2014}, Table 3 \\
\hline
3.55&	0.35&	0.024047&	0.000084& \textit{Spitzer} IRAC Ch1 & \citet{Pont2013}, Table 6 \\
4.5&	0.5&	0.024155&	0.000109& \textit{Spitzer} IRAC Ch2 & \citet{Pont2013}, Table 6 \\
5.7&	0.7&	0.023951&	0.000207& \textit{Spitzer} IRAC Ch3 & \citet{Pont2013}, Table 6 \\
7.85&	1.45&	0.024056&	0.000105& \textit{Spitzer} IRAC Ch4 & \citet{Pont2013}, Table 6 \\
24&	0.5&	0.023898&	0.000291& \textit{Spitzer} MIPS & \citet{Pont2013}, Table 6
\enddata 
\tablecomments{Central wavelength (1), bin half width (2), transit depth (3), and transit depth error (4) used in the transmission retrievals. Instrument (5) indicates which instrumental sensitivity function was used to convolve forward models for the retrieval analysis. Note that the \citet{Pont2013} 0.345-1.085 data is multiple instruments over multiple observations averaged over \nt{passbands, and therefore that dataset is convolved with a generic box function.} All transmission data is adapted from \citet{Zhang2020}, Table 8 with direct sources listed in (6).}
\end{deluxetable*}

\begin{deluxetable*}{llllll}
\colnumbers
\tablecolumns{6}
\tablewidth{0pt}
\tabletypesize{\footnotesize}
\tablecaption{Emission Data \label{table:emission_data}}
\tablehead{Wavelength (\textmu m) & Half-Width (\textmu m) & (F$_p$/F$_s$) & Error & Instrument & Source}
\startdata
0.370& 0.08& 0.000126& 0.0000365& HST STIS (\textit{G430L}) & \citet{Evans2013}, Table 1 \\
0.510& 0.06& 0.000001& 0.0000335& HST STIS (\textit{G430L}) & \citet{Evans2013}, Table 1 \\
\hline
1.1279&	0.0095&	0&	0.000047 & HST WFC3 (\textit{G141}) & \citet{Crouzet2014}, Table 2 \\
1.1467&	0.0095&	0.000078&	0.00005 & HST WFC3 (\textit{G141}) & \citet{Crouzet2014}, Table 2 \\
1.1655&	0.0095&	0.000124&	0.000045 & HST WFC3 (\textit{G141}) & \citet{Crouzet2014}, Table 2 \\
1.1843&	0.0095&	0.000093&	0.000044 & HST WFC3 (\textit{G141}) & \citet{Crouzet2014}, Table 2 \\
1.2031&	0.0095&	0.000089&	0.000043 & HST WFC3 (\textit{G141}) & \citet{Crouzet2014}, Table 2 \\
1.2218&	0.0095&	0.000051&	0.00005 & HST WFC3 (\textit{G141}) & \citet{Crouzet2014}, Table 2 \\
1.2406&	0.0095&	0.000064&	0.000042 & HST WFC3 (\textit{G141}) & \citet{Crouzet2014}, Table 2 \\
1.2594&	0.0095&	0.000099&	0.000042 & HST WFC3 (\textit{G141}) & \citet{Crouzet2014}, Table 2 \\
1.2782&	0.0095&	0.00008&	0.000042 & HST WFC3 (\textit{G141}) & \citet{Crouzet2014}, Table 2 \\
1.2969&	0.0095&	0.000149&	0.000041 & HST WFC3 (\textit{G141}) & \citet{Crouzet2014}, Table 2 \\
1.3157&	0.0095&	0.000127&	0.000041 & HST WFC3 (\textit{G141}) & \citet{Crouzet2014}, Table 2 \\
1.3345&	0.0095&	0.000108&	0.000041 & HST WFC3 (\textit{G141}) & \citet{Crouzet2014}, Table 2 \\
1.3533&	0.0095&	0.000075&	0.000041 & HST WFC3 (\textit{G141}) & \citet{Crouzet2014}, Table 2 \\
1.3721&	0.0095&	0.000069&	0.000041 & HST WFC3 (\textit{G141}) & \citet{Crouzet2014}, Table 2 \\
1.3908&	0.0095&	0.000056&	0.000042 & HST WFC3 (\textit{G141}) & \citet{Crouzet2014}, Table 2 \\
1.4096&	0.0095&	0.000104&	0.000042 & HST WFC3 (\textit{G141}) & \citet{Crouzet2014}, Table 2 \\
1.4284&	0.0095&	0.000094&	0.000042 & HST WFC3 (\textit{G141}) & \citet{Crouzet2014}, Table 2 \\
1.4472&	0.0095&	0.000067&	0.000042 & HST WFC3 (\textit{G141}) & \citet{Crouzet2014}, Table 2 \\
1.466&	0.0095&	0.000032&	0.000043 & HST WFC3 (\textit{G141}) & \citet{Crouzet2014}, Table 2 \\
1.4848&	0.0095&	0.000105&	0.000043 & HST WFC3 (\textit{G141}) & \citet{Crouzet2014}, Table 2 \\
1.5035&	0.0095&	0.000228&	0.000043 & HST WFC3 (\textit{G141}) & \citet{Crouzet2014}, Table 2 \\
1.5223&	0.0095&	0.000211&	0.000043 & HST WFC3 (\textit{G141}) & \citet{Crouzet2014}, Table 2 \\
1.5411&	0.0095&	0.000143&	0.000044 & HST WFC3 (\textit{G141}) & \citet{Crouzet2014}, Table 2 \\
1.5599&	0.0095&	0.000099&	0.000045 & HST WFC3 (\textit{G141}) & \citet{Crouzet2014}, Table 2 \\
1.5786&	0.0095&	0.000135&	0.000045 & HST WFC3 (\textit{G141}) & \citet{Crouzet2014}, Table 2 \\
1.5974&	0.0095&	0.00008&	0.000046 & HST WFC3 (\textit{G141}) & \citet{Crouzet2014}, Table 2 \\
1.6162&	0.0095&	0.000032&	0.000046 & HST WFC3 (\textit{G141}) & \citet{Crouzet2014}, Table 2 \\
1.635&	0.0095&	0.00011&	0.000074 & HST WFC3 (\textit{G141}) & \citet{Crouzet2014}, Table 2 \\
\hline
3.6&	0.4&	0.001481&	0.000034 & \textit{Spitzer} IRAC Ch1 & \citet{Kilapatrick2020}, Table 4 \\
4.5&	0.5&	0.001827&	0.000033 & \textit{Spitzer} IRAC Ch2 & \citet{Kilapatrick2020}, Table 4 \\
5.7&	0.6&	0.0031& 	0.00034 & \textit{Spitzer} IRAC Ch3 & \citet{Charbonneau2008}, Table 1 \\
7.8&	1.2&	0.00344& 	0.000036 & \textit{Spitzer} IRAC Ch4 & \citet{Agol2010}, Table 4 \\
16&   	2.5&	0.00519&	0.00022 & \textit{Spitzer} IRS & \citet{Charbonneau2008}, Table 1 \\
23.45&	2.65&	0.00598&	0.00038 & \textit{Spitzer} MIPS & \citet{Charbonneau2008}, Table 1 \\
\enddata
\tablecomments{Central wavelength (1), bin half width (2), eclipse depth (3), and eclipse depth error (4) used in the emission + reflection retrievals. Instrument (5) indicates which instrumental sensitivity function was used to convolve forward models for the retrieval analysis. All emission data, with the exception of HST STIS \textit{G430L}, is adapted from \citet{Zhang2020}, Table 9 with direct sources listed in (6).}
\end{deluxetable*}

\begin{deluxetable*}{lllllllllll}
\colnumbers
\tablecolumns{11}
\tablewidth{0pt}
\tabletypesize{\tiny}
\tablecaption{Transmission Priors and Results \label{table:transmission_results}}
\tablehead{ & Variable & Prior & Clear & Gray & MgSiO$_3$ & SiO$_2$ & MnS & Na$_2$S & Soot & Tholin}
\startdata
Base & R$_{\rm{p,ref}}$ [R$_{\rm{J}}$] & $\mathcal{U}$(1.11, 1.13) & $1.130^{+0.0}_{-0.0}$ & $1.127^{+0.0}_{-0.0}$ &  $1.129^{+0.0}_{-0.0}$ &  $1.129^{+0.0}_{-0.0}$ &  $1.126^{+0.0}_{-0.0}$ &  $1.129^{+0.0}_{-0.0}$ &  $1.123^{+0.0}_{-0.0}$ &  $1.129^{+0.0}_{-0.0}$ \\
& T [K] & $\mathcal{U}$(500, 1300) & $984.5^{+35.2}_{-46.3}$ & $700.2^{+149.5}_{-99.1}$ &  $695.5^{+94.4}_{-65.7}$ &  $679.9^{+83.4}_{-58.3}$ &  $775.1^{+139.9}_{-108.8}$ &  $641.2^{+75.9}_{-54.2}$ &  $1023.1^{+63.2}_{-56.2}$ &  $776.6^{+110.3}_{-84.9}$ \\
& $\log$\, Na & $\mathcal{U}$(-12, -1) & $-1.0^{+0.0}_{-0.1}$ & $-7.9^{+2.7}_{-2.6}$ &  $-10.4^{+1.1}_{-1.0}$ &  $-10.3^{+1.1}_{-1.0}$ &  $-9.7^{+1.5}_{-1.4}$ &  $-10.1^{+1.3}_{-1.2}$ &  $-9.8^{+0.9}_{-1.0}$ &  $-10.2^{+1.0}_{-0.9}$\\
& $\log$\, K & $\mathcal{U}$(-12, -1) & $-7.3^{+2.0}_{-3.0}$ & $-6.2^{+1.6}_{-3.4}$ &  $-9.1^{+0.7}_{-0.9}$ &  $-8.9^{+0.7}_{-0.8}$ &  $-8.8^{+0.8}_{-1.0}$ &  $-8.8^{+0.9}_{-1.0}$ &  $-9.5^{+0.4}_{-0.5}$ &  $-8.7^{+0.5}_{-0.6}$ \\
& $\log$\, H$_2$O & $\mathcal{U}$(-12, -1) & $-3.5^{+0.2}_{-0.2}$ &  $-2.5^{+0.5}_{-0.7}$ &  $-3.8^{+0.5}_{-0.5}$ &  $-3.7^{+0.4}_{-0.4}$ &  $-3.9^{+0.5}_{-0.5}$ &  $-3.5^{+0.5}_{-0.5}$ &  $-4.9^{+0.2}_{-0.2}$ &  $-4.0^{+0.4}_{-0.4}$ \\
& $\log$\, CO & $\mathcal{U}$(-12, -1) & $-7.3^{+3.3}_{-3.1}$ & $-6.9^{+3.2}_{-3.2}$ &  $-6.5^{+3.2}_{-3.4}$ &  $-6.7^{+3.2}_{-3.4}$ &  $-6.7^{+3.2}_{-3.3}$ &  $-6.7^{+3.2}_{-3.4}$ &  $-6.5^{+2.4}_{-2.9}$ &  $-4.8^{+1.8}_{-2.9}$\\
& $\log$\, CO$_2$ & $\mathcal{U}$(-12, -1) & $-8.0^{+2.5}_{-2.5}$ & $-7.0^{+2.8}_{-3.1}$ &  $-7.5^{+2.7}_{-2.8}$ &  $-7.7^{+2.6}_{-2.6}$ &  $-6.9^{+2.5}_{-3.1}$ &  $-7.0^{+2.9}_{-3.0}$ &  $-6.9^{+1.2}_{-2.1}$ &  $-9.1^{+2.1}_{-1.7}$ \\
& $\log$\, CH$_4$ & $\mathcal{U}$(-12, -1) & $-9.0^{+2.1}_{-1.9}$ & $-8.5^{+2.3}_{-2.2}$ &  $-8.8^{+1.9}_{-2.0}$ &  $-8.9^{+2.0}_{-1.9}$ &  $-8.6^{+2.2}_{-2.1}$ &  $-8.6^{+2.1}_{-2.1}$ &  $-8.9^{+1.9}_{-1.8}$ &  $-9.2^{+1.6}_{-1.5}$\\
& $\log$\, HCN & $\mathcal{U}$(-12, -1) & $-8.1^{+2.6}_{-2.4}$ & $-5.8^{+2.2}_{-3.9}$ &  $-4.9^{+1.0}_{-2.5}$ &  $-5.0^{+0.9}_{-2.6}$ &  $-5.0^{+1.1}_{-3.1}$ &  $-4.7^{+1.0}_{-2.6}$ &  $-7.0^{+1.6}_{-3.0}$ &  $-8.4^{+3.0}_{-2.2}$\\
& $\log$\, NH$_3$ & $\mathcal{U}$(-12, -1) & $-8.8^{+1.9}_{-2.0}$ & $-8.6^{+2.3}_{-2.2}$ &  $-9.2^{+2.1}_{-1.8}$ &  $-9.2^{+2.1}_{-1.8}$ &  $-9.1^{+2.0}_{-1.9}$ &  $-8.9^{+2.0}_{-1.9}$ &  $-10.0^{+1.3}_{-1.1}$ &  $-9.2^{+1.8}_{-1.5}$\\
 & $\delta_{\rm{rel},WFC3}$ [ppm]& $\mathcal{U}$(-100, 100) & $-87.4^{+13.1}_{-8.5}$ & $-63.2^{+47.3}_{-25.7}$ &  $-59.7^{+23.1}_{-21.1}$ &  $-59.4^{+23.2}_{-20.4}$ &  $-81.7^{+17.7}_{-12.1}$ &  $-77.8^{+19.5}_{-14.3}$ &  $-88.7^{+10.2}_{-6.8}$ &  $-73.7^{+17.4}_{-14.0}$\\
\hline
Cloud Profiles & & & \\
\hline
Gray & log a & $\mathcal{U}$(-4, 8) & $\cdots$ & $5.6^{+0.7}_{-0.8}$ & & & & & & \\
& $\gamma$ & $\mathcal{U}$(-20, 2) & $\cdots$ & $-12.2^{+2.2}_{-2.4}$ & & & & & &  \\
& $\log$\,$P_{\rm{cloud}}$\, [bars] & $\mathcal{U}$(-8, 2) & $\cdots$ & $-1.4^{+1.6}_{-0.9}$ & & & & & &  \\
\hline
Slab &$\log$\,Y & $\mathcal{U}$(-30, -1) & $\cdots$ & $\cdots$ & $-11.1^{+0.8}_{-1.0}$ &  $-11.0^{+0.7}_{-0.9}$ &  $-11.8^{+0.9}_{-0.9}$ &  $-8.4^{+1.1}_{-1.2}$ &  $-10.9^{+0.6}_{-0.6}$ &  $-12.2^{+0.8}_{-0.9}$\\
&$\log$\,$r_{\rm{m}}$\,Y [\textmu m] & $\mathcal{U}$(-3, -1) & $\cdots$ & $\cdots$ & $-1.3^{+0.1}_{-0.1}$ &  $-1.4^{+0.1}_{-0.1}$ &  $-1.3^{+0.1}_{-0.1}$ &  $-2.0^{+0.2}_{-0.1}$ &  $-1.7^{+0.1}_{-0.1}$ &  $-1.4^{+0.1}_{-0.1}$ \\
&$\log$\,$P_{\rm{top,slab}}$\,Y [bars] & $\mathcal{U}$(-8, 2) & $\cdots$ & $\cdots$ & $-7.6^{+0.5}_{-0.3}$ &  $-7.6^{+0.5}_{-0.3}$ &  $-7.2^{+0.8}_{-0.5}$ &  $-7.5^{+0.6}_{-0.4}$ &  $-7.3^{+0.5}_{-0.4}$ &  $-7.1^{+0.8}_{-0.5}$\\
&$\Delta\log$\,$P$\,Y [bars] & $\mathcal{U}$(0, 10) & $\cdots$ & $\cdots$ & $1.3^{+0.7}_{-0.5}$ &  $1.3^{+0.6}_{-0.5}$ &  $3.0^{+0.8}_{-0.7}$ &  $1.6^{+0.8}_{-0.7}$ &  $1.6^{+0.5}_{-0.4}$ &  $2.5^{+0.8}_{-0.7}$  \\
\enddata
\tablecomments{Transmission retrieval priors and results discussed in \S~\ref{sec:hd189-transmission-retrievals}. Base (1) represents retrieved parameters common to all retrievals, Cloud Profiles (1) represents parameters specific to certain columns. Variable name (2), prior (3), and retrieved 1 sigma results for the clear and cloudy models (4-11). Pressure grid for transmission taken from 100 to 1e-8 bars, with a reference pressure of 1 bar. Y refers to pertinent condensate species. Retrievals were run at a resolution R = 10,000 with 1000 live points.}
\end{deluxetable*}

\begin{deluxetable*}{llll}
\colnumbers
\tablecolumns{4}
\tablewidth{0pt}
\tabletypesize{\footnotesize}
\tablecaption{Transmission Aerosol Model Exploration Priors and Results \label{table:transmission_cloud_model_results}}
\tablehead{ & Variable & Prior & MgSiO$_3$}
\startdata
Patchy Slab &$\log$\,Y & $\mathcal{U}$(-30, -1) & $-13.0^{+0.6}_{-0.5}$ \\
&$\log$\,$r_{\rm{m}}$\,Y [\textmu m] & $\mathcal{U}$(-3, -1) & $-1.4^{+0.1}_{-0.1}$ \\
&$\log$\,$P_{\rm{top,slab}}$\,Y [bars] & $\mathcal{U}$(-8, 2) & $-6.6^{+0.8}_{-0.8}$ \\
&$\Delta\log$\,$P$\,Y [bars] & $\mathcal{U}$(0, 10) & $2.7^{+0.7}_{-0.7}$ \\
&$f_{\rm{cloud}}$ [$\%$] & $\mathcal{U}$(0, 1) & $0.9^{+0.1}_{-0.1}$ \\
\hline
Slab + Deck &$\log$\,Y & $\mathcal{U}$(-30, -1) & $-12.1^{+0.6}_{-0.8}$ \\
&$\log$\,$r_{\rm{m}}$\,Y [\textmu m] & $\mathcal{U}$(-3, -1) & $-1.5^{+0.1}_{-0.1}$ \\
&$\log$\,$P_{\rm{top,slab}}$\,Y [bars] & $\mathcal{U}$(-8, 2) & $-7.0^{+0.8}_{-0.6}$ \\
&$\Delta\log$\,$P$\,Y [bars] & $\mathcal{U}$(0, 10) & $2.9^{+0.8}_{-0.6}$ \\
&$\log$\,$P_{\rm{top,deck}}$\,Y [bars] & $\mathcal{U}$(-8, 2) & $-1.5^{+0.5}_{-0.7}$ \\
\hline
Fuzzy Deck &$\log$\,$P_{\rm{top,deck}}$\,Y [bars] & $\mathcal{U}$(-8, 2) & $1.2^{+0.5}_{-1.1}$ \\
&$\log$\,$r_{\rm{m}}$\,Y [\textmu m] & $\mathcal{U}$(-3, -1) & $-1.8^{+0.1}_{-0.1}$\\
&$\log$\,$n_{\rm{max}}$\,Y [m-3] & $\mathcal{U}$(0.01, 100) & $14.7^{+0.8}_{-1.2}$ \\
&f$_{\rm{H}}$\, Y  & $\mathcal{U}$(0.01, 1) &  $1.0^{+0.0}_{-0.0}$ \\
\hline
Fuzzy Deck &$\log$\,$P_{\rm{top,deck}}$\,Y [bars] & $\mathcal{U}$(-8, 2) & $-2.3^{+0.8}_{-0.7}$ \\
Larger Prior &$\log$\,$r_{\rm{m}}$\,Y [\textmu m] & $\mathcal{U}$(-3, -1) & $-1.4^{+0.1}_{-0.1}$ \\
&$\log$\,$n_{\rm{max}}$\,Y [m-3] & $\mathcal{U}$(0.01, 100) & $7.0^{+0.5}_{-0.5}$ \\
&f$_{\rm{H}}$\, Y  & $\mathcal{U}$(0.01, 100) & $59.8^{+24.0}_{-25.8}$\\
\hline
Uniform X &$\log$\,Y & $\mathcal{U}$(-30, -1) & $-12.2^{+1.2}_{-0.8}$ \\
&$\log$\,$r_{\rm{m}}$\,Y [\textmu m] & $\mathcal{U}$(-3, -1) & $-1.9^{+0.1}_{-0.1}$ \\
\hline
Patchy &$\log$\,Y & $\mathcal{U}$(-30, -1) & $-12.5^{+0.7}_{-0.5}$ \\
Uniform X &$\log$\,$r_{\rm{m}}$\,Y [\textmu m] & $\mathcal{U}$(-3, -1) & $-1.9^{+0.1}_{-0.1}$ \\
&$f_{\rm{cloud}}$ [$\%$] & $\mathcal{U}$(0, 1) & $1.0^{+0.0}_{-0.1}$ \\
\hline 
Uniform X &$\log$\,Y & $\mathcal{U}$(-30, -1) & $-12.2^{+1.2}_{-0.7}$ \\
Two Species &$\log$\,$r_{\rm{m}}$\,Y [\textmu m] & $\mathcal{U}$(-3, -1) & $-1.9^{+0.1}_{-0.1}$ \\
 &$\log$\,Z & $\mathcal{U}$(-30, -1) & $-23.7^{+5.1}_{-4.3}$ \\
 &$\log$\,$r_{\rm{m}}$\,Z [\textmu m] & $\mathcal{U}$(-3, -1) & $-1.5^{+1.5}_{-1.0}$ \\
\hline 
Uniform X + Deck &$\log$\,Y & $\mathcal{U}$(-30, -1) & $-11.6^{+0.9}_{-1.1}$ \\
&$\log$\,$r_{\rm{m}}$\,Y [\textmu m] & $\mathcal{U}$(-3, -1) & $-1.9^{+0.1}_{-0.1}$ \\
&$\log$\,$P_{\rm{top,deck}}$\,Y [bars] & $\mathcal{U}$(-8, 2) & $-0.5^{+1.6}_{-1.7}$ \\
\enddata
\tablecomments{Transmission retrieval priors and results for the aerosol model exploration discussed in \S~\ref{sec:different_cloud_models}. aerosol model name (1), with variable names (2), priors (3), and retrieved 1 sigma results (4). Pressure grid for transmission taken from 100 to 1e-8 bars, with a reference pressure of 1 bar. Y and Z refer to condensate species MgSiO$_3$ and SiO$_2$, respectably. Retrievals were run at a resolution R = 5000 with 500 live points. }
\end{deluxetable*}

\begin{deluxetable*}{llllll}
\colnumbers
\tablecolumns{4}
\tablewidth{0pt}
\tabletypesize{\footnotesize}
\tablecaption{Transmission Statistics \label{table:transmission_stats}}
\tablehead{Model & lnZ & DoF & $\chi^2$ & $\chi^2_{red}$ & $\sigma$ Significance}
\startdata
Clear & $238.78^{+0.13}_{-0.13}$ & 41 & 388.29 & 9.47 & 19.1 \\
Gray & $418.05^{+0.11}_{-0.11}$ & 38 & 36.34 & 0.96 & --- \\
\hline
MgSiO$_3$ Slab & $404.48^{+0.15}_{-0.15}$ & 37 & 35.09 & 0.95 & 2.8\\
SiO$_2$ Slab & $405.88^{+0.15}_{-0.15}$ & 37 & 38.86 & 1.05 & 3.7 \\
MnS Slab & $411.18^{+0.14}_{-0.14}$ & 37 & 33.21 & 0.90 & ---\\
Na$_2$S Slab & $410.62^{+0.14}_{-0.14}$ & 37 & 34.59 & 0.93 & 1.7 \\
Soot Slab & $406.58^{+0.17}_{-0.17}$ & 37 & 27.09 & 0.73 & 3.5 \\
Tholin Slab & $399.91^{+0.17}_{-0.17}$ & 37 & 38.56 & 1.04 & 5.1\\
\hline
MgSiO$_3$ Slab & $404.48^{+0.15}_{-0.15}$ & 37 & 35.09 & 0.95 & 3.1 \\
MgSiO$_3$ Slab (Patchy) & $394.88^{+0.24}_{-0.24}$ & 36 & 43.44 & 1.21 & 6.2 \\
MgSiO$_3$ Slab + Deck & $399.87^{+0.24}_{-0.24}$ & 36 & 38.01 & 1.06 & 5.3 \\
MgSiO$_3$ Fuzzy Deck & $378.16^{+0.21}_{-0.21}$ & 37 & 93.47 & 2.53 & 8.5 \\
MgSiO$_3$ Fuzzy Deck (larger prior) & $412.01^{+0.20}_{-0.20}$ & 37 & 34.49 & 0.93 & --- \\
MgSiO$_3$ Uniform X & $379.21^{+0.18}_{-0.18}$ & 39 & 103.54 & 2.65 & 8.4 \\
MgSiO$_3$ Uniform X (Patchy) & $376.04^{+0.20}_{-0.20}$ & 38 & 104.79 & 2.76 & 8.7 \\
MgSiO$_3$ Uniform X (+SiO$_2$) & $378.64^{+0.19}_{-0.19}$ & 37 & 104.30 & 2.82 & 8.4 \\
MgSiO$_3$ Uniform X + Deck & $378.63^{+0.19}_{-0.19}$ & 38 & 103.93 & 2.74 & 8.4 \\
\enddata
\tablecomments{Statistics for the transmission retrievals shown in Tables \ref{table:transmission_results} and \ref{table:transmission_cloud_model_results}. Model name (1), evidence (2), degrees of freedom (data points - number of parameters) (3), chi-squared (4), reduced chi squared (5), and sigma significance (6). Sigma significances are computed between clear vs gray, aerosol species, and aerosol models and are computed with respect to the model denoted with ---. The MgSiO3 Slab in the third group is identical to the retrieval performed in the second group was run at a different resolution than the rest of the cloud exploration retrievals, and is included to provide a sigma significance.}
\end{deluxetable*}

\clearpage
\begin{turnpage}
\startlongtable
\begin{deluxetable*}{lllllllll}
\colnumbers
\tablecolumns{9}
\tablewidth{0pt}
\tabletypesize{\tiny}
\tablecaption{Emission/Reflection Priors and Results \label{table:emission_results}}
\tablehead{ & Variable & Prior & Clear & + Reflection & Gray & + Reflection & MgSiO$_3$ & + Reflection}
\startdata
Base & R$_{\rm{p,ref}}$ [R$_{\rm{J}}$] & $\mathcal{U}$(1.11, 1.13) & $1.120^{+0.0}_{-0.0}$ &  $1.120^{+0.0}_{-0.0}$ &  $1.120^{+0.0}_{-0.0}$ & $1.120^{+0.0}_{-0.0}$ & $1.120^{+0.0}_{-0.0}$ & $1.120^{+0.0}_{-0.0}$ \\
& $\log$\, Na & $\mathcal{U}$(-12, -1) & $-8.4^{+2.6}_{-2.3}$ &  $-6.6^{+0.7}_{-0.7}$ &  $-8.2^{+2.5}_{-2.4}$ & $-6.9^{+0.9}_{-1.4}$ & $-8.3^{+2.4}_{-2.3}$ & $-6.6^{+0.7}_{-0.7}$ \\
& $\rightarrow$ $\log$\, K & $\mathcal{U}$(-13, -2) & $\cdots$ & & & & &\\
& $\log$\, H$_2$O & $\mathcal{U}$(-12, -1) & $-5.3^{+0.6}_{-1.1}$ &  $-5.4^{+0.6}_{-1.5}$ &  $-5.2^{+0.8}_{-1.2}$ & $-5.3^{+0.6}_{-1.3}$ & $-5.3^{+0.5}_{-1.0}$ &  $-5.3^{+0.6}_{-1.0}$  \\
& $\log$\, CO & $\mathcal{U}$(-12, -1) & $-5.1^{+3.3}_{-4.6}$ &  $-4.9^{+3.1}_{-4.7}$ &  $-5.7^{+3.6}_{-4.1}$ & $-5.4^{+3.5}_{-4.3}$ & $-4.7^{+3.0}_{-4.8}$ & $-5.3^{+3.4}_{-4.3}$  \\
& $\log$\, CO$_2$ & $\mathcal{U}$(-12, -1) & $-3.1^{+0.5}_{-1.5}$ &  $-3.1^{+0.5}_{-2.7}$ &  $-2.9^{+0.7}_{-0.8}$ & $-3.0^{+0.6}_{-1.4}$ & $-3.2^{+0.6}_{-2.9}$ & $-3.1^{+0.5}_{-1.2}$   \\
& $\log$\, CH$_4$ & $\mathcal{U}$(-12, -1) & $-9.1^{+1.9}_{-1.9}$ &  $-9.2^{+1.9}_{-1.8}$ &  $-8.9^{+1.9}_{-2.0}$ & $-9.1^{+1.9}_{-1.9}$ & $-9.2^{+1.9}_{-1.8}$ & $-9.1^{+1.8}_{-1.8}$  \\
& $\log$\, HCN & $\mathcal{U}$(-12, -1) & $-8.4^{+2.4}_{-2.4}$ &  $-8.5^{+2.4}_{-2.3}$ &  $-7.8^{+2.4}_{-2.7}$ & $-8.3^{+2.4}_{-2.4}$ & $-8.5^{+2.3}_{-2.3}$ & $-8.4^{+2.4}_{-2.3}$   \\
& $\log$\, NH$_3$ & $\mathcal{U}$(-12, -1) & $-9.0^{+2.0}_{-2.0}$ &  $-9.0^{+2.1}_{-2.0}$ &  $-8.8^{+2.0}_{-2.0}$ & $-8.9^{+2.0}_{-2.0}$ & $-8.9^{+2.0}_{-2.0}$ & $-8.9^{+2.0}_{-2.0}$   \\
\hline
Cloud Profiles & & & \\
\hline
Gray & log a & $\mathcal{U}$(-4, 8) & $\cdots$ & $\cdots$ & $0.2^{+3.6}_{-2.7}$ & $-2.0^{+1.2}_{-1.3}$ & $\cdots$ & $\cdots$\\
& $\gamma$ & $\mathcal{U}$(-20, 2) & $\cdots$ & $\cdots$ & $-12.3^{+6.9}_{-5.2}$ & $-9.9^{+7.2}_{-6.7}$ & $\cdots$ & $\cdots$ \\
& $\log$\,$P_{\rm{cloud}}$\, [bars] & $\mathcal{U}$(-6, 2) & $\cdots$ & $\cdots$ & $1.0^{+0.6}_{-0.8}$ &  $1.1^{+0.6}_{-0.5}$ & $\cdots$ & $\cdots$ \\
\hline
Slab &$\log$\,Y & $\mathcal{U}$(-30, -1) & $\cdots$ & $\cdots$ & $\cdots$ & $\cdots$ & $-21.6^{+7.8}_{-5.6}$ & $-21.6^{+7.6}_{-5.5}$    \\
&$\log$\,$r_{\rm{m}}$\,Y [\textmu m] & $\mathcal{U}$(-3, -1) & $\cdots$ & $\cdots$ & $\cdots$ & $\cdots$ & $-1.4^{+1.4}_{-1.1}$ & $-1.3^{+1.4}_{-1.1}$   \\
&$\log$\,$P_{\rm{top,slab}}$\,Y [bars] & $\mathcal{U}$(-6, 2) & $\cdots$ & $\cdots$ & $\cdots$ & $\cdots$ & $-1.2^{+2.4}_{-3.1}$ & $4.0^{+2.5}_{-2.6}$    \\
&$\Delta\log$\,$P$\,Y [bars] & $\mathcal{U}$(0, 8) & $\cdots$ & $\cdots$ & $\cdots$ & $\cdots$ & $3.8^{+2.8}_{-2.5}$ & $-1.4^{+2.5}_{-3.0}$   \\
\hline
Patchy Slab &$\log$\,Y & $\mathcal{U}$(-30, -1) & $\cdots$ & $\cdots$ & $\cdots$ & $\cdots$ & $-20.6^{+9.1}_{-6.2}$ & $-20.5^{+9.4}_{-6.2}$   \\
&$\log$\,$r_{\rm{m}}$\,Y [\textmu m] & $\mathcal{U}$(-3, -1) & $\cdots$ & $\cdots$ & $\cdots$ & $\cdots$ & $-1.4^{+1.3}_{-1.0}$ & $-1.4^{+1.4}_{-1.1}$  \\
&$\log$\,$P_{\rm{top,slab}}$\,Y [bars] & $\mathcal{U}$(-6, 2) & $\cdots$ & $\cdots$ & $\cdots$ & $\cdots$ & $-1.5^{+2.5}_{-2.9}$ & $-1.3^{+2.4}_{-3.0}$   \\
&$\Delta\log$\,$P$\,Y [bars] & $\mathcal{U}$(0, 8) & $\cdots$ & $\cdots$ & $\cdots$ & $\cdots$ & $3.8^{+2.6}_{-2.5}$ & $3.9^{+2.6}_{-2.6}$   \\
&$f_{\rm{cloud}}$ [$\%$] & $\mathcal{U}$(0, 1) & $\cdots$ & $\cdots$ & $\cdots$ & $\cdots$ & $0.4^{+0.3}_{-0.3}$ & $0.5^{+0.3}_{-0.3}$  \\
\hline
Uniform X &$\log$\,Y & $\mathcal{U}$(-30, -1) & $\cdots$ & $\cdots$ & $\cdots$ & $\cdots$ & $-23.1^{+5.6}_{-4.5}$ & $-22.9^{+5.5}_{-4.7}$ \\
&$\log$\,$r_{\rm{m}}$\,Y [\textmu m] & $\mathcal{U}$(-3, -1) & $\cdots$ & $\cdots$ & $\cdots$ & $\cdots$ & $-1.5^{+1.5}_{-1.0}$ & $-1.5^{+1.5}_{-1.0}$ \\
\hline 
Patchy &$\log$\,Y & $\mathcal{U}$(-30, -1) & $\cdots$ & $\cdots$ & $\cdots$ & $\cdots$ & $-20.8^{+11.8}_{-6.1}$ & $-21.5^{+8.6}_{-5.7}$ \\
Uniform X &$\log$\,$r_{\rm{m}}$\,Y [\textmu m] & $\mathcal{U}$(-3, -1) & $\cdots$ & $\cdots$ & $\cdots$ & $\cdots$ & $-1.3^{+1.4}_{-1.1}$ & $-1.4^{+1.4}_{-1.1}$ \\
&$f_{\rm{cloud}}$ [$\%$] & $\mathcal{U}$(0, 1) & $\cdots$ & $\cdots$ & $\cdots$ & $\cdots$ & $0.4^{+0.3}_{-0.3}$ & $0.4^{+0.4}_{-0.3}$ \\
\hline 
P-T Profile & & & \\
\hline
Dayside Guillot & $\log$\,$\kappa_{\rm{IR}}$ & $\mathcal{U}$(-5, 0) & $-4.8^{+0.3}_{-0.1}$ &  $-4.8^{+0.3}_{-0.1}$ &  $-4.8^{+0.3}_{-0.1}$ & $-4.8^{+0.3}_{-0.1}$ & $-4.8^{+0.3}_{-0.1}$ &  $-4.8^{+0.2}_{-0.1}$ \\
& $\log$\,$\gamma$ & $\mathcal{U}$(-4, 1) & $-0.4^{+0.1}_{-0.1}$ &  $-0.4^{+0.1}_{-0.1}$ &  $-0.5^{+0.1}_{-0.1}$ & $-0.4^{+0.1}_{-0.1}$ & $-0.5^{+0.1}_{-0.1}$ & $-0.4^{+0.1}_{-0.1}$ \\
& T$_{\rm{eq}}$ [K] & $\mathcal{U}$(600, 2400) & $1160.5^{+15.3}_{-21.0}$ &  $1158.0^{+15.8}_{-22.3}$ &  $1165.3^{+26.0}_{-20.9}$ & $1160.6^{+16.6}_{-19.8}$ & $1160.1^{+15.9}_{-19.7}$ & $1159.5^{+14.4}_{-19.9}$ \\
& T$_{\rm{int}}$ [K] & $\mathcal{U}$(0, 500) & $256.9^{+165.5}_{-168.0}$ &  $262.8^{+165.0}_{-173.2}$ &  $251.1^{+159.3}_{-159.3}$ &  $249.3^{+167.7}_{-162.3}$ & $259.6^{+159.4}_{-162.3}$ & $255.7^{+157.9}_{-159.6}$ \\
\enddata
\tablecomments{Emission and Emission + reflection retrieval priors and results discussed in \S~\ref{sec:hd189-emission-retrievals}. Base (1) represents retrieved parameters common to all retrievals, Cloud Profiles (1) represents parameters specific to certain columns. All retrievals utilized the Dayside Guillot P-T profile. Variable name (2), prior (3), and retrieved 1 sigma results for the clear retrievals with emission (4) and emission + reflection (5). Repeated for gray cloud and MgSiO$_3$ models (6-9). Gaseous abundances and P-T profile are only reported for MgSiO3 Patchy UX (emission) and Patchy Slab (emission + reflection) aerosol models, as those were the most preferred. Pressure grid for emission taken from 100 to 1e-6 bars, with a reference pressure of 1 bar. Y refers to MgSiO$_3$. Note that Na + K are one variable in emission and reflection retrievals, with log K = 0.1 log Na. For all emission and reflection retrievals, multiple scattering is turned on. For all reflection retrievals, reflection is only computed up to 5 \textmu m. Guillot profile priors are taken from \citet{Zhang2020}, Table 5 where T$_{eq} = \beta$ 1200K.  Retrievals were run at a resolution R = 10,000 with 1000 live points.}
\end{deluxetable*}
\end{turnpage}

\clearpage
\begin{deluxetable*}{lllll}
\colnumbers
\tablecolumns{11}
\tablewidth{0pt}
\tabletypesize{\footnotesize}
\tablecaption{Emission + Reflection Cloud Species Exploration Priors and Results \label{table:emission_cloud_model_results}}
\tablehead{ & Variable & Prior & SiO$_2$ & MnS}
\startdata
Patchy Slab &$\log$\,Y & $\mathcal{U}$(-30, -1) & $-19.8^{+9.7}_{-6.6}$ &  $-20.8^{+8.3}_{-6.2}$ \\
&$\log$\,$r_{\rm{m}}$\,Y [\textmu m] & $\mathcal{U}$(-3, -1) & $-1.3^{+1.3}_{-1.1}$ &  $-1.4^{+1.4}_{-1.1}$ \\
&$\log$\,$P_{\rm{top,slab}}$\,Y [bars] & $\mathcal{U}$(-8, 2) & $-1.3^{+2.2}_{-2.9}$ &  $-1.3^{+2.3}_{-3.1}$ \\
&$\Delta\log$\,$P$\,Y [bars] & $\mathcal{U}$(0, 10) &  $4.0^{+2.6}_{-2.5}$ &  $3.9^{+2.7}_{-2.5}$ \\
&$f_{\rm{cloud}}$ [$\%$] & $\mathcal{U}$(0, 1) & $0.4^{+0.3}_{-0.3}$ &  $0.5^{+0.3}_{-0.3}$ \\
\enddata
\tablecomments{Emission + Reflection retrieval priors and results for the aerosol model exploration discussed in \S~\ref{er_cloud_species}. aerosol model name (1), with variables (2), priors (3), and retrieved 1 sigma results (4-5).}
\end{deluxetable*}

\startlongtable
\begin{deluxetable*}{llll}
\colnumbers
\tablecolumns{11}
\tablewidth{0pt}
\tabletypesize{\footnotesize}
\tablecaption{Emission+Reflection Clear P-T Profile Exploration Priors and Results \label{table:pt_results}}
\tablehead{ & Variable & Prior & Clear}
\startdata
\hline 
Dayside Guillot & $\log$\,$\kappa_{\rm{IR}}$ & $\mathcal{U}$(-12, 0) & $-5.4^{+0.1}_{-0.1}$ \\
Larger Prior & $\log$\,$\gamma$ & $\mathcal{U}$(-10, 5) & $-5.9^{+4.0}_{-4.0}$ \\
& T$_{\rm{eq}}$ [K] & $\mathcal{U}$(600, 2400) & $1264.4^{+8.8}_{-8.8}$ \\
& T$_{\rm{int}}$ [K] & $\mathcal{U}$(0, 500) & $257.8^{+149.9}_{-161.2}$ \\
\hline
Line & $\log$\,$\kappa_{\rm{IR}}$ & $\mathcal{U}$(-5, 0) & $-4.7^{+0.3}_{-0.2}$ \\
& $\log$\,$\gamma$ & $\mathcal{U}$(-4, 1) & $-0.3^{+0.1}_{-0.1}$  \\
& $\log$\,$\gamma_{\rm{2}}$ & $\mathcal{U}$(-4, 1) &  $-1.4^{+1.0}_{-1.6}$  \\
& $\alpha$ & $\mathcal{U}$(0, 0.5) & $0.1^{+0.1}_{-0.1}$ \\
& $\beta$ & $\mathcal{U}$(0.5, 2) & $1.1^{+0.0}_{-0.0}$ \\
& T$_{\rm{int}}$ [K] & $\mathcal{U}$(0, 500) & $258.8^{+158.8}_{-165.0}$\\
\hline
Madhusudhan \& & a$_1$ & $\mathcal{U}$(0.02, 2) & $1.1^{+0.5}_{-0.5}$ \\
Seager & a$_2$ & $\mathcal{U}$(0.02, 2) & $0.2^{+0.1}_{-0.0}$\\
& $\log$\,$P_{\rm{1}}$ [bars]  & $\mathcal{U}$(-6, 2) & $-2.0^{+1.4}_{-2.6}$ \\
& $\log$\,$P_{\rm{2}}$ [bars]  & $\mathcal{U}$(-6, 2) & $-3.4^{+1.1}_{-1.5}$ \\
& $\log$\,$P_{\rm{3}}$ [bars]  & $\mathcal{U}$(-2, 2) & $1.3^{+0.4}_{-0.4}$ \\
& T$_{\rm{ref}}$ [K] & $\mathcal{U}$(600, 2400) & $1221.6^{+37.2}_{-39.3}$\\
\hline
Pelletier (150) & T$_{\rm{1}}$ [K] & $\mathcal{U}$(100, 3000) & $1222.7^{+1004.1}_{-716.4}$\\
& T$_{\rm{2}}$ [K] & $\mathcal{U}$(100, 3000) & $1176.1^{+615.7}_{-470.2}$\\
& T$_{\rm{3}}$ [K] & $\mathcal{U}$(100, 3000) & $1172.5^{+331.9}_{-389.1}$\\
& T$_{\rm{4}}$ [K] & $\mathcal{U}$(100, 3000) & $1162.1^{+204.2}_{-269.5}$\\
& T$_{\rm{5}}$ [K] & $\mathcal{U}$(100, 3000) & $1179.5^{+115.2}_{-139.0}$ \\
& T$_{\rm{6}}$ [K] & $\mathcal{U}$(100, 3000) & $1258.0^{+39.5}_{-61.4}$\\
& T$_{\rm{7}}$ [K] & $\mathcal{U}$(100, 3000) & $1368.8^{+22.2}_{-20.4}$\\
& T$_{\rm{8}}$ [K] & $\mathcal{U}$(100, 3000) & $1564.1^{+58.5}_{-58.9}$\\
& T$_{\rm{9}}$ [K] & $\mathcal{U}$(100, 3000) & $1775.1^{+427.1}_{-436.8}$\\
& $\sigma_{\rm{s}}$ & $\mathcal{U}$(130, 170) & $155.5^{+10.0}_{-14.1}$\\
\hline
Pelletier (550) & T$_{\rm{1}}$ [K] & $\mathcal{U}$(100, 3000) & $835.0^{+561.0}_{-463.6}$\\
& T$_{\rm{2}}$ [K] & $\mathcal{U}$(100, 3000) & $732.9^{+344.6}_{-362.5}$\\
& T$_{\rm{3}}$ [K] & $\mathcal{U}$(100, 3000) & $982.5^{+278.9}_{-364.5}$\\
& T$_{\rm{4}}$ [K] & $\mathcal{U}$(100, 3000) & $1631.1^{+44.1}_{-73.4}$\\
& T$_{\rm{5}}$ [K] & $\mathcal{U}$(100, 3000) & $540.1^{+255.3}_{-230.1}$ \\
& T$_{\rm{6}}$ [K] & $\mathcal{U}$(100, 3000) & $1378.0^{+179.2}_{-150.5}$\\
& T$_{\rm{7}}$ [K] & $\mathcal{U}$(100, 3000) & $2213.9^{+399.5}_{-353.9}$\\
& T$_{\rm{8}}$ [K] & $\mathcal{U}$(100, 3000) & $1957.9^{+615.1}_{-639.9}$\\
& T$_{\rm{9}}$ [K] & $\mathcal{U}$(100, 3000) & $1624.7^{+843.3}_{-889.4}$\\
& $\sigma_{\rm{s}}$ & $\mathcal{U}$(530, 570) & $552.5^{+11.6}_{-13.4}$ \\
\hline
Slope & T$_{\rm{phot}}$ [K] & $\mathcal{U}$(600, 2400) & $1281.8^{+30.6}_{-41.4}$ \\
& $\Delta T_{\rm{1}}$ [K] & $\mathcal{U}$(0, 1000) & $228.4^{+265.7}_{-162.3}$\\
& $\Delta T_{\rm{2}}$ [K] & $\mathcal{U}$(0, 1000) & $189.0^{+218.0}_{-133.9}$ \\
& $\Delta T_{\rm{3}}$ [K] & $\mathcal{U}$(0, 1000) & $190.0^{+209.2}_{-134.6}$\\
& $\Delta T_{\rm{4}}$ [K] & $\mathcal{U}$(0, 1000) & $137.9^{+164.4}_{-97.6}$\\
& $\Delta T_{\rm{5}}$ [K] & $\mathcal{U}$(0, 1000) & $67.2^{+105.8}_{-47.4}$\\
& $\Delta T_{\rm{6}}$ [K] & $\mathcal{U}$(0, 1000) & $81.7^{+57.8}_{-46.2}$\\
& $\Delta T_{\rm{7}}$ [K] & $\mathcal{U}$(0, 1000) & $221.3^{+75.5}_{-82.7}$\\
& $\Delta T_{\rm{8}}$ [K] & $\mathcal{U}$(0, 1000) & $477.7^{+339.9}_{-313.5}$\\
\enddata
\tablecomments{Emission + Reflection retrieval priors and results for the P-T exploration discussed in \S~\ref{er_pt_profiles}. P-T profile name (1), with variables (2), priors (3), and retrieved 1 sigma results (4). Retrievals were run at a resolution R = 5000 with 500 live points. Line profile priors are taken from \citet{Zhang2020}, Table 5. For Pelletier profile: T$_1$ represents 1e-6 bars, T$_9$ represented 1e2 bars. For Slope profile: photosphere defined at 1e-1 bars with $\Delta$ T$_1$ from 1e-6 to 1e-5 bars, $\Delta$ T$_2$ from 1e-5 to 1e-4 bars, and so forth.}
\end{deluxetable*}

\startlongtable
\begin{deluxetable*}{llllll}
\colnumbers
\tablecolumns{4}
\tablewidth{0pt}
\tabletypesize{\footnotesize}
\tablecaption{Emission Statistics \label{table:emission_stats}}
\tablehead{Model & lnZ & DoF & $\chi^2$ & $\chi^2_{red}$ & $\sigma$ Significance}
\startdata
Clear & $275.44^{+0.11}_{-0.11}$ & 22 & 26.12 & 1.19 & --- \\
Gray & $274.26^{+0.12}_{-0.12}$ & 19 & 26.11 & 1.37 & 2.1 \\
\hline
MgSiO$_3$ Slab & $275.44^{+0.11}_{-0.11}$ & 18 & 25.66 & 1.43 & 1.1\\
MgSiO$_3$ Slab (Patchy) & $275.43^{+0.11}_{-0.11}$ & 17 & 25.57 & 1.50 & 1.1 \\
MgSiO$_3$ Uniform X & $275.48^{+0.11}_{-0.11}$ & 19 & 25.72 & 1.35 & 1.8 \\
MgSiO$_3$ Uniform X (Patchy) & $274.84^{+0.11}_{-0.11}$ & 20 & 25.82 & 1.29 & --- \\
\hline
\enddata
\tablecomments{Statistics for the emission retrievals shown in Tables \ref{table:emission_results}. Model name (1), evidence (2), degrees of freedom (data points - number of parameters) (3), chi-squared (4), reduced chi squared (5), and sigma significance (6). Sigma significances are computed between clear vs gray, and MgSiO$_3$ aerosol models and are computed with respect to the model denoted with ---.}
\end{deluxetable*}

\startlongtable
\begin{deluxetable*}{llllll}
\colnumbers
\tablecolumns{4}
\tablewidth{0pt}
\tabletypesize{\footnotesize}
\tablecaption{Emission + Reflection Statistics \label{table:reflection_stats}}
\tablehead{Model & lnZ & DoF & $\chi^2$ & $\chi^2_{red}$ & $\sigma$ Significance}
\startdata
Clear & $292.31^{+0.11}_{-0.11}$ & 24 & 27.22 & 1.13 & --- \\
Gray & $290.51^{+0.12}_{-0.12}$ & 21 & 27.90 & 1.33 & 2.4\\
\hline
MgSiO$_3$ Slab & $292.33^{+0.11}_{-0.11}$ & 20 & 27.87 & 1.39 & 1.5 \\
MgSiO$_3$ Slab (Patchy) & $292.62^{+0.11}_{-0.11}$ & 19 & 27.90 & 1.47 & ---\\
MgSiO$_3$ Uniform X & $291.95^{+0.12}_{-0.12}$ & 22 & 27.02 & 1.23 & 1.8\\
MgSiO$_3$ Uniform X (Patchy) & $292.40^{+0.11}_{-0.11}$ & 21 & 27.62 & 1.32 & 1.4\\
\hline
SiO$_2$ Slab (Patchy) & $293.04^{+0.16}_{-0.16}$ & 19 & 26.91 & 1.42 & ---  \\
MnS Slab (Patchy)  & $293.05^{+0.16}_{-0.16}$ & 19 & 27.47 & 1.45 & 1.0 \\
\hline
Clear Dayside Guillot & $292.31^{+0.11}_{-0.11}$ & 24 & 27.22 & 1.13 & 2.0 \\
Clear Dayside Guillot (Larger Prior) & $286.11^{+0.17}_{-0.17}$ & 24 & 35.81 & 1.49 & 4.2 \\
Clear Line &  $293.25^{+0.16}_{-0.16}$ & 22 & 27.71 & 1.24 & --- \\
Clear Madhu & $290.47^{+0.17}_{-0.17}$ & 22 & 26.83 & 1.22 & 2.9\\
Clear Pelletier (150) & $284.43^{+0.19}_{-0.19}$ & 18 & 28.09 & 1.56 & 4.6 \\
Clear Pelletier (550) & $283.88^{+0.17}_{-0.17}$ & 18 & 43.88 & 2.44 & 4.7\\
Clear Slope & $290.50^{+0.17}_{-0.17}$ & 19 & 27.13 & 1.43 & 2.9\\
\enddata
\tablecomments{Statistics for the emission + reflection retrievals shown in Tables \ref{table:emission_results}, \ref{table:emission_cloud_model_results}, and \ref{table:pt_results}. Model name (1), evidence (2), degrees of freedom (data points - number of parameters) (3), chi-squared (4), reduced chi squared (5), and sigma significance (6). Sigma significances are computed between clear vs gray, MgSiO$_3$ aerosol models, SiO$_2$ vs MnS, and P-T profiles and are computed with respect to the model denoted with ---. Note that Clear Dayside Guillot in group 4 was run at a different resolution than the other P-T profile explorations and is equivalent to Clear in group 1, and is included to provide a sigma significance.}
\end{deluxetable*}

\startlongtable
\begin{deluxetable*}{ll}
\colnumbers
    \renewcommand{\arraystretch}{1.1}
    \tabletypesize{\footnotesize}
    \tablecolumns{2} 
    \tablecaption{Chemical Inventory \label{table:chemical_inventory}}
    \tablehead{Molecule \phantom{space} & \phantom{space}Opacity References\phantom{space}}
    \startdata
    \hline
    Na & \citet{Barklem2016} \\
    K & \citet{Barklem2016}  \\
    H$_2$O & \citet{Polyansky2018} \\
    CO & \citet{Rothman2010}\\
    CO$_2$ & \citet{Tashkun2011}\\
    CH$_4$ & \citet{Yurchenko2017}\\
    HCN & \citet{Barber2014}\\
    NH$_3$ & \citet{Coles2019}\\
    \hline
    H$_2$-H$_2$ CIA & \citet{Karman2019}\\
    H$_2$-He CIA & \citet{Karman2019}\\
    H$_2$-CH$_4$ CIA & \citet{Karman2019}\\
    H$_2$-CO$_2$ CIA & \citet{Karman2019}\\
    CO$_2$-CH$_4$ CIA & \citet{Karman2019}\\
    CO$_2$-CO$_2$ CIA & \citet{Karman2019}\\
    H$_2$ Rayeligh & \citet{Hohm1994}\\
    \enddata
    \label{tab:line-lists}
    \tablecomments{Chemical inventory used in the transmission and emission/reflection retrieval analysis. Molecule name (1), and opacity reference (2). Na and K no longer have a wing cutoff, and are modeled in \citet{Marley2021}. Transmission: Opacities loaded in from 0.2 to 30 \textmu m at R = 10,000, T ranging from 500 to 1300K. Emission/Reflection: Opacities loaded in from 0.2 to 1 \textmu m linearly spaced with 1000 points, and then 1 to 30 \textmu m at R = 10,000, T ranging from 100 to 3000K.}
\end{deluxetable*}


\clearpage
\bibliographystyle{aasjournal}
\bibliography{sample.bib}

\end{document}